\documentclass[a4paper, 11 pt]{article}
\pdfoutput = 1

\usepackage{jheppub, bm}
\usepackage[utf8]{inputenc}
\usepackage{xcolor}

\DeclareMathOperator*{\SumInt}{%
\mathchoice%
  {\ooalign{$\displaystyle\sum$\cr\hidewidth$\displaystyle\int$\hidewidth\cr}}
  {\ooalign{\raisebox{.14\height}{\scalebox{.7}{$\textstyle\sum$}}\cr\hidewidth$\textstyle\int$\hidewidth\cr}}
  {\ooalign{\raisebox{.2\height}{\scalebox{.6}{$\scriptstyle\sum$}}\cr$\scriptstyle\int$\cr}}
  {\ooalign{\raisebox{.2\height}{\scalebox{.6}{$\scriptstyle\sum$}}\cr$\scriptstyle\int$\cr}}
}

\usepackage{graphicx} 
\usepackage[export]{adjustbox}
\usepackage{subcaption} 
\usepackage{comment}

\usepackage{amssymb,mathrsfs}
\usepackage{slashed} 
\usepackage{physics} 
\usepackage[capitalise]{cleveref} 

\usepackage{hyperref}

\newcommand{\nn}{\nonumber}
\newcommand{\bef}{\begin{figure}[t]\centering}
\newcommand{\eef}{\end{figure}}
\def\bea#1\eea{\begin{align}#1\end{align}}
\def \be  {\begin{equation}}
\def \ee  {\end{equation}}
\def \ba  {\begin{eqnarray}}
\def \ea  {\end{eqnarray}}

\allowdisplaybreaks[2]
	
\title{Anisotropic jet broadening and jet shape}
	
\author[a]{Weiyao Ke,} 
\author[b]{John Terry,}
\author[b]{and Ivan Vitev}

\affiliation[a]{Institute of Particle Physics and Key Laboratory of Quark and Lepton Physics (MOE), Central China Normal University, Wuhan, Hubei, 430079, China}
\affiliation[b]{Theoretical Division, Los Alamos National Laboratory, Los Alamos, New Mexico 87545, USA}

\emailAdd{weiyaoke@ccnu.edu.cn}
\emailAdd{jdterry@lanl.gov}
\emailAdd{ivitev@lanl.gov}

\abstract{In this paper, we explore the use of jet substructure as a way of probing phenomena which break the isotropic behavior of jets, such as jet propagation through an anisotropically flowing quark-gluon plasma or spin correlations. We introduce two novel observables for this purpose: the azimuthal-dependent jet broadening and the azimuthal-dependent jet shape, which generalize the traditional isotropic substructure studies. Using Soft-Collinear Effective Theory, we explicitly calculate the jet functions associated with these observables with a standard jet axis and with a Winner-Take-All jet axis in both the resummed and fixed order limits. While our analysis first and foremost establishes the formalism for the azimuthal-dependent jet substructure, it  also brings to light new results for jet substructure in the azimuthally integrated case, such as the semi-inclusive jet function  and the exclusive jet shape for the Winner-Take-All axis, and the jet broadening in the fixed order region. As an illustrative example for the new formalism we demonstrate that the azimuthal-dependent jet broadening can be used as a direct probe of the transversity parton distribution function in deep inelastic scattering.
}

\begin{document}
\preprint{
LA-UR-24-28428\\
\makebox[\textwidth][r]{INT-PUB-24-035}
}
\maketitle

\section{Introduction}\label{Introduction}

Jet substructure observables correlate the pattern of hadronic activity contained within the jet with the quark and gluon dynamics \cite{Seymour:1993mx,Seymour:1994by,Butterworth:2002tt}, providing a method to uncover the properties of the partonic interaction. They can be used to test beyond the Standard Model theories~\cite{Soper:2010xk, Godbole:2014cfa, Chen:2014dma, Adams:2015hiv} and serve as a probe of the fundamental strong interaction dynamics~\cite{Britzger:2017maj, CMS:2013vbb, ATLAS:2017qir, ATLAS:2015yaa, CMS:2014mna,ATLAS:2021qnl, ATLAS:2013pbc, CMS:2014qtp, CMS:2016lna, AbdulKhalek:2020jut, Harland-Lang:2017ytb, Pumplin:2009nk, Watt:2013oha}. Various jet substructure observables have been proposed in the literature, see for instance \cite{Marzani:2019hun} for a review, but in this paper we focus on jet angularities and jet broadening~\cite{Berger:2003iw, Almeida:2008yp, Larkoski:2014pca} and the jet shape~\cite{Ellis:1992qq}. Jet angularities allow for a  smooth transition between jet mass and jet broadening studies, while the jet shape yields information for the energy profile of the jet. While these observables contain a wealth of information on the momenta of jet constituents, both observables integrate over the azimuthal dynamics of the jet, providing only isotropic information on its substructure.

In Ref.~\cite{Collins:1992kk} it was demonstrated that non-perturbative effects correlate the transverse spin of a quark with the distribution of final-state hadrons. Namely, fewer hadrons with a higher average transverse momentum relative to the parent quark are produced in one direction, while a higher multiplicity of hadrons which carry on average smaller transverse momentum are produced in the opposite direction in the azimuthal plane. In this way, the transverse spin of the initial state quark breaks the isotropic symmetry of hadronization and results in an anisotropic distribution of hadrons in the jet. This Collins effect and the resulting distribution of hadrons are governed by the non-perturbative fragmentation function (FF) known as the Collins Fragmentation Function. More recently, perturbative computations associated with the correlation of the quark spin and final-state hadron transverse momentum have also been performed~\cite{Arratia:2020nxw}.  Di-hadron fragmentation functions~\cite{Pitonyak:2023gjx,Cocuzza:2023oam,Cocuzza:2023vqs} and correlations associated with the spin of the final-state hadron can also be shown to give rise to similar effects. With the ongoing interest in mapping out the spin structure of nucleons and understanding polarization dependent fragmentation function, see for instance~\cite{AbdulKhalek:2021gbh,Bacchetta:2006tn,Callos:2020qtu,Echevarria:2020hpy,Kang:2023big}, jet substructure measurements which are sensitive to the transverse spin of the jet can offer novel insight into spin-dependent dynamics. Within a transverse momentum dependent framework this has been studied for instance in~\cite{Kang:2020xyq}, where azimuthal asymmetries around the jet are considered for a single hadron close to the jet axis. Lastly in Refs.~\cite{Lai:2022aly,Liu:2021ewb}, it was demonstrated that hadronization effects lead to correlations between the transverse spin and the direction of the Winner-Take-All (WTA) jet axis. 

Anisotropic jet substructure can also arise due to color flow in a preferential direction, such as in the case of jet pull~\cite{Gallicchio:2010sw,Larkoski:2019fsm,Larkoski:2019urm}. In~\cite{Gallicchio:2010sw} the authors introduced the idea of jet pull as a means of discriminating between Higgs and gluon initiated $b$ di-jets in hadronic collisions. In the case of Higgs initiated b-dijets, the tree level process generates a di-jet pair which is a color dipole. This color dipole generates a preferential direction for radiative emissions, resulting in an anisotropic contribution to the jet substructure. In the gluon initiated dijets however, color dipoles are generated between the jets and the incoming hadronic beams, results in a two separate pulls between the beam remnants and the jets.

In reactions with nuclei, parton showers are modified via radiative and collisional processes in matter. Qualitatively, they are broader and softer than the ones in the vacuum. Quantitatively, jet observables in such mesoscopic collisions, were shown to be sensitive diagnostics of the transport properties of matter and non-local quantum coherence effects~\cite{Landau:1953um,Migdal:1956tc}. Jet substructure can be further sensitive to redistribution of the energy deposited in the medium by hard quarks and gluons. This has motivated studies of several substructure observables in heavy ion reactions~\cite{Vitev:2008rz,Chien:2015hda,Salgado:2003rv,Cao:2020wlm,Brewer:2018mpk,Li:2019dre,Chien:2016led,Mehtar-Tani:2016aco,Barata:2023bhh,Wang:2020ukj,Li:2022tcr,Li:2017wwc,Devereaux:2023vjz,Budhraja:2023rgo}.          
Lastly, there has been recent interest in using jet substructure to probe the flow of the quark-gluon plasma (QGP). Off-center collisions of energetic nuclei result in  pressure gradients and velocities of the QGP that drive anisotropic collisional interactions and bremsstrahlung  patterns that have been derived theoretically~\cite{Sadofyev:2021ohn} and visualized at the parton branching level~\cite{Barata:2023zqg}. These theoretical developments offer a promising avenue for exploring how energetic particles interact with flowing, strongly coupled plasmas~\cite{Antiporda:2021hpk,Bahder:2024jpa}. Thus far, a first-principles approach for azimuthal-anisotropic jet substructure that can be generalized to reactions with nuclei has not been developed. 

In this paper, we address this deficiency and define azimuthal-dependent generalizations of jet shape and jet angularities. Using Soft-Collinear Effective Theory (SCET), we derive the factorization and resummation of the jet functions associated with experimental observables and demonstrate that they share the same anomalous dimension as the exclusive and semi-inclusive jet functions. This allows for novel jet substructure studies to be performed across a large number of processes. While our analysis focuses on defining the azimuthal-dependent jet functions, it also brings to light new insights into jet substructure in the azimuthally integrated case. Specifically, we derive the semi-inclusive jet function for the Winner-Take-All axis, calculate jet broadening in the fixed order region, and develop the exclusive jet shape with the WTA axis. In Sec.~\ref{sec:ang}, we calculate the azimuthal-dependent jet broadening in the fixed order and resummed limits using a Standard Jet Axis (SJA) and a WTA jet axis. In Sec.~\ref{sec:shape}, we calculate the azimuthal-dependent jet shape in the resummed and fixed order limits. In Sec.~\ref{sec:Model}, we demonstrate that anisotropic effects of the jet substructure allow us to directly probe the transversity PDF in DIS. We conclude in Sec.~\ref{sec:Conclusions}. Selected results on jet functions, anomalous dimensions, and partonic cross sections are collected in the Appendix.

\section{Azimuthal-dependent jet broadening}\label{sec:ang}
\begin{figure}
    \centering
    \includegraphics[width = 0.3\textwidth]{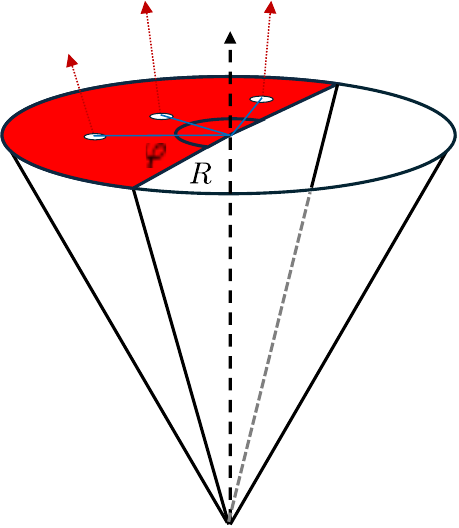}
    \caption{The azimuthal-dependent jet broadening. Emissions which pass through the red surface are constituents of $J_{\varphi}$ and thus contribute to the observable.}
    \label{fig:Azi-Ang}
\end{figure}

Traditional jet angularities in hadronic colliders \cite{Berger:2003iw,Almeida:2008yp,Ellis:2010rwa,Hornig:2016ahz} are defined by
\begin{align}
    \bar{\tau}_a = \frac{1}{p_T} \sum_{i \in J} p_T^i \left|\Delta \mathcal{R}_{iJ}\right|^{2-a}\,,
\end{align}
where the bar is used to denote that azimuthal averaging has been performed, $p_T^i$ is the transverse momentum of particle $i$ with respect to the beam axis, $p_T$ denotes the transverse momentum of the jet, and the inter-particle distance is defined as
\begin{align}
    \Delta \mathcal{R}_{ij} = \sqrt{\Delta y_{ij}^2+\Delta \phi_{ij}^2}\,.
\end{align}
The final-state particles are collected into the jet if they satisfy the condition that $\Delta \mathcal{R}_{ij} < \mathcal{R}$, where $\mathcal{R}$ is defined in terms of the jet pseudo-rapidity and jet radius as
\begin{align}
    \mathcal{R} = \frac{R}{\cosh{\eta}}\,.
\end{align}
The parameter $a$ in the definition of the angularity controls the exact measurement that is performed on the jet. Thus, for different values of $a$ the relevant EFT that is used for the factorization and resummation is altered. In the region $a \lesssim 2$ the dominant contribution to the angularity comes from the large light-cone momentum component of the constituent particles. In this case, the angularity provides information on their  energy distribution  and  we have the power counting that $\bar{\tau}_a \sim 1$, indicating that SCET I is the proper EFT. Alternatively, in the region where $a \sim 1$ the particles measure the distribution of momentum that is transverse to the direction of the jet. Thus, the proper EFT in this region is SCET II. Lastly, in the region where $a \sim 0 $ the measurement is dominated by the small light-cone momentum component of the particles of the jet, proving information for the mass distribution of the jet. Thus, this subset of the angularities has a power counting that $\bar{\tau}_a \sim \lambda^2$, once again requiring the use of SCET I.

In this paper, we generalize the jet angularities to provide additional sensitivity in the azimuthal plane. In the traditional definition of  jet angularities we see that while the jet broadening characterizes the mean width of a jet, this measurement cannot be used to determined if the hadrons in the jet are more spread out in one direction or another. Thus, while jet angularities can be used to address the question as to how effects such as flavor can alter the overall width of a jet, they are not ideal in obtaining information associated with the direction of the effect in anisotropic scenarios. To address this issue, we define the set of jet angularities
\begin{align}
    \tau_a = \frac{1}{p_T} \sum_{i \in J_\varphi} p_T^i \left|\Delta \mathcal{R}_{iJ}\right|^{2-a}\,,
    \qquad
    \upsilon_a\left(\varphi\right) = \frac{\partial}{\partial \varphi}\tau_a\,,
\end{align}
where $J_\varphi$ defines a wedge of the jet which is specified with the azimuthal angles $\varphi_i$ and $\varphi$, see Fig.~\ref{fig:Azi-Ang}. Here $\tau_a$ represents the jet angularity while $\upsilon_a$ represents the differential jet angularity. In this paper, we focus on the jet broadening, which has $a = 1$. For simplicity, we will take the notation $\tau_1 \equiv \tau$. 

We now note that there are two ways to define a jet wedge, these can either be iterative or by slicing the jet. In the iterative approach, the jet is reconstructed. At each point in the history of the jet, if the azimuthal angle of an emission falls within a range that is determined by the wedge, the particle remains in the jet wedge. The jet slicing definition of the wedge is much simpler. In this case, a jet is constructed and all particles that fall within an azimuthal range are contained within the wedge. For the jet broadening, we will use the slicing method while for the jet shape, we will use the iterative method.

\subsection{Power counting the jet broadening}
\begin{figure}
\centering
\includegraphics[width=\linewidth]{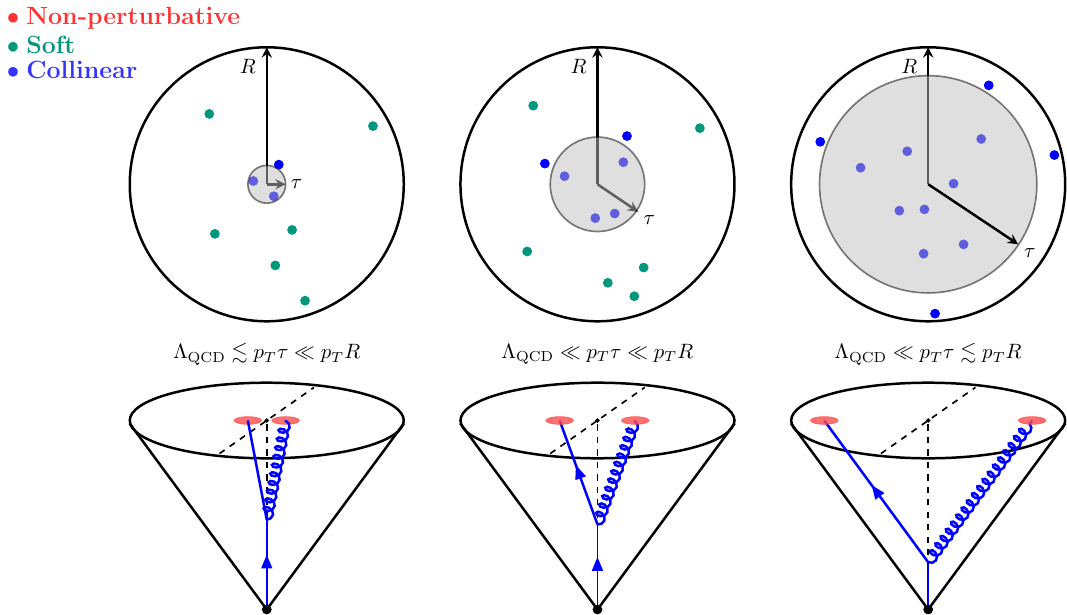}
\caption{Illustration of the EFT modes that obey different power counting and contribute to the non-perturbative, resummed, and fixed order
regions.}
\label{fig:cartoon-jet-broadening}
\end{figure}

The precise definition of the jet broadening requires careful consideration of several scales,  including the hard scale for the process $p_T$, the jet scale $p_T R$, the jet broadening scale $p_T \tau$, and $\Lambda_{\rm QCD}$. Focusing on $p_T \tau$, in this section we'll cover three regions, where $\Lambda_{\rm QCD} \lesssim p_T \tau \ll p_T R \ll p_T$, where $\Lambda_{\rm QCD} \ll p_T \tau \ll p_T R \ll p_T$, and where $\Lambda_{\rm QCD} \ll p_T \tau \lesssim p_T R \ll p_T$. We will refer to these three regions as the non-perturbative, resummed, and fixed order regions, respectively, and illustrate them in Fig.~\ref{fig:cartoon-jet-broadening}. For $\tau \ll R$ logarithms of $\tau/R$ are resummed by re-factorizing the jet function into hyper-collinear and soft-collinear functions and solving the rapidity renormalization group evolution equations. 

Events in which no collinear emissions occur outside of the jet are characterized as `exclusive', see for instance Ref.~\cite{Ellis:2010rwa,Hornig:2016ahz} for the computation of jet angularities with $a < 1$, while events in which collinear emissions occur outside of the jet are characterized as `semi-inclusive', see for instance \cite{Kang:2016mcy,Kang:2016ehg,Kang:2018qra}.
The exclusive jet broadening functions are given by the SCET I matrix elements as
\begin{align}
\hat{G}_{q\, {\rm alg}}^{\rm axis}\left(\varphi, \tau, R,\omega_J,\mu, \zeta \right) &= \frac{1}{2N_c}{\rm Tr} \left[\frac{\slashed{\bar n}_J}{2}
\langle 0| \delta\left(\omega_J - \bar n_J\cdot {\mathcal P} \right)\delta\left(\tau - \hat \tau\right) \chi_{n_J}(0)  |J_\varphi\rangle \right. \nn \\
&  \quad \times \langle J_\varphi|\bar \chi_{n_J}(0) |0\rangle \Big] \,, \\
\hat{G}_{g\, {\rm alg}}^{\rm axis}\left(\varphi, \tau, R,\omega_J,\mu, \zeta \right) &= - \frac{\omega_J}{2(N_c^2-1)}
\langle 0| \delta\left(\omega_J - \bar n_J\cdot {\mathcal P} \right)\delta\left(\tau - \hat \tau\right) {\mathcal B}_{n_J \perp \mu}(0) 
 |J_\varphi\rangle  \nn \\
 & \quad \times \langle J_\varphi|{\mathcal B}_{n_J \perp}^\mu(0)  |0\rangle\,,  
\end{align}
where $\mu$ is the renormalization group scale, $\zeta$ is the Collins-Soper parameter, the wedge of the jet runs from some initial value to $\varphi$, and the hat is used to denote that the function is bare. We note that the Collins-Soper scale enters only in the resummed and non-perturbative regions; however, to simplify the notation, we include the Collins-Soper parameter in the functional dependence in all regions. The jet boundary here depends on the definition of the jet axis, which may be either a Standard Jet Axis (SJA) or a Winner-Take-All axis (WTA), and the algorithm used, which may be either anti-$\rm{k_T}$ or cone. Additionally, in these expressions, the measurement operator for the jet broadening is defined by 
\begin{align}
    \hat{\tau} |J_\varphi, X\rangle = \frac{1}{\omega_J} \sum_{i \in J_\varphi} p_{i\perp} |J_\varphi,X\rangle \,,
\end{align}
where $p_{i \perp}$ denotes the magnitude of the transverse momentum of particle $i$ with respect to the direction of the jet, and the $\hat{\tau}$ operator acts only on the sub-space of the jet wedge.

For the semi-inclusive jet broadening functions there are two characteristic collinear momenta, the large component of the parton $\omega$ and the large component of the jet $\omega_J$. The semi-inclusive jet broadening functions are given by \cite{Kang:2018qra}
\begin{align}
\hat{G}_{q\, {\rm alg}}^{\rm axis}\left(z,\varphi, \tau, R,\omega_J,\mu, \zeta \right) &= \frac{z}{2N_c} \SumInt_X {\rm Tr} \Big[\frac{\slashed{\bar n}_J}{2}
\langle 0| \delta\left(\omega - \bar n_J\cdot {\mathcal P} \right)\delta\left(\tau - \hat \tau\right) \chi_{n_J}(0)  |J_\varphi, X\rangle \nn \\
&\quad \times \langle J_\varphi, X|\bar \chi_{n_J}(0) |0\rangle \Big]  \,,
\\
\hat{G}_{g\, {\rm alg}}^{\rm axis}\left(z,\varphi, \tau, R,\omega_J,\mu, \zeta \right) &= - \frac{z\,\omega}{2(N_c^2-1)} \SumInt_X 
\langle 0|  \delta\left(\omega - \bar n_J\cdot {\mathcal P} \right)\delta\left(\tau - \hat \tau\right) {\mathcal B}_{n_J\perp \mu}(0) 
 |J_\varphi, X\rangle \nn \\
 &\quad \times \langle J_\varphi, X|{\mathcal B}_{n_J\perp}^\mu(0)  |0\rangle\,, 
\end{align}
where the exclusive and semi-inclusive functions are differentiated from one another by the $z$ argument, which is the energy fraction of the jet with respect to the parent parton.

In the region where $\tau \ll R$, there are two modes which contribute to the jet broadening, namely the hyper-collinear and soft-collinear modes that have the power counting
\begin{align}
    p_{c}^\mu \sim \omega_J\left(\lambda^2, 1, \lambda\right)\,,
    \qquad
    p_{s}^\mu \sim \omega_J\frac{\lambda}{R}\left(R^2, 1, R\right)\,,
\end{align}
where $\lambda \sim \tau$ is our power counting parameter.  Based on this power counting, we see that these modes have the natural scales 
\begin{align}
    \mu_c \sim \mu_s \sim \omega_J \lambda\,,
    \qquad
    \nu_c \sim \omega_J\,,
    \qquad
    \nu_s \sim \omega_J \frac{\lambda}{R}\,,
\end{align}
where  $\mu$ denotes the usual renormalization group scale while $\nu$ is the rapidity renormalization group scale. Under a BPS field redefinition, the momentum space jet function in the resummed and non-perturbative regions can be written as 
\begin{align}
    \hat{G}_{i\, {\rm alg}}^{\rm axis}\left(\varphi, \tau, R, \omega_J, \mu, \zeta\right) & = H_{i\,{\rm alg}}^{\rm axis}\left(\nu_s, \nu_c\right) \mathscr{C}^{\rm axis}\left[\hat{C}_i^{\rm axis}\, \hat{S}_i^{\rm axis}\right] \left(\varphi, \tau, R, \omega_J, \mu, \zeta\right)\,,
    \\
    \hat{G}_{i\, {\rm alg}}^{\rm axis}\left(\varphi, \tau, R, \omega_J, \mu, \zeta\right) & = \hat{H}_{i\,{\rm alg}}^{\rm axis}\left(\nu_s, \nu_c\right) \mathscr{C}^{\rm axis}\left[\hat{C}_i^{\rm axis}\, \hat{S}_i^{\rm axis}\, F_i^{\rm axis}\right] \left(\varphi, \tau, R, \omega_J, \mu, \zeta\right)\,,
\end{align}
where $i$ denotes the flavor of the parton that initiates the jet and we have dropped power correction terms for simplicity. The four functions that enter into these expressions are the matching coefficient function $H$, the hyper-collinear function $C$, the soft-collinear function $S$, and the non-perturbative contribution $F$. The hyper-collinear and soft-collinear functions depend trivially on the choice of the axis, namely they will contain subtleties associated with factors of 2. In using the fixed order expressions for the hyper-collinear and soft-collinear functions in the asymptotic region, the perturbative expression contains logs of the rapidity scales $\nu_c$ and $\nu_s$. The matching function removes these logarithms such that the resummed and fixed order regions match. Lastly, the non-perturbative function does not depend on the jet constraint based on the power counting. However, the non-perturbative effects do depend on the jet axis used. In this expression, the dependence on the jet algorithm is encoded in the power suppressed terms. In the next section we will provide details as to why this occurs. In what follows we will  use $\bm{k}_\perp$ to denote the transverse momentum of the initial-state hyper-collinear parton radiation in the jet. In this paper, we consider this momentum to be in $2-2\epsilon$ dimensions. The integration in this variable accounts for the recoil of the collinear mode due to soft radiation~\cite{Dokshitzer:1998kz, Kang:2017mda} and largely complicates the full computation of the jet function due to interplay between the polar angle of the collinear mode and the jet constraint (aka geometric effects), see for instance~\cite{Cal:2019hjc} where the recoil effects were considered for the jet shape. The Winner-Take-All (WTA) jet axis greatly simplifies the expression since the direction of the jet axis is insensitive to soft emission \cite{Larkoski:2014uqa}. Thus, the expression for a WTA jet axis is the same as that for the SJA except that the soft-collinear function for the WTA axis will always be proportional to a delta function in the transverse momentum so that the integration can be performed trivially. Lastly, we have used the short-hand notation that
\begin{align}
     \mathscr{C}^{\rm SJA}& \left[\hat{C}_i^{\rm SJA}\, \hat{S}_i^{\rm SJA}\right] \left(\varphi, \tau, R, \omega_J, \mu, \zeta\right) = \int d^{2-2\epsilon}k_\perp\, \int d\tau_c\, d\tau_s\,  \delta\left(\tau -\tau_c-\tau_s\right)  \nn \\
    & \times \hat{C}_i^{\rm SJA}\left(\varphi, \tau_c, \bm{k}_\perp, \omega_J, \mu, \frac{\zeta}{\nu^2}\right)\, \hat{S}_i^{\rm SJA} \left(\varphi, \tau_s, -\bm{k}_\perp, R, \omega_J, \mu, \nu\right) \,,
    \\
     \mathscr{C}^{\rm SJA} & \left[\hat{C}_i^{\rm SJA}\, \hat{S}_i^{\rm SJA}\, F_i\right] \left(\varphi, \tau, R, \omega_J, \mu, \zeta\right) = \int d^{2-2\epsilon}k_\perp\, \int d\tau_c\, d\tau_s\, d\tau_{\rm NP} \delta\left(\tau -\tau_c-\tau_s-\tau_{\rm NP}\right) \nn \\
    & \times \hat{C}_i^{\rm SJA}\left(\varphi, \tau_c, \bm{k}_\perp, \omega_J, \mu, \frac{\zeta}{\nu^2}\right)\, \hat{S}_i^{\rm SJA} \left(\varphi, \tau_s, -\bm{k}_\perp, R, \omega_J, \mu, \nu\right)\, F_i^{\rm SJA}\left(\tau_{\rm NP}, \zeta_0, \zeta\right) \,.
\end{align}
For the WTA axis, we have
\begin{align}
    \mathscr{C}^{\rm WTA} & \left[\hat{C}_i^{\rm WTA}\, \hat{S}_i^{\rm WTA}\right] \left(\varphi, \tau, R, \omega_J, \mu, \zeta\right) = \int d\tau_c\, d\tau_s\,  \delta\left(\tau -\tau_c-\tau_s\right) \nn \\
    &\times  \hat{C}_i^{\rm WTA}\left(\varphi, \tau_c, \omega_J, \mu, \frac{\zeta}{\nu^2}\right)\, \hat{S}_i^{\rm WTA} \left(\varphi, \tau_s, R, \omega_J, \mu, \nu\right) \,,
    \\
    \mathscr{C}^{\rm WTA} & \left[\hat{C}_i^{\rm WTA}\, \hat{S}_i^{\rm WTA}\, F_i\right] \left(\varphi, \tau, R, \omega_J, \mu, \zeta\right) = \int d\tau_c\, d\tau_s\,d\tau_{\rm NP}\,  \delta\left(\tau -\tau_c-\tau_s-\tau_{\rm NP}\right) \nn \\
    &\times  \hat{C}_i^{\rm WTA}\left(\varphi, \tau_c, \omega_J, \mu, \frac{\zeta}{\nu^2}\right)\, \hat{S}_i^{\rm WTA} \left(\varphi, \tau_s, R, \omega_J, \mu, \nu\right)\, F_i^{\rm WTA}\left(\tau_{\rm NP}, \zeta_0, \zeta\right) \,.
\end{align}
In performing this refactorization, the measurement function in the non-perturbative region factorizes as
\begin{align}
    \hat{\tau} | J \rangle = \hat{\tau}_c | J_c, J_s, J_{\rm NP} \rangle + \hat{\tau}_s | J_c, J_s, J_{\rm NP} \rangle + \hat{\tau}_{\rm NP} | J_c, J_s, J_{\rm NP} \rangle\,,
\end{align}
where $J_c$, $J_s$, and $J_{\rm NP}$ represent the sub-spaces of the hyper-collinear, soft-collinear, and non-perturbative contributions to the jet, while the operators $\hat{\tau}_{c/s/{\rm NP}}$ act on those sub-spaces. The refactorization in the resummed region is similar except that one takes $\hat{\tau}_{\rm NP} \rightarrow 0$.

The matrix elements that define the hyper-collinear function is given by
\begin{align}
\hat{C}_{q}^{\rm SJA} \left(\varphi, \tau,\bm{k}_\perp,\omega_J,\mu, \frac{\zeta}{\nu^2} \right) = & \frac{1}{2N_c} {\rm Tr} \Bigg[\frac{\slashed{\bar n}_J}{2}
\langle 0| \delta^{2-2\epsilon}\left( \bm{\mathcal{P}}_\perp - \bm{k}_\perp\right)\,\delta\left(\omega_J - \bar n_J\cdot {\mathcal P} \right)\,\delta\left(\tau - \hat \tau\right) \nn \\
&  \times  \chi_{n_J}(0)  |J_\varphi\rangle \langle J_\varphi|\bar \chi_{n_J}(0) |0\rangle \Bigg] \,, \\
\hat{C}_{g}^{\rm SJA} \left(\varphi, \tau,\bm{k}_\perp,\omega_J,\mu, , \frac{\zeta}{\nu^2} \right) = & - \frac{\omega_J}{2(N_c^2-1)} \langle 0| \delta^{2-2\epsilon}\left( \bm{\mathcal{P}}_\perp - \bm{k}_\perp\right)\,\delta\left(\omega_J - \bar n_J\cdot {\mathcal P} \right)\,\delta\left(\tau - \hat \tau\right) \nn \\
& \times {\mathcal B}_{n_J \perp \mu}(0) 
 |J_\varphi\rangle \langle J_\varphi|{\mathcal B}_{n_J \perp}^\mu(0)  |0\rangle\,, 
 \\
 \hat{C}_{q}^{\rm WTA} \left(\varphi, \tau,\omega_J,\mu, \frac{\zeta}{\nu^2} \right) = & \frac{1}{2N_c} {\rm Tr} \Bigg[\frac{\slashed{\bar n}_J}{2}
\langle 0| \delta^{2-2\epsilon}\left( \bm{\mathcal{P}}_\perp\right)\,\delta\left(\omega_J - \bar n_J\cdot {\mathcal P} \right)\,\delta\left(\tau - \hat \tau\right) \nn \\
& \times \chi_{n_J}(0)  |J_\varphi\rangle \langle J_\varphi|\bar \chi_{n_J}(0) |0\rangle \Bigg] \,, \\
\hat{C}_{g}^{\rm WTA} \left(\varphi, \tau,\omega_J,\mu, , \frac{\zeta}{\nu^2} \right) = & - \frac{\omega_J}{2(N_c^2-1)} \langle 0| \delta^{2-2\epsilon}\left( \bm{\mathcal{P}}_\perp\right)\,\delta\left(\omega_J - \bar n_J\cdot {\mathcal P} \right)\,\delta\left(\tau - \hat \tau\right) \nn \\
& \times {\mathcal B}_{n_J \perp \mu}(0) 
 |J_\varphi\rangle \langle J_\varphi|{\mathcal B}_{n_J \perp}^\mu(0)  |0\rangle\,, 
\end{align}
where the collinear fields are those of SCET II. Note that there is no $k_T$ dependence in the WTA case.
$S$ denotes the soft-collinear function which is constructed from the vacuum expectation values of SCET II  Wilson lines as 
\begin{align}
    \hat{S}_q ^{\rm SJA}\left(\varphi, \tau_s, \bm{k}_\perp, R, \omega_J, \mu, \nu\right) = & \frac{1}{N_c} \SumInt_{J_s, X_s} \operatorname{Tr} \left \langle 0 \left | S_{n_J}^\dagger(0) S_{\bar{n}_J}(0) \delta\left(\tau_s- \hat{\tau}_s\right)\right | J_s, X_s\right \rangle \nn \\
    &\times \left \langle J_s, X_s \left | S_{\bar{n}_J}^\dagger(0) S_{n_J}(0) \right | 0 \right \rangle \delta^{2-2\epsilon}\left(\bm{k}_\perp - \sum_{i \in J} \bm{k}_{i, \perp}\right)\,, 
    \\
    \hat{S}_q ^{\rm WTA}\left(\varphi, \tau_s, R, \omega_J, \mu, \nu\right) = & \frac{1}{N_c} \SumInt_{J_s, X_s} \operatorname{Tr} \left \langle 0 \left | S_{n_J}^\dagger(0) S_{\bar{n}_J}(0) \delta\left(\tau_s- \hat{\tau}_s\right)\right | J_s, X_s\right \rangle \nn \\
    &\times \left \langle J_s, X_s \left | S_{\bar{n}_J}^\dagger(0) S_{n_J}(0) \right | 0 \right \rangle \,, 
\end{align}
while there are analogous expressions for the gluon soft-collinear function with Wilson lines in the adjoint representation of SU(3).

The jet function involves a convolution of the hyper-collinear and soft-collinear contributions to the total jet broadening. This convolution is simplified by working in Laplace space. The exclusive jet broadening function becomes
\begin{align}
    \hat{G}_{i\, {\rm alg}}^{\rm axis} & \left(\varphi, \tau, R, \omega_J, \mu, \zeta\right)  = \hat{H}_{i\,{\rm alg}}^{\rm axis}(\nu_s, \nu_c)\int d^{2-2\epsilon}k_\perp \int_{c-i\infty}^{c+i\infty} \frac{d\kappa}{2\pi i} \exp\left(\frac{\kappa \tau}{e^{\gamma_E}}\right) \nn \\
    \label{eq:BR-fact-Lap}
    &\times \hat{\mathcal{C}}_i^{\rm axis}\left(\varphi, \kappa, \bm{k}_\perp, \omega_J, \mu,\frac{\zeta}{\nu^2}\right) \hat{\mathcal{S}}_i^{\rm axis}\left(\varphi, \kappa, -\bm{k}_\perp, R, \omega_J, \mu,\nu\right)\,,
\end{align}
while there is an analogous expression for the WTA jet function, but without the transverse momentum integration. Here $\kappa$ denotes the Laplace conjugate variable to $\tau$. In this expression, the hyper-collinear and soft-collinear functions can be written as Laplace transforms of the momentum space distributions
\begin{align}
    \hat{\mathcal{C}}_i^{\rm axis}\left(\varphi, \kappa, \bm{k}_\perp, \omega_J, \mu,\frac{\zeta}{\nu^2}\right)  = &\int_0^\infty d\tau \exp\left(-\frac{\kappa \tau}{e^{\gamma_E}}\right) \hat{C}_i^{\rm axis}\left(\varphi, \tau, \bm{k}_\perp, \omega_J, \mu,\frac{\zeta}{\nu^2}\right)\,, 
    \\
    \hat{\mathcal{S}}_i^{\rm axis}\left(\varphi, \kappa, \bm{k}_\perp, R, \omega_J, \mu,\nu\right)  = &\int_0^\infty d\tau \exp\left(-\frac{\kappa \tau}{e^{\gamma_E}}\right) \hat{S}_i^{\rm axis}\left(\varphi, \tau, \bm{k}_\perp, R, \omega_J, \mu,\nu\right)\,.
\end{align}

In our definition of the jet function, we omitted emissions outside of the jet. These additional emissions result in the jet carrying a portion $z$ of the parent parton's momentum. As discussed in~\cite{Kang:2018qra}, from an EFT perspective, these contributions can be considered by accounting for hard-collinear emissions, which have a power counting $p_{hc}\sim \omega_J (R^2, 1, R)$ and thus are wide angled enough to occur outside of the jet and can contribute to energy loss of the fragmenting parton. In this case the jet function can be written as
\begin{align}
    \hat{G}_{i\, {\rm alg}}^{\rm axis} & \left(z, \varphi, \tau, R, \omega_J, \mu, \zeta\right) = \hat{h}_{ij}\left(z, R, \omega_J, \mu\right)\, \hat{G}_{j\, {\rm alg}}^{\rm axis}\left(\varphi, \tau, R, \omega_J, \mu, \zeta\right)\,, 
\end{align}
where $i$ represents the flavor of the parent parton while $j$ represents the flavor of the jet. The function $h$ is the hard-collinear function, which does not depend on the jet algorithm as we will later discuss.

\subsection{Kinematics at one-loop}\label{subsec:kinematics}
In a rotationally invariant about an axis system, the tree level functions are simply given by
\begin{align}
    \hat{G}^{{\rm axis}\, (0)}_{i\, {\rm alg}}\left(\varphi, \tau, R, \omega_J, \mu, \zeta\right) &= \frac{\varphi}{2\pi}\delta\left(\tau\right)\,,
    \\
    \hat{G}^{{\rm axis}\, (0)}_{i\, {\rm alg}}\left(z, \varphi, \tau, R, \omega_J, \mu, \zeta\right) &= \frac{\varphi}{2\pi}\delta\left(\tau\right)\delta\left(1-z\right)\,.
\end{align}
The interpretation of this expression is that a parton that moves in the direction of the jet axis has a probability $\varphi/2\pi$ of entering into the jet wedge. 

Before we begin computing the perturbative ingredients at one-loop, we first touch on a subtlety in the definition of the jet direction. In our matrix elements for the jet function, we enforce that the jet has momentum 
\begin{align}
    \label{eq:frame}
    P_J^\mu = \omega_J \frac{\bar{n}^\mu}{2}\,.
\end{align}
At one-loop, if an emission is inside the jet, the parent parton then has a momentum
\begin{align}
    l^\mu = \omega_J \frac{n^\mu}{2} + n\cdot l \frac{\bar{n}^\mu}{2}\,,
\end{align}
where $n\cdot l \sim \lambda^2 \omega_J$. This constraint is fixed by Eq.~\eqref{eq:frame}. However, if a particle with momentum $q$ leaves the jet, the momentum of the particle which remained in the jet carries the jet momentum and we have
\begin{align}
    (l-q)^\mu = \omega_J \frac{n^\mu}{2}\,,
\end{align}
which is once again fixes the frame and is enforced by Eq.~\eqref{eq:frame}. The exclusive jet broadening function contains emissions only within the jet and, therefore, we use only the first frame discussed. The computation of the semi-inclusive jet broadening with a SJA however requires the use of both frames. For simplicity, we first discuss the exclusive jet broadening function and then discuss the semi-inclusive jet broadening function.

The expression for the exclusive jet broadening function can be written in terms of the Lorentz invariant amplitudes as
\begin{align}
    \hat{G}^{\rm SJA\, (1)}_{i\, {\rm alg}} \left(\varphi, \tau, R, \omega_J, \mu, \zeta\right)  = &2 \int dx\, d\Phi_{l q}\, (2\pi)^{d-1}\delta\left(\omega_J-\bar{n}_J\cdot l\right)\,\delta^{d-2}\left(\bm{l}_\perp\right) \nn \\
    & \times \left | \overline{\mathcal{M}}\right|_{i}^2\,  \Theta_{(jk)\, {\rm alg}}^{\rm SJA}\, \delta_\tau^{\rm SJA}\, \delta\left(1-x - \frac{\bar{n}\cdot q}{\bar{n}\cdot l}\right)\,,
\end{align}
where $x$ and $(1-x)$ are the momentum fractions of the two partons in the jet, the $\delta_\tau$ term measures the broadening of the jet wedge and it's precise form will be discussed later, and $\Theta_{\rm alg}^{\rm SJA}$ is the jet constraint. Here we use the index $i$ to denote the parent parton while we use $j$ and $k$ to denote the final-state partons. We use the $(jk)$ subscript to mean that both partons are restricted to be within the jet. The anti-${\rm k_T}$ algorithm measures the angle of the final-state partons with respect to one another while the cone algorithm measures the angle of each parton with respect to the jet. The form of these restrictions are known to be \cite{Ellis:2010rwa}
\begin{align}
    \Theta_{{\rm cone} (jk)}^{\rm SJA} &=  \Theta\left(\omega_J\, x\, \tan{\frac{R}{2}}-q_\perp\right)\,\Theta\left(\omega_J\, (1-x)\tan{\frac{R}{2}} - q_\perp\right)\,,
    \\
    \Theta_{{\rm{k_T}} (jk)}^{\rm SJA} &= \Theta\left(\omega_J\, x\, (1-x)\tan{\frac{R}{2}} - q_\perp\right)\,,
\end{align}
where $q_\perp$ represents the transverse momentum of the partons with respect to the parent parton. The frame is fixed in this expression by the delta function in the transverse momentum of the initial-state parton and the delta function that fixes the initial-state energy of the incoming parton. In this expression, we have used the short-hand notation for the phase space measure
\begin{align}
    d\Phi_{lq} = \frac{d^d l}{(2\pi)^{d-1}} \frac{d^d q}{(2\pi)^{d-1}} \delta\left(q^2\right) \delta\left(\left(l-q\right)^2\right)\,.
\end{align}
The squared amplitudes read
\begin{align}
    \label{eq:Mq}
    \left | \overline{\mathcal{M}}\right|_{q}^2  = &\; \frac{g_s^2}{2}C_F  \left(\frac{\mu^2 e^{\gamma_E}}{4\pi}\right)^\epsilon\, \frac{1}{l^4}\, \operatorname{Tr}\left[\frac{\slashed{\bar{n}}_J}{2}\,\slashed{l}\,\gamma_\alpha \,\slashed{k} \,\gamma_\beta \, \slashed{l} \right] \Delta^{\alpha \beta}\left(q\right) \,,
    \\
    \left | \overline{\mathcal{M}}\right|_{g}^2 = &\; \omega_J \frac{g_s^2}{2} \left(\frac{\mu^2 e^{\gamma_E}}{4\pi}\right)^\epsilon\, \frac{1}{l^4}\, \frac{1}{2-2\epsilon}\,\Delta_{\mu\nu}(l) \, \Bigg \{ n_f \operatorname{Tr}\left[\slashed{k}\gamma^\mu \slashed{q} \gamma^\nu \right]   \nn \\
    & 
    \label{eq:Mg}
    \; + C_A V^{\nu \sigma \beta}\left(l, -q, -k\right)\, V^{\rho \mu \alpha}\left(q, -l, k\right)\, \Delta_{\rho \sigma}\left(q\right)\, \Delta_{\alpha \beta}\left(k\right)\Bigg \} \,,
\end{align}
where $k = l-q$, and the Feynman rules are
\begin{align}
    \Delta_{\mu\nu}\left(q\right) & = -g_{\mu \nu}+\frac{q_\mu \bar{n}_\nu+q_\nu \bar{n}_\mu}{\bar{n}\cdot q}-\frac{q^2}{\left(\bar{n}\cdot q\right)^2}\bar{n}_\mu \bar{n}_\nu\,,
    \\
    V^{\mu \nu \rho}\left(p,q,r\right) & = \left(p-q\right)^\rho g^{\mu\nu}+\left(q-r\right)^\mu g^{\nu \rho}+\left(r-p\right)^\nu g^{\rho\mu}\,.
\end{align}
To simplify discussion on the semi-inclusive jet function, we now decompose it in terms of the contribution inside and outside of the jet as
\begin{align}
    \hat{G}^{\rm SJA\, (1)}_{i\, {\rm alg}} \left(z, \varphi, \tau, R, \omega_J, \mu, \zeta\right) = & \hat{G}^{\rm SJA\, (1)}_{i\, {\rm alg\, in}} \left(z, \varphi, \tau, R, \omega_J, \mu, \zeta\right) \nn \\
    & +\hat{G}^{(1)}_{i\, {\rm out}} \left(z, \tau, R, \omega_J, \mu, \zeta\right)\,,
\end{align}
where we have purposefully dropped the dependence on the jet axis and the algorithm dependence for emissions outside of the jet. We will discuss this point later in this section. The expression for the contribution inside the jet can be related to the exclusive jet broadening through the relation
\begin{align}
    \hat{G}^{\rm SJA\, (1)}_{i\, {\rm alg\, in}} \left(z, \varphi, \tau, R, \omega_J, \mu, \zeta\right) & =  \delta\left(1-z\right)\,\hat{G}^{\rm SJA\, (1)}_{i\, {\rm alg}} \left(\varphi, \tau, R, \omega_J, \mu, \zeta\right)\,.
\end{align}
For the case of an emission outside of the jet, one of the final-state partons points in the direction of the jet. Since the anti-$\rm k_T$ algorithm measures the angle relative to the final-state partons while the cone algorithm measures the angle with respect to the jet axis, these two restrictions are identical to one another for an emission outside of the jet.  We then have
\begin{align}
    \Theta_{{\rm alg} (j)}^{\rm SJA} & = \Theta_{{\rm alg} (k)}^{\rm SJA} =  \Theta\left(q_\perp - \omega_J \left(1-z\right)\, \tan{\frac{R}{2}} \right)\,,
\end{align}
such that the `out' terms do not depend on the jet algorithm. When we discuss the WTA jet axis we will demonstrate that these contributions are also axis independent. The expression for the emissions outside of the jet then take on the form 
\begin{align}
    \hat{G}^{(1)}_{i\, {\rm out}} & \left(z, \varphi, \tau, R, \omega_J, \mu, \zeta\right) = 2\int dx\, d\Phi_{l q}\, (2\pi)^{d-1}  \delta\left(z - \frac{\omega_J}{\omega}\right) \,\delta^{d-2}\left(\bm{l}_\perp-\bm{q}_\perp\right) \nn \\
    &\times \left | \overline{\mathcal{M}}\right|_i^2\,  \left[\Theta^{\rm SJA}_{{\rm alg}\, (j)} +\Theta^{\rm SJA}_{{\rm alg}\, (k)} \right]\,\delta\left(\omega_J-\bar{n}_J\cdot (l-q)\right)\,  \delta\left(\tau\right)\, \delta\left(1-x-\frac{\bar{n}\cdot q}{\bar{n}\cdot l}\right)\,.
\end{align}
In this expression $q_\perp$ represents the transverse momentum of the parton leaving the jet with respect to the parton that remains in the jet. Thus, the cone and anti-${\rm k_T}$ jet constraints have the same meaning and we have the jet constraints
\begin{align}
    \Theta_{{\rm alg} (j)}^{\rm SJA} & = \Theta_{{\rm alg} (k)}^{\rm SJA} =  \Theta\left(q_\perp - \omega_J \left(1-z\right)\, \tan{\frac{R}{2}} \right)\,.
\end{align}
For the WTA scheme, the jet direction is determined by the most energetic particle. This will result in an identical frame being used for both emissions inside and outside the jet as well as an independence on the jet algorithm. The jet constraint for the WTA jet is given by
\begin{align}
    \Theta_{(jk)}^{\rm WTA} & = \Theta\left(\omega_J\, \left(1-x\right)\, \tan{\frac{R}{2}} - q_\perp\right)\, \Theta\left(x-\frac{1}{2}\right)\,.
\end{align}
In this expression, we enforce that the parton with momentum fraction $x$ is more energetic. Because the jet broadening is only sensitive to the transverse component of the momentum, we will then have to sum over the cases that distinguish which particle is more energetic. Taking all of this into account, we have the expression for the exclusive WTA jet broadening
\begin{align}
    \hat{G}^{\rm WTA\, (1)}_{i} \left(\varphi, \tau, R, \omega_J, \mu, \zeta\right) = & 2 \int dx\, d\Phi_{l q}\, (2\pi)^{d-1}\delta\left(\omega_J-\bar{n}_J\cdot l\right)\,\delta^{d-2}\left(\bm{l}_\perp-\bm{q}_\perp\right) \nn \\
    &\times \left | \overline{\mathcal{M}}\right|_i^2\,  \Theta^{\rm WTA}\, \delta_\tau^{\rm WTA}+ x \rightarrow (1-x)\,,
\end{align}
where the term associated with switching $x$ and $1-x$ accounts for interchanging which parton is more energetic. 

The semi-inclusive jet broadening function can be written in terms of contributions within and outside of the jet as
\begin{align}
    \hat{G}^{\rm WTA\, (1)}_{i\, {\rm alg}} \left(z, \varphi, \tau, R, \omega_J, \mu, \zeta\right) = & \hat{G}^{\rm WTA\, (1)}_{i\, {\rm alg\, in}} \left(z, \varphi, \tau, R, \omega_J, \mu, \zeta\right) \nn \\
    & +\hat{G}^{(1)}_{i\, \rm{out}} \left(z, \tau, R, \omega_J, \mu, \zeta\right)\,.
\end{align}
Once again, the emissions inside the jet can be related to the exclusive jet broadening function through the relation
\begin{align}
    \hat{G}^{\rm WTA\, (1)}_{i\, {\rm in}} \left(z, \varphi, \tau, R, \omega_J, \mu, \zeta\right) & = \delta\left(1-z\right)\hat{G}^{\rm WTA\, (1)}_{i} \left(\varphi, \tau, R, \omega_J, \mu, \zeta\right)\,,
\end{align}
while the emissions outside are the same as those with a SJA. This can be seen simply since the frame used for both a SJA and a WTA axis are identical if the emission is outside of the jet. 

\begin{figure}
    \centering
    \includegraphics[width = \textwidth]{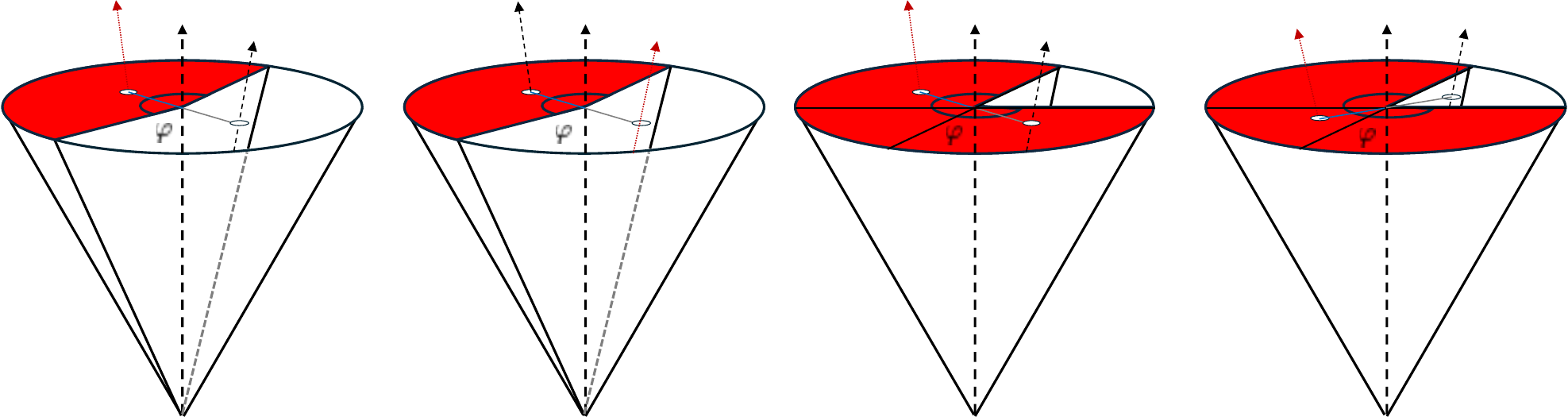}
    \caption{Left two diagrams represent the contributions with $\varphi < \pi$ while the right two diagrams correspond to $\varphi > \pi$. The red and black lines exiting the jet represent the two partons at NLO.}
    \label{fig:shape-graphs}
\end{figure}

After accounting for these details, we find that the exclusive jet broadening functions take on the form
\begin{align}
    \hat{G}^{{\rm axis}\, (1)}_{i\, {\rm alg}}\left(\varphi, \tau, R, \omega_J, \mu, \zeta\right) &= \sum_{j,k} \int d\Phi \left[ \hat{P}_{(jk)i}\left(x,q_\perp,\epsilon\right)\right] \delta_{\tau}^{\rm axis} \, \Theta_{\rm alg}^{\rm axis}\,,
\end{align}
where all of the axis and algorithm dependence is encoded in the measurement function and the jet algorithm constraint, with the exception of the interchange of $x$ and $1-x$ that must be performed for the WTA case. In this expression, the splitting functions are given in the Appendix and we have use the short-hand notation for the phase space integral 
\begin{align}
    \int d\Phi = \int d^{d-2}q_\perp\,  dx\,.
\end{align}
For the semi-inclusive jet broadening functions, we have 
\begin{align}
    \hat{G}^{{\rm axis}\, (1)}_{i\, {\rm alg}}\left(z, \varphi, \tau, R, \omega_J, \mu, \zeta\right) = & \delta\left(1-z\right)\, \hat{G}^{{\rm axis}\, (1)}_{i\, {\rm alg}}\left(\varphi, \tau, R, \omega_J, \mu, \zeta\right) \nn \\
    & +\hat{G}^{(1)}_{i\, \rm{out}} \left(z, \tau, R, \omega_J, \mu, \zeta\right)\,.
\end{align}
The last thing that we now touch on are the measurement functions. For a SJA, we find
\begin{align}
    \delta_{\tau}^{\rm SJA} = & \Theta\left(\pi-\varphi\right) \left[\Theta\left(\varphi>\varphi_q>0\right)+ \Theta\left(\pi+\varphi> \varphi_q > \pi\right)\right]\, \delta\left(\tau-\frac{q_\perp}{\omega_J}\right) \nn \\
    & +\Theta\left(\varphi-\pi\right)\left[\Theta\left(\varphi-\pi>\varphi_q>0\right) + \Theta\left(\varphi > \varphi_q>\pi\right) \right] \delta\left(\tau-\frac{2q_\perp}{\omega_J}\right) \nn \\
    & + \Theta\left(\varphi-\pi\right) \left[ \Theta\left(\pi > \varphi_q > \varphi-\pi\right) + \Theta\left(2\pi > \varphi_q > \varphi\right) \right] \delta\left(\tau-\frac{q_\perp}{\omega_J}\right)  \,.
\end{align}
The logic behind this expression is that for a SJA if $\varphi < \pi$ then only a single parton can enter the jet wedge. As a result, the jet broadening is simply given by $q_\perp/\omega_J$. However, there are two contributions for $\varphi > \pi$ since we must sum over which parton enters the wedge. These contribution are the left two subfigures of Fig.~\ref{fig:shape-graphs}. For $\varphi > \pi$, we can have either that one or two partons enters into the jet wedge. These contributions are represented by the right two figures of Fig.~\ref{fig:shape-graphs}. In the next section, we will perform the integration in detail.

For a WTA jet, one parton is always along the direction of the jet and thus will not contribute to the jet broadening. In this case, the measurement function for the jet angularity is simply given by the expression
\begin{align}
    \delta_{\tau}^{\rm WTA} = \Theta\left(\varphi-\varphi_q\right)\, \delta\left(\tau-\frac{q_\perp}{\omega_J}\right)\,.
\end{align}

\subsection{The fixed order region}
We will now derive the jet broadening in both the fixed order region for both the azimuthal-dependent and azimuthaly integrated cases in this subsection.

For the SJA, we saw that either one or two partons can enter into the jet wedge while in the case of the WTA jet axis, we can either have a single parton within the wedge along the jet direction or we can one parton along the direction of the jet and the other parton within the jet wedge. In this section, we will begin by computing the exclusive jet broadening function for the case of the SJA with a single parton and then with both partons. We then compute the exclusive jet broadening functions with a single parton and then with both partons. Lastly, we compute the expressions for the semi-inclusive jet broadening functions.

To perform the integration of the jet function in the fixed order region, we employ the identity
\begin{align}
    x^{-1-2\epsilon} \ln(x) = -\frac{1}{4\, \epsilon^2}\delta(x)+\left(\frac{\ln{x}}{x}\right)_+-2\epsilon \left(\frac{\ln^2{x}}{x}\right)_+ + \mathcal{O}\left(\epsilon^2\right)\,,
\end{align}
where $x \rightarrow 0$ is regularized by the plus distributions. Next, we note that if one parton is in the jet, the jet constraint restricts
\begin{align}
    \tau_{\rm max}^{\rm k_T} = \frac{1}{2}\tan{\frac{R}{2}}\,,
    \qquad
    \tau_{\rm max}^{\rm cone} = \tau_{\rm max}^{\rm WTA} = \tan{\frac{R}{2}}\,.
\end{align}

Taking this into consideration, for the standard jet axis with an anti-${\rm k_T}$ algorithm and a two partons, we find that
\begin{align}
    \hat{G}^{{\rm SJA}\, (1)}_{q\, {\rm k_T}} & \left(\tau, R,\omega_J, \mu, \zeta\right) = \frac{\alpha_s C_F}{2\pi}\frac{1}{\tau_{\rm max}^{\rm k_T}}\Bigg[ \nn \\
    & -\left(\frac{3}{\tau_{\rm k}}\right)_+\sqrt{1- \tau_{\rm k}}+\left(\frac{8}{\tau_{\rm k}}\right)_+\ln \left(\sqrt{1-\tau_{\rm k}}+1\right)-\left(\frac{4}{\tau_{\rm k}}\ln \left(\tau_{\rm k}\right)\right)_+ \nn \\
   & +\delta (\tau_{\rm k}) \left(\frac{1}{\epsilon ^2}-\frac{L_R}{\epsilon }+\frac{3}{2 \epsilon
   }+\frac{L_R^2}{2}-\frac{3 L_R}{2}-\frac{\pi ^2}{12}+\frac{1}{2}-8 \ln ^2(2)+6 \ln (2)\right)\Bigg]\,, \\
    \hat{G}^{{\rm SJA}\, (1)}_{g\, {\rm k_T}} & \left(\tau, R,\omega_J, \mu, \zeta\right) = \frac{\alpha_s}{2\pi}\frac{1}{ \tau_{\rm max}^{\rm k_T}}\Bigg\{ \frac{\beta_0}{2} \left(\frac{1}{2}\sqrt{1-\tau_{\rm k}}-\left(\frac{2}{\tau_{\rm k}}\right)_+\sqrt{1-\tau_{\rm k}}\right) \nn \\
    & +C_A\left(-\frac{3}{4}\sqrt{1-\tau_{\rm k}} + 8\left(\frac{1}{\tau_{\rm k}}\right)_+\ln\left(1+\sqrt{1-\tau_{\rm k}}\right)-4\left(\frac{\ln{\tau_{\rm k}}}{\tau_{\rm k}}\right)_+ \right) \nn \\
    & +\delta\left(\tau_{\rm k}\right) \Bigg[C_A\left(\frac{1}{\epsilon^2}-\frac{L_R}{\epsilon}+\frac{L_R^2}{2}-\frac{\pi^2}{12}+\frac{11}{12}-8\ln^2(2)\right) \nn \\
    &  +\frac{\beta_0}{2}\left(\frac{1}{\epsilon}-L_R-\frac{1}{2}+4\ln(2)\right)\Bigg] 
    \Bigg\} \,,
\end{align}
where we have use the short hand that $\tau_{\rm k} = \tau/\tau_{\rm max}^{\rm k_T}$ and we note that the divergences at the origin are regulated by the plus prescription. Here we have also used the short-hand that $L_R = 2 \ln\left(\frac{\omega}{\mu} \tan{\frac{R}{2}} \right)$. The results for a single partons entering into the jet wedge can be obtained from this expression simply by taking $\tau_{\rm k} \rightarrow 2 \tau_{\rm k}$ in the finite terms. By studying these expressions, one can show that by taking $\tau_k \rightarrow 1$, the jet broadening always vanishes, as expected. 

For cone jets, we find
\begin{align}
    \hat{G}^{{\rm SJA}\, (1)}_{q\, {\rm Cone}} & \left(\tau, R,\omega_J, \mu, \zeta\right) = \frac{\alpha_s C_F}{2\pi}\frac{1}{\tau^{\rm{Cone}}_{\rm max}}\Bigg[-3 \left(\frac{1}{\tau_{\rm c}}\right)_++8\left(\frac{\tanh^{-1}(1-\tau_{\rm c})}{\tau_{\rm c}}\right)_++3 \nn \\
   & +\delta (\tau_{\rm c}) \left(\frac{1}{\epsilon
   ^2}-\frac{L_R}{\epsilon}+\frac{3}{2 \epsilon}+\frac{L_R^2}{2}-\frac{3 L_R}{2}-\frac{\pi ^2}{12}+\frac{1}{2}-2\ln^2(2)+3 \ln(2)\right)\Bigg] \,, \\
    \hat{G}^{{\rm SJA}\, (1)}_{g\, {\rm Cone}} & \left(\tau, R,\omega_J, \mu, \zeta\right) = \frac{\alpha_s}{2\pi}\frac{1}{ \tau_{\rm max}^{\rm Cone}}\Bigg\{ \frac{\beta_0}{2} \left(\frac{\tau_c^2}{2}-\frac{3}{2}\tau_c+3-\left(\frac{2}{\tau_c}\right)_+\right) \nn \\
    & +C_A\left(-\frac{3}{4}\tau_c^2+\frac{9}{4}\tau_c-\frac{3}{2}+8\left(\frac{\tanh^{-1}(1-\tau_c)}{\tau_c}\right)_+ \right) \nn \\
    & +\delta\left(\tau_{\rm c}\right) \Bigg[C_A\left(\frac{1}{\epsilon^2}-\frac{L_R}{\epsilon}+\frac{L_R^2}{2}-\frac{\pi^2}{12}+\frac{11}{12}-2\ln^2(2)\right) +\frac{\beta_0}{2}\left(\frac{1}{\epsilon}-L_R-\frac{1}{2}+2\ln(2)\right)\Bigg] 
    \Bigg\} \,,
\end{align}
where for cone jets we have $\tau_{\rm c} = \tau/\tau_{\rm max}^{\rm cone}$. Lastly, for WTA jets, we find
\begin{align}
    \hat{G}^{{\rm WTA}\, (1)}_i \left(\tau, R,\omega_J, \mu, \zeta\right) = \hat{G}^{{\rm SJA}\, (1)}_{i\, {\rm Cone}} \left(\tau, R,\omega_J, \mu, \zeta\right)\,.
\end{align}
To compute the azimuthal-dependent jet broadening, we now note that the result for the SJA can be obtained from the above expressions simply by considering the dependence on the azimuthal angle and through the redefinition that $\tau^{\rm max} \rightarrow \tau^{\rm max}/2$. Thus,  we have
\begin{align}
    & \hat{G}^{{\rm SJA}\, (1)}_{i\, {\rm alg}} \left(\varphi, \tau, R,\omega_J, \mu, \zeta\right) = \frac{\varphi}{2\pi}\Theta\left(\pi-\varphi\right) \hat{G}^{{\rm SJA}\, (1)}_{i\, {\rm alg}} \left(\frac{\tau}{2}, R,\omega_J, \mu, \zeta\right) + \Theta\left(\varphi-\pi\right) \nn \\
    & \times \left[\left(\frac{2\pi-\varphi}{2\pi}\right) \hat{G}^{{\rm SJA}\, (1)}_{i\, {\rm alg}} \left(\frac{\tau}{2}, R,\omega_J, \mu, \zeta\right)+\left(\frac{\varphi-\pi}{\pi}\right) \hat{G}^{{\rm SJA}\, (1)}_{i\, {\rm alg}} \left(\tau, R,\omega_J, \mu, \zeta\right)\right]\,.
\end{align}
Due to our treatment of the emissions along the jet direction, the terms that are proportional to a delta function are suppressed by a factor of $\varphi/2\pi$ for any value of $\varphi$. This will play a role in our discussion on renormalization group consistency of the jet wedges.

Due to the simplicity of the WTA measurement function, for this definition of the jet axis, we have trivially that
\begin{align}
    \hat{G}^{{\rm WTA}\, (1)}_{i\, {\rm alg}} \left(\varphi, \tau, R,\omega_J, \mu, \zeta\right) = \frac{\varphi}{2\pi}\hat{G}^{{\rm WTA}\, (1)}_{i\, {\rm alg}} \left(\tau, R,\omega_J, \mu, \zeta\right)\,.
\end{align}

For the semi-inclusive jet broadening function, we have to consider emissions outside of the jet. For the SJA with the anti-${\rm k_T}$ we have
\begin{align}
    \hat{G}^{{\rm SJA}\, (1)}_{q\, {\rm k_T}} & \left(z, \tau, R,\omega_J, \mu, \zeta\right) = \frac{\alpha_s C_F}{2\pi}\frac{1}{\tau_{\rm max}^{\rm k_T}}\Bigg[\left(P_{qq}(z)+P_{gq}(z)\right)\left(\frac{1}{\epsilon}-L_R-2 \ln\left(1-z\right)_+\right) \delta\left(\tau_{\rm k}\right) \nn \\
    &+\delta\left(1-z\right)\left(-\left(\frac{3}{\tau_{\rm k}}\right)_+\sqrt{1-\tau_{\rm k}}-4\left(\frac{\ln{\tau_{\rm k}}}{\tau_{\rm k}}\right)_++ \left(\frac{8}{\tau_{\rm k}}\right)_+\ln\left(1+\sqrt{1-\tau_{\rm k}}\right)\right) \nn \\
    & + \delta\left(\tau_{\rm k}\right) \delta\left(1-z\right) \left(\frac{1}{2}-8\ln^2(2)+6 \ln(2)\right) -\delta\left(\tau_{\rm k}\right) \Bigg] \,, 
    \\
    \hat{G}^{{\rm SJA}\, (1)}_{g\, {\rm k_T}} & \left(z, \tau, R,\omega_J, \mu, \zeta\right) = \frac{\alpha_s}{2\pi}\frac{1}{\tau_{\rm max}^{\rm k_T}}\Bigg[\left(C_A P_{gg}(z)+n_fP_{qg}(z)\right)\left(\frac{1}{\epsilon}-L_R-2 \ln\left(1-z\right)_+\right) \delta\left(\tau_{\rm k}\right) \nn \\
    &+C_A\delta\left(1-z\right)\left(\frac{1}{6}\sqrt{1-\tau_{\rm k}}-\frac{11}{3}\left(\frac{1}{\tau_{\rm k}}\right)_+\sqrt{1-\tau_k}-4\left(\frac{\ln{\tau_{\rm k}}}{\tau_{\rm k}}\right)_+ \right.\nn \\
    & \left. +\left(\frac{8}{\tau_{\rm k}}\right)_+\ln\left(1+\sqrt{1-\tau_{\rm k}}\right)\right) \nn \\
    & +n_f\left(\frac{2}{3}\left(\frac{1}{\tau_{\rm k}}\right)_+\sqrt{1-\tau_{\rm k}}-\frac{1}{6}\sqrt{1-\tau_{\rm k}}\right)\delta\left(1-z\right) -2n_f z(1-z)\delta\left(\tau_{\rm k}\right)\nn \\
    & + \delta\left(\tau_{\rm k}\right) \delta\left(1-z\right) \left(C_A\left(\frac{22}{3}\ln(2)-8\ln^2(2)\right)+n_f\left(\frac{1}{6}-\frac{4}{3}\ln(2)\right)\right) \Bigg] \,, 
\end{align}
For a cone jet, we have
\begin{align}
    \hat{G}^{{\rm SJA}\, (1)}_{q\, {\rm cone}} & \left(z, \tau, R,\omega_J, \mu, \zeta\right) = \frac{\alpha_s C_F}{2\pi}\frac{1}{\tau_{\rm max}^{\rm cone}}\Bigg[\left(P_{qq}(z)+P_{gq}(z)\right)\left(\frac{1}{\epsilon}-L_R-2 \ln\left(1-z\right)_+\right) \delta\left(\tau_{\rm c}\right) \nn \\
    &+\delta\left(1-z\right)\left(-\left(\frac{3}{\tau_{\rm c}}\right)_+-4\left(\frac{\ln{\tau_{\rm c}}}{\tau_{\rm c}}\right)_++ \left(\frac{4}{\tau_{\rm c}}\right)_+\ln\left(2-\tau_{\rm c} \right) +3 \right) \nn \\
    & + \delta\left(\tau_{\rm c}\right) \delta\left(1-z\right) \left(\frac{1}{2}-2\ln^2(2)+3 \ln(2)\right) -\delta\left(\tau_{\rm c}\right) \Bigg] \,, 
    \\
    \hat{G}^{{\rm SJA}\, (1)}_{g\, {\rm cone}} & \left(z, \tau, R,\omega_J, \mu, \zeta\right) = \frac{\alpha_s}{2\pi}\frac{1}{\tau_{\rm max}^{\rm cone}}\Bigg[\left(C_A P_{gg}(z)+n_fP_{qg}(z)\right)\left(\frac{1}{\epsilon}-L_R-2 \ln\left(1-z\right)_+\right) \delta\left(\tau_{\rm c}\right) \nn \\
    &+C_A\delta\left(1-z\right)\left(\frac{\tau_c^2}{6}-\frac{\tau_c}{2}-\frac{11}{3}\left(\frac{1}{\tau_c}\right)_++4+8\left(\frac{\tanh^{-1}(1-\tau_c)}{\tau_c}\right)_+\right) \nn \\
    & +n_f\left(-\frac{\tau_c^2}{6}+\frac{\tau_c}{2}+\frac{2}{3}\left(\frac{1}{\tau_c}\right)_+-1\right)\delta\left(1-z\right) -2n_f z(1-z)\delta\left(\tau_{\rm c}\right)\nn \\
    & + \delta\left(\tau_{\rm c}\right) \delta\left(1-z\right) \left(C_A\left(\frac{11}{3}\ln(2)-2\ln^2(2)\right)+n_f\left(\frac{1}{6}-\frac{2}{3}\ln(2)\right)\right) \Bigg] \,, 
\end{align}
Lastly, for WTA jets, we find
\begin{align}
    \hat{G}^{{\rm WTA}\, (1)}_i \left(z, \tau, R,\omega_J, \mu, \zeta\right) = \hat{G}^{{\rm SJA}\, (1)}_{i\, {\rm Cone}} \left(z, \tau, R,\omega_J, \mu, \zeta\right)\,.
\end{align}

\subsection{The soft-collinear contribution}
In this section, we will now compute the soft-collinear function in Eq.~\eqref{eq:BR-fact-Lap}.

In defining a wedge in the jet, emissions along the jet axis can be treated as either being partially or totally contained within the wedge. The scheme for the treatment however, must respect the matching relations between the regions of different power counting. At tree-level this reads
\begin{align}
    \hat{G}_{i\, {\rm alg}}^{\rm{axis}\, (0)}\left(\varphi, \tau, R, \omega_J, \mu, \zeta\right) = & \hat{H}_{i\, {\rm alg}}^{{\rm axis}\, (0)}(\nu_s, \nu_c) \nn \\
    &\times \mathscr{C}^{\rm axis} \left[\hat{C}_i^{\rm{axis}\, (0)}\, \hat{S}_i^{\rm{axis}\, (0)}\right] \left(\varphi, \tau, R, \omega_J, \mu, \zeta\right) \,.
\end{align}
In this paper, we will take the scheme where any emissions from the jet and hyper-collinear sectors along the jet direction are suppressed by a factor of $\varphi/2\pi$ while any soft-collinear emissions along the direction of the jet are not suppressed by this factor. Under this scheme, we have the tree level expressions for the jet and hyper-collinear functions
\begin{align}
    \hat{G}_i^{\rm{axis}\, (0)}\left(\varphi, \tau, R, \omega_J, \mu, \zeta\right) & = \frac{\varphi}{2\pi}\delta\left(\tau\right)\,,
    \\
    \hat{H}_{i\, {\rm alg}}^{{\rm axis}\, (0)}(\nu_s, \nu_c) & = 1\,,  
    \\
    \hat{C}_i^{\rm{axis}\, (0)}\left(\varphi, \tau_c, \bm{0}_\perp, R, \omega_J, \mu, \frac{\zeta}{\nu^2}\right) & = \frac{\varphi}{2\pi}\delta\left(\tau_c\right)\,, 
\end{align}
where we have used the fact that the tree level soft-collinear function should scale like a delta function in the transverse momentum. From these expressions, the tree level soft-collinear function is given by
\begin{align}
    \hat{S}_i^{\rm{axis}\, {(0)}} \left(\varphi, \tau_s, \bm{k}_\perp, R, \omega_J, \mu, \nu\right) = \delta\left(\tau_s\right)\, \delta^{2-2\epsilon}\left(\bm{k}_\perp\right)\,.
\end{align}
The above expression is motivated by the fact that the tree level soft-collinear function is associated with no emissions and, thus, cannot be suppressed by the angular suppression factor.

In this section, we begin by deriving the soft-collinear function for the azimuthally integrated case. We then use this result to construct the azimuthal-dependent one. Accounting for the effects of soft-recoil, the one-loop expression for the jet function can be written as
\begin{align}
    \label{eq:G-one-loop}
    \hat{G}_{i\, {\rm alg}}^{\rm{SJA}\, (1)} \left(\tau, R, \omega_J, \mu, \zeta\right) = & \hat{H}_{i\, {\rm alg}}^{{\rm SJA}\, (1)}(\nu_s, \nu_c)\, \delta\left(\tau\right) + \hat{C}_i^{\rm{SJA}\, (1)} \left(\tau, \bm{0}_\perp, \omega_J, \mu, \frac{\zeta}{\nu^2}\right) \nn \\
    &+ \mathscr{C}^{\rm SJA} \left[\hat{C}_i^{{\rm SJA}\, (0)}\, \hat{S}_i^{{\rm SJA}\, (1)}\right] \left(\tau, R, \omega_J, \mu, \nu\right)\,.
\end{align}
For simplicity, we will now express the soft contribution as
\begin{align}
    \hat{\mathscr{S}}^{{\rm SJA}\, (1)}\left(\tau, R, \omega_J, \mu, \nu\right) = \mathscr{C}^{\rm SJA} \left[\hat{C}_i^{{\rm SJA}\, (0)}\, \hat{S}_i^{{\rm SJA}\, (1)}\right] \left(\tau, R, \omega_J, \mu, \nu\right)\,.
\end{align}
We note that from this expression, we will obtain the hard matching coefficient functions. Additionally, we remark that the third term in this expression accounts for the soft-recoil of the collinear function. From Eq.~\eqref{eq:G-one-loop}, we see the issue associated with the soft-recoil effects. The third term  has a natural rapidity scale of $\tau/R \omega_J$ but will also contain a contribution from the hyper-collinear sector, making a clear separation of contributions from the soft-collinear and hyper-collinear sectors challenging. The expression for the WTA axis  is relatively more simple and is given by
\begin{align}
    \hat{G}_{i\, {\rm alg}}^{\rm{WTA}\, (1)} \left(\tau, R, \omega_J, \mu, \zeta\right) = & \hat{H}_{i\, {\rm alg}}^{{\rm WTA}\, (1)}(\nu_s, \nu_c)\, \delta\left(\tau\right) + \hat{C}_i^{\rm{WTA}\, (1)} \left(\tau, \omega_J, \mu, \frac{\zeta}{\nu^2}\right) \nn \\
    &+ \hat{S}_i^{\rm{WTA}\, {(1)}}\left(\tau, R, \omega_J, \mu, \nu\right)\,.
\end{align}
The bare soft-collinear functions for the SJA and WTA axes at one-loop are given by the expressions
\begin{align}
    \hat{\mathscr{S}}^{{\rm SJA}\, (1)} \left(\tau, R, \omega_J, \mu, \nu\right) = & 2 g^2 C_i \left(\frac{\mu^2 e^{\gamma_E}}{4\pi}\right)^\epsilon \int \frac{d^dl}{\left(2\pi\right)^d} \, \left(2\pi\right) \delta\left(l^2\right) \frac{n_J \cdot \bar{n}_J}{n_J\cdot l\, \bar{n}_J\cdot l} \nn \\
    & \times \left(\frac{\bar{n}_J\cdot l}{\nu}\right)^{\eta}\, \delta \left(\tau-\frac{2 l_\perp}{\omega_J}\right)  \, \Theta \left(\tan^2{\frac{R}{2}}-\frac{n_J\cdot l}{\bar{n}_J\cdot l}\right)\,, 
    \\
    \hat{S}_i^{\rm{WTA}\, {(1)}} \left(\tau, R, \omega_J, \mu, \nu\right) = & 2 g^2 C_i \left(\frac{\mu^2 e^{\gamma_E}}{4\pi}\right)^\epsilon \int \frac{d^dl}{\left(2\pi\right)^d} \, \left(2\pi\right) \delta\left(l^2\right) \frac{n_J \cdot \bar{n}_J}{n_J\cdot l\, \bar{n}_J\cdot l} \nn \\
    & \times \left(\frac{\bar{n}_J\cdot l}{\nu}\right)^{\eta}\, \delta \left(\tau-\frac{l_\perp}{\omega_J}\right)  \, \Theta \left(\tan^2{\frac{R}{2}}-\frac{n_J\cdot l}{\bar{n}_J\cdot l}\right)\,, 
\end{align}
where the color factor is  $C_i = C_F$ for a quark and $C_i = C_A$ for a gluon. The factor of $2$ in the measurement function enters from the soft-recoil.

For the SJA, the soft factor is complicated due to the recoil effects, but the exact expressions for the soft-collinear function are given by
\begin{align}
    \hat{\mathscr{S}}^{{\rm SJA}\, (1)} \left(\tau, R, \omega_J, \mu, \nu\right) = \frac{\alpha_s C_i}{\pi^2}\frac{1}{\eta}\left(\mu^2 \pi e^{\gamma_E}\right)^\epsilon \frac{\left(\nu \tan{\frac{R}{2}}\right)^\eta}{k_\perp^{2+\eta}} \delta\left(\tau-2\frac{k_\perp}{\omega}\right)\,,
    \\
    \hat{S}_i^{{\rm WTA}\, (1)} \left(\tau, R, \omega_J, \mu, \nu\right) = \frac{\alpha_s C_i}{\pi^2}\frac{1}{\eta}\left(\mu^2 \pi e^{\gamma_E}\right)^\epsilon \frac{\left(\nu \tan{\frac{R}{2}}\right)^\eta}{k_\perp^{2+\eta}} \delta\left(\tau-\frac{k_\perp}{\omega}\right)\,.
\end{align}
Expanding these expressions in $\eta$ and $\epsilon$, we find that momentum space expressions
\begin{align}
    \hat{\mathscr{S}}^{{\rm SJA}\, (1)} \left(\tau, R, \omega_J, \mu, \nu\right)  = & \frac{\alpha_s C_i}{2\pi}\Bigg\{ \delta(\tau) \Bigg [\frac{1}{\epsilon^2}-\frac{2}{\eta \epsilon}-\frac{2}{\eta}L_{\mu s}^{\rm SJA}+\frac{1}{\epsilon}L_{\nu s}-\frac{1}{2}{L_{\mu s}^{\rm SJA}}^2 + {L_{\mu s}^{\rm SJA}}L_{\nu s}-\frac{\pi^2}{12} \Bigg] \nn \\
   &+\left(\frac{4}{\eta }-2L_{\nu s}+2{L_{\mu s}^{\rm SJA}}\right)\left(\frac{1}{\tau}\right)_+-4 \left(\frac{\ln (\tau )}{\tau}\right)_+ \Bigg\} \,,
    \\
    \hat{S}^{{\rm WTA}\, (1)} \left(\tau, R, \omega_J, \mu, \nu\right) & = \hat{\mathscr{S}}^{{\rm SJA}\, (1)} \left(\tau, R, \omega_J, \mu, \nu\right)|_{L_{\mu s}^{\rm SJA} \rightarrow L_{\mu s}^{\rm WTA}} \,,
\end{align}
where we have introduced the momentum space expressions for the logs
\begin{align}
    L_{\mu s}^{\rm SJA} = 2 \ln\left(\frac{2 \mu}{\omega_J}\right)\,,
    \qquad
    L_{\mu s}^{\rm WTA} = 2 \ln\left(\frac{\mu}{\omega_J}\right)\,,
    \qquad
    L_{\nu s} = -2 \ln\left(\frac{\nu \tan{\frac{R}{2}}}{\mu}\right)\,.
\end{align}
To obtain the anomalous dimension of the soft factor, we take the Laplace transform of the soft-collinear function yielding
\begin{align}
    \hat{\mathcal{S}}_i^{\rm SJA} \left(\kappa, \omega_J, R, \mu, \nu\right) & = \frac{\alpha_s C_i}{2\pi}\left[\frac{1}{\epsilon^2}-\frac{2}{\eta}\left(\frac{1}{\epsilon}-\mathcal{L}_{\mu s}^{\rm SJA}\right)+\frac{1}{\epsilon}\mathcal{L}_{\nu s}+\frac{1}{2}{\mathcal{L}_{\mu s}^{\rm SJA}}^2-\mathcal{L}_{\mu s}^{\rm SJA} \mathcal{L}_{\nu s} - \frac{5\pi^2}{12}\right]  \,,
    \\
    \hat{\mathcal{S}}_i^{\rm WTA} \left(\kappa, \omega_J, R, \mu, \nu\right) & = \hat{\mathcal{S}}_i^{\rm SJA} \left(\kappa, \omega_J, R, \mu, \nu\right)|_{\mathcal{L}_{\mu s}^{\rm SJA} \rightarrow \mathcal{L}_{\mu s}^{\rm WTA}}\,,
\end{align}
where the logs are defined as
\begin{align}
    \mathcal{L}_{\mu s}^{\rm SJA} = \ln\left(\frac{\omega^2}{4 \kappa^2 \mu^2}\right)\,,
    \qquad
    \mathcal{L}_{\mu s}^{\rm WTA} = \ln\left(\frac{\omega^2}{\kappa^2 \mu^2}\right)\,,
    \qquad
    \mathcal{L}_{\nu s} = \ln\left(\frac{\mu^2}{\nu^2 \tan^2{\frac{R}{2}}}\right)\,.
\end{align}
The renormalized $k_\perp$ integrated soft-collinear function for the SJA then satisfies the renormalization group equations
\begin{align}
    \frac{d}{d \ln \mu} \ln \mathcal{S}_i^{\rm axis}\left(\kappa, \omega_J, R, \mu, \nu\right) & = \gamma_{S\,i}^{\mu\, {\rm axis}}\left(\mu, \frac{\mu}{\nu}\right)\,,
    \\
    \frac{d}{d \ln \nu} \ln \mathcal{S}_i^{\rm axis}\left(\kappa, \omega_J, R, \mu, \nu\right) & = \gamma_{S\,i}^{\nu\, {\rm axis}}\left(\mu\right) \, 
\end{align}
and we note that
\begin{align}
        \gamma_{S\,i}^{\mu\, {\rm SJA}}\left(\mu, \frac{\mu}{\nu}\right) & =\Gamma^{\rm cusp}_i \mathcal{L}_{\nu s}\,,
        \qquad
        \gamma_{S\,i}^{\nu\, {\rm SJA}}\left(\mu\right) =\Gamma^{\rm cusp}_i \mathcal{L}_{\mu s}^{\rm axis}\,,
\end{align}
The expression for the evolution of the soft contribution for the WTA jet is identical, except that the 
soft-collinear function is then given by the solution
\begin{align}
    \mathcal{S}_i^{\rm axis}\left(\kappa, \omega_J, R, \mu, \nu\right) = \mathcal{S}_i^{\rm axis}\left(\kappa, \omega_J, R, \mu_i, \nu_i\right)\, U_{S_i}^{\rm axis}\left(\mu_i, \mu; \nu\right)\, V_{S_i}^{\rm axis}\left(\nu_i, \nu; \mu\right)\,,
\end{align}
where the initial scales $\mu_i$ and $\nu_i$ are taken to minimize the logarithms in the fixed order term and the Sudakov exponents are given by
\begin{align}
    U_{S_i}^{\rm axis}\left(\mu_i, \mu; \nu\right) & = \exp\left[ \int_{\mu_i}^\mu \frac{d\mu'}{\mu'} \gamma^{\mu\, {\rm axis}}_{S_i}\left(\mu', \frac{\mu'}{\nu}\right)\right] \, , 
    \\
    V_{S_i}^{\rm axis}\left(\nu_i, \nu; \mu_i\right) & = \left(\frac{\nu}{\nu_i}\right)^{\gamma_{S_i}^{\nu\, {\rm axis}}(\mu_i)}\,.
\end{align}
The construction of the azimuthal-dependent soft-collinear function can be written as
\begin{align}
     & \hat{\mathscr{S}}^{{\rm SJA}\, (1)} \left(\varphi, \tau, R, \omega_J, \mu, \nu\right) = \Theta\left(\pi-\varphi\right)\, \frac{\varphi}{\pi}\, \hat{S}_i^{{\rm WTA}\, (1)} \left(\tau, R, \omega_J, \mu, \nu\right) + \Theta\left(\varphi-\pi\right)\nn \\
     & \times \left[\left(\frac{2\pi-\varphi}{\pi}\right)\hat{S}_i^{{\rm WTA}\, (1)} \left(\tau, R, \omega_J, \mu, \nu\right) + \left(\frac{\varphi-\pi}{\pi}\right) \hat{\mathscr{S}}^{{\rm SJA}\, (1)} \left(\tau, R, \omega_J, \mu, \nu\right)\right]\,.
\end{align}
For the WTA axis, we have
\begin{align}
    \hat{S}_i^{{\rm WTA}\, (1)} \left(\varphi, \tau, R, \omega_J, \mu, \nu\right) = \frac{\varphi}{2\pi}\hat{S}_i^{{\rm WTA}\, (1)} \left(\tau, R, \omega_J, \mu, \nu\right)\,.
\end{align}

We now note a subtlety in the soft-collinear function. The finite terms of the soft-collinear function will be suppressed by $\varphi/2\pi$. This can be seen from the matching relation between the jet, hyper-collinear, and soft-collinear functions. However, we note that the rapidity divergence of this function occurs where $\bar{n}_J\cdot l$ goes to infinity and is, therefore, collinear to the direction of the jet axis. Similarly, the UV divergences of the soft factor occur where $l_\perp = 0$. As a result, all divergences that enter into the soft-collinear function are associated with emissions along the jet axis. Consequently, under our scheme for the soft-collinear emissions the divergences  will not contain the angular suppression factor. As a result, we find that the finite terms of the soft-collinear functions are suppressed by a factor of $\varphi/2\pi$ while the anomalous dimensions are unchanged.

\subsection{The hyper-collinear functions}
The hyper-collinear function enters into the resummed expression for the jet function in Eq.~\eqref{eq:BR-fact-Lap}. The tree level expressions for the hyper-collinear functions are given by
\begin{align}
    \hat{C}^{\rm axis\, (0)}_{i}\left(\varphi, \tau, \bm{0}_\perp, \omega_J, \mu,\frac{\zeta}{\nu^2}\right) &= \frac{\varphi}{2\pi}\delta\left(1-x\right)\,,
\end{align}
Beyond tree level, however, additional subtleties enter into the computation including soft-recoil and the number of partons. The expression for the collinear function is then given by
\begin{align}
    \hat{C}^{{\rm axis}\, (1)}_i\left(\varphi, \tau, \bm{0}_\perp, \omega_J, \mu,\frac{\zeta}{\nu^2}\right) = \int d\Phi \, \sum_j \hat{P}_{ji}\left(x,q_\perp, \epsilon\right)\, \left(\frac{\nu^2}{(1-x)^2\zeta}\right)^{\eta/2}  \delta_{\tau}^{\rm axis}\,, 
\end{align}
where we note that this expression is identical to that for the fixed order region except that we have dropped the dependence on the jet algorithm since $\tau \ll R$ in this region. 

To obtain the momentum space expression for the hyper-collinear functions, we perform the expansion
\begin{align}
    \tau^{-1-c} = -\frac{1}{c}\delta\left(\tau\right)+\left(\frac{1}{\tau}\right)_+^{1+c}-c \left(\frac{\ln{\tau}}{\tau}\right)_+^{1+c} + \mathcal{O}\left(c^2\right)\,,
\end{align}
where $c$ is some regularization parameter. Performing this expansion, we find that the hyper-collinear function for the azimuthally integrated jet broadening is given by
\begin{align}
    \hat{C}_q^{(1)\, {\rm axis}} \left(\tau, \bm{0}_\perp,\omega_J, \mu,\frac{\zeta}{\nu^2}\right) = & \frac{\alpha_s C_F}{\pi}\Bigg\{ \delta\left(\tau\right) \nn \\
    & \times \Bigg[ \frac{1}{\eta}\left(\frac{1}{\epsilon}+L_c^{\rm axis}\right)-\frac{L_c^{\rm axis} L_{\nu c}}{2}+\frac{3 L_c^{\rm axis}}{4}-\frac{L_{\nu c}}{2 \epsilon
   }+\frac{3}{4 \epsilon }+\frac{1}{4} \Bigg] \nn \\
    & +\left(-\frac{2}{\eta}+ L_{\nu c}-\frac{3}{2}\right)\left(\frac{1}{\tau}\right)_+\Bigg\} \,, 
    \\
    \hat{C}_g^{(1)\, {\rm axis}} \left(\tau, \bm{0}_\perp,\omega_J, \mu,\frac{\zeta}{\nu^2}\right) = & \frac{\alpha_s C_A}{\pi}\Bigg\{ \delta\left(\tau \right) \nn \\
    &\times \Bigg[\frac{1}{\eta}\left(\frac{1}{\epsilon }-L_c^{\rm axis} \right)-\frac{L_c^{\rm axis} L_{\nu c}}{2}+\frac{11 L_c^{\rm axis}}{12}-\frac{L_{\nu c}}{2 \epsilon
   }+\frac{11}{12 \epsilon }\Bigg] \nn \\
    & +\left(-\frac{2}{\eta}+ L_{\nu c}-\frac{11}{6}\right)\left(\frac{1}{ \tau}\right)_+\Bigg\} \nn \\
   & + \frac{\alpha_s n_f}{\pi}  \Bigg\{  \delta\left(\tau\right) \Bigg[-\frac{L_c^{\rm axis}}{6}-\frac{1}{6\epsilon}+\frac{1}{12}\Bigg] +\frac{1}{3}\left(\frac{1}{\tau}\right)_+\Bigg\} \, ,
\end{align}
where the logarithms in momentum space are given by
\begin{align}
    L_c^{\rm SJA} = \ln\left(\frac{4 \mu^2}{\omega_J^2}\right)\,,
    \qquad
    L_c^{\rm WTA} = \ln\left(\frac{\mu^2}{\omega_J^2}\right)\,,
    \qquad
    L_{\nu c} = \ln\left(\frac{\omega^2}{\nu^2}\right)\,.
\end{align}
The Laplace space expression for the hyper-collinear function reads 
\begin{align}
    \hat{\mathcal{C}}_q^{{\rm axis}\, (1)} & \left(\kappa, \bm{0}_\perp,\omega_J, \mu, \frac{\zeta}{\nu^2}\right) = \frac{\alpha_s C_F}{2\pi} \nn \\
    &\times \left[\frac{2}{\eta}\left(\frac{1}{\epsilon}+\mathcal{L}_{\mu c}^{\rm axis}\right)+\frac{\mathcal{L}_\zeta}{\epsilon} +\frac{3}{2\epsilon}+\mathcal{L}_\zeta \mathcal{L}_{\mu c}^{\rm axis} + \frac{3}{2}\mathcal{L}_{\mu c}^{\rm axis} +\frac{1}{2}\right]\,, \nn \\
    \\
    \hat{\mathcal{C}}_g^{{\rm axis}\, (1)} & \left(\kappa, \bm{0}_\perp,\omega_J, \mu, \frac{\zeta}{\nu^2}\right) = \frac{\alpha_s C_A}{2\pi} \nn \\
    &\times \left[\frac{2}{\eta}\left(\frac{1}{\epsilon}+\mathcal{L}_{\mu c}^{\rm axis}\right)+\frac{\mathcal{L}_\zeta}{\epsilon} +\frac{\beta_0}{2C_A\epsilon}+\mathcal{L}_\zeta \mathcal{L}_{\mu c}^{\rm axis} + \frac{\beta_0\mathcal{L}_{\mu c}^{\rm axis}}{2C_A} +\frac{n_f}{6 C_A}\right]\,,
\end{align}
where the Laplace space logs are given by
\begin{align}
    \mathcal{L}_{\zeta} = \ln\left(\frac{\nu^2}{\zeta}\right)\,,
    \qquad
    \mathcal{L}_{\mu c}^{\rm SJA} = \ln\left(\frac{4 \kappa^2 \mu^2}{\omega_J^2}\right)\,,
    \qquad
    \mathcal{L}_{\mu c}^{\rm WTA} = \ln\left(\frac{\kappa^2 \mu^2}{\omega_J^2}\right)\,.
\end{align}

These functions obey the evolution equations
\begin{align}
    \frac{d}{d\ln\mu} \ln \mathcal{C}_i^{\rm axis}\left(\kappa, \bm{0}_\perp,\omega_J, \mu, \nu\right) & = \gamma_{C i}^{\mu\, {\rm axis}}\left(\mu, \frac{\mu}{\nu}\right)\,,
    \\
    \frac{d}{d\ln\nu} \ln\mathcal{C}_i^{\rm axis}\left(\kappa, \bm{0}_\perp,\omega_J, \mu, \nu\right) & = \gamma_{C i}^{\nu\, {\rm axis}}\left(\kappa, \mu\right)\,, 
\end{align}
where the anomalous dimensions are given by
\begin{align}
    \gamma_{C i}^{\mu\, {\rm axis}}\left(\mu, \frac{\mu}{\nu}\right) & = C_i \gamma^{\rm cusp} \mathcal{L}_{\nu c}+\gamma^i \,, 
    \qquad
    \gamma_{C i}^{\nu\, {\rm axis}}\left(\kappa, \mu\right) = C_i \gamma^{\rm cusp} \mathcal{L}_{\mu c}^{\rm axis} \,,
\end{align}
The collinear function is then obtained by the solution
\begin{align}
    \mathcal{C}_i^{\rm axis}\left(\kappa, \bm{0}_\perp,\omega_J, R, \mu, \frac{\zeta}{\nu^2}\right) = & \mathcal{C}_i^{\rm axis}\left(\kappa, \bm{0}_\perp,\omega_J, R, \mu_i, \frac{\zeta}{\nu_i^2}\right) \nn \\
    & \times U_{C_i}^{\rm axis}\left(\mu_i, \mu; \nu_i\right)\, V_{C_i}^{\rm axis}\left(\nu_i, \nu; \mu_i\right)\,,
\end{align}
where the initial scales $\mu_i$ and $\nu_i$ are taken to minimize the logarithms in the fixed order term and
\begin{align}
    U_{C_i}^{\rm axis}\left(\mu_i, \mu, \nu\right) &= \exp\left[ \int_{\mu_i}^{\mu} \frac{d\mu'}{\mu'} \gamma^{\mu\, {\rm axis}}_{C_i}\left(\mu, \mu_i\right)\right] \, , 
    \\
    V_{C_i}^{\rm axis}\left(\nu_i, \nu, \mu_i\right) &= \left(\frac{\nu}{\nu_i}\right)^{\gamma_{C_i}^{\nu\, {\rm axis}}(\mu_i)}\,.
\end{align}
The hyper-collinear function for the azimuthal-dependent jet broadening can be expressed in terms of the azimuthally integrated jet shape through 
\begin{align}
    & \hat{C}^{{\rm SJA}\, (1)}_{i} \left(\varphi, \tau, \bm{0}_\perp, \omega_J, \mu, \frac{\zeta}{\nu^2}\right) = \frac{\varphi}{\pi}\Theta\left(\pi-\varphi\right) \hat{C}^{{\rm WTA}\, (1)}_{i} \left(\tau,\bm{0}_\perp, \omega_J, \mu, \frac{\zeta}{\nu^2}\right) + \Theta\left(\varphi-\pi\right)\nn \\
    & \times \left[\left(\frac{2\pi-\varphi}{\pi}\right) \hat{C}^{{\rm WTA}\, (1)}_{i} \left(\tau,\bm{0}_\perp, \omega_J, \mu, \frac{\zeta}{\nu^2}\right)+\left(\frac{\varphi-\pi}{\pi}\right) \hat{C}^{{\rm SJA}\, (1)}_{i} \left(\tau,\bm{0}_\perp, \omega_J, \mu, \frac{\zeta}{\nu^2}\right)\right] \,. \nn \\
\end{align}
For the WTA axis, the hyper-collinear function is simply given by
\begin{align}
    & \hat{C}^{{\rm WTA}\, (1)}_{i} \left(\varphi, \tau, \bm{0}_\perp, \omega_J, \mu, \zeta\right) = \frac{\varphi}{\pi} \hat{C}^{{\rm SJA}\, (1)}_{i} \left(\tau,\bm{0}_\perp, \omega_J, \mu\right)\,.
\end{align}
We find that the anomalous dimension of the azimuthal-dependent hyper-collinear functions are identical to those of the azimuthally integrated case.
\subsection{The asymptotic region}
The contribution in the asymptotic region can be obtained by considering only the terms in the fixed order region which contribute the most in the small $\tau$ limit. Furthermore, by expanding the result from the resummed region and comparing with the jet functions in the asymptotic region, we can obtain the expression for the matching functions.

Combining the soft-collinear and soft modes, we arrive at the expressions for the asymptotic region
\begin{align}
    \hat{G}_{q\, {\rm alg}}^{{\rm SJA}\, (1)}\left(\tau, \omega_J, R, \mu, \zeta\right)  = & \hat{H}_{q\, {\rm alg}}^{{\rm SJA}\, (1)}\left(\nu_s, \nu_c\right)\delta(\tau) \nn \\
    & + \frac{\alpha_s C_F}{2\pi}\Bigg\{ \delta(\tau) \Bigg[\frac{1}{ \epsilon ^2}-\frac{L_R}{\epsilon}+\frac{3}{2 \epsilon }+\frac{L_R^2}{2}-\frac{3 L_R}{2} \nn \\
    & -2\ln^2\left(2 \tan{\frac{R}{2}}\right)+3\ln\left(2 \tan{\frac{R}{2}}\right)+\frac{1}{2}-\frac{\pi^2}{12}\Bigg] \nn 
   \\
   & +4 \ln\left(2 \tan{\frac{R}{2}}\right) \left(\frac{1}{\tau}\right)_+- \left(\frac{3}{\tau}\right)_+- 4\left(\frac{\ln{\tau}}{\tau}\right)_+\Bigg\}\,,
   \\
    \hat{G}_{g\, {\rm alg}}^{{\rm SJA}\, (1)}\left(\tau, \omega_J, R, \mu, \zeta\right)  = & \hat{H}_{g\, {\rm alg}}^{{\rm SJA}\, (1)}\left(\nu_s, \nu_c\right)\delta(\tau) \nn \\
    & + \frac{\alpha_s C_A}{2\pi}\Bigg\{\delta(\tau)\Bigg[\frac{1}{\epsilon^2}-\frac{L_R}{\epsilon}+\frac{11}{6\epsilon}+\frac{L_R^2}{2}-\frac{11}{6}L_R \nn \\
    & -2 \ln^2\left(2 \tan{\frac{R}{2}}\right)+\frac{11}{3}\ln\left(2 \tan{\frac{R}{2}}\right) -\frac{\pi^2}{12} \Bigg]\nn 
   \\
   & +4 \ln\left(2 \tan{\frac{R}{2}}\right) \left(\frac{1}{\tau}\right)_+- \frac{11}{3}\left(\frac{1}{\tau}\right)_+- 4\left(\frac{\ln{\tau}}{\tau}\right)_+  \nn \\
   & +\frac{\alpha_s n_f}{2\pi}\Bigg\{ \delta(\tau)\left[-\frac{1}{3 \epsilon}+\frac{L_R}{3}-\frac{2}{3}\ln\left(2 \tan{\frac{R}{2}}\right)+\frac{1}{6} \right] +\frac{2}{3}\left(\frac{1}{\tau}\right)_+\Bigg\} \,,
\end{align}
while the expression for the WTA jet function we replace
\begin{align}
    \ln\left(2 \tan{\frac{R}{2}}\right) \rightarrow \ln\left(\tan{\frac{R}{2}}\right)\,.
\end{align}
From these expressions, we can obtain the matching functions, which are UV finite, to be
\begin{align}
    H_{q\, {\rm Cone}}^{{\rm SJA}\, (1)}\left(\nu_s, \nu_c\right) =& \frac{\alpha_s C_F}{\pi}\left[\ln^2\left(\tan{\frac{R}{2}}\right)+2\ln(2) \ln\left(\tan{\frac{R}{2}}\right)-\frac{3}{2}\ln\left(\tan{\frac{R}{2}}\right)\right]\,,
    \\
    H_{q\,{\rm k_T}}^{{\rm SJA}\, (1)}\left(\nu_s, \nu_c\right) =& H_{q\,{\rm Cone}}^{{\rm SJA}\, (1)}\left(\nu_s, \nu_c\right)+\frac{\alpha_s C_F}{\pi}\left[-3 \ln^2(2)+\frac{3}{2}\ln(2)\right]\,,
    \\
    H_q^{{\rm WTA}\, (1)}\left(\nu_s, \nu_c\right) =& H^{{\rm SJA}\, (1)}_{q\,{\rm Cone}}\left(\nu_s, \nu_c\right)\,,
    \\
    H_{g\, {\rm Cone}}^{{\rm SJA}\, (1)}\left(\nu_s, \nu_c\right) =& \frac{\alpha_s C_A}{\pi}\Bigg[\frac{n_f}{3 C_A}\ln \left(\tan \left(\frac{R}{2}\right)\right) +2 \ln ^2\left(\tan \left(\frac{R}{2}\right)\right) \nn \\
    & +2
   \ln (2) \ln \left(\tan \left(\frac{R}{2}\right)\right)-\frac{11}{6} \ln \left(\tan \left(\frac{R}{2}\right)\right)\Bigg]\,,
    \\
    H_{g\,{\rm k_T}}^{{\rm SJA}\, (1)}\left(\nu_s, \nu_c\right) =& H_{g\,{\rm Cone}}^{{\rm SJA}\, (1)}\left(\nu_s, \nu_c\right) \nn \\
    & +\frac{\alpha_s C_A}{\pi}\left[\frac{n_f \ln (2)}{3 C_A}+\ln ^2\left(\tan \left(\frac{R}{2}\right)\right)+3 \ln ^2(2)-\frac{11 \ln (2)}{6}\right]\,,
    \\
    H_g^{{\rm WTA}\, (1)}\left(\nu_s, \nu_c\right) =& H^{{\rm SJA}\, (1)}_{g\,{\rm Cone}}\left(\nu_s, \nu_c\right)\,,
\end{align}
where $2\tan{R/2} = \nu_s/\nu_c$.

\subsection{Evolution of the exclusive jet broadening function}
\begin{figure}
    \centering
    \hspace{0.5in}
    \includegraphics[width = 0.5\textwidth]{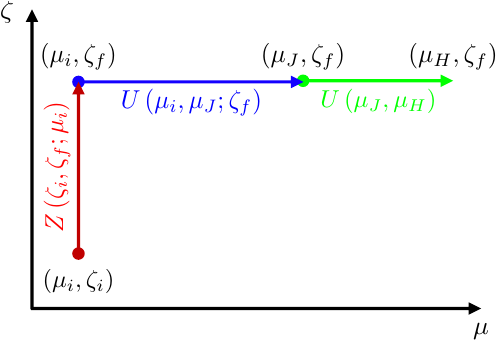}
    \hspace{0.5in}
    \caption{The evolution for the jet broadening functions. The final step of the evolution in green is performed either via a DGLAP evolution equation or a Sudakov resummation.}
    \label{fig:Jet-evo}
\end{figure}
The exclusive jet angularities satisfy the evolution equations that
\begin{align}
    \frac{d}{d\ln \mu} \ln G^{\rm axis}_{i\, {\rm alg}} \left(\tau,\omega_J, R, \mu\right) = \gamma_i^J\left[\alpha_s(\mu)\right]\,,
\end{align}
where $\gamma_i^J$ is the anomalous dimension of the exclusive jet function that is given in Eq.~\eqref{eq:anom-jet}.

Based on the computation in the previous section, we can see that the anomalous dimensions of the jet function and the soft-collinear function are opposite to one another. This result is expected and serves as a benchmark of our formalism. In the Collins-Soper formalism, these functions can then be combined to a single jet substructure function
\begin{align}
    \mathcal{G}_i^{\rm axis}\left(\kappa, \omega_J, R, \mu, \zeta\right) = \mathcal{C}_i^{\rm axis}\left(\kappa, \omega_J, R, \mu,\frac{\zeta}{\nu^2}\right)\, \mathcal{S}_i^{\rm axis}\left(\kappa, \omega_J, R, \mu,\nu\right)\,.
\end{align}
This combined  function will contain rapidity logs of the form $\ln(\kappa \tan{\frac{R}{2}})$ that are resummed in the Collins-Soper equation. Thus the evolution of this combined jet substructure function takes on the form 
\begin{align}
    \frac{d}{d\ln\mu} \ln \mathcal{G}_i^{\rm axis}\left(\kappa, \omega_J, R, \mu, \zeta\right) & = \gamma^{\mu\, {\rm axis}}_{C_i}\left(\mu, \frac{\zeta}{\nu^2}\right)+\gamma^{\mu\, {\rm axis}}_{S_i}\left(\mu, \frac{\mu}{\nu}\right) = \gamma^{\mu\, {\rm axis}}_{J_i}(\mu,\zeta)\,,
    \\
    \frac{d}{d\ln\zeta} \ln \mathcal{G}_i^{\rm axis}\left(\kappa, \omega_J, R, \mu, \zeta\right) & = -\frac{1}{2}\gamma^{\nu\, {\rm axis}}_{C_i}(\kappa, \mu) \,.
\end{align}
Thus, the solution is given by
\begin{align}
    \mathcal{G}_i^{\rm axis}\left(\kappa, \omega_J, R, \mu, \zeta\right) = \mathcal{G}_i^{\rm axis}\left(\kappa, \omega_J, R, \mu_i, \zeta_i\right)\, U_{\mathcal{G}\, i}^{\rm axis}\left(\mu_i, \mu;\zeta\right)\,  Z_{\mathcal{G}\, i}^{\rm axis}\left(\zeta_i, \zeta;\mu_i\right)\,,
\end{align}
where the initial scales which minimize the large logs in the expressions are
\begin{align}
    \mu_i = \frac{\omega_J}{\kappa}\,,
    \qquad
    \zeta_i = \frac{\omega_J^2}{\kappa^2 \tan^2{R/2}}\, ,
\end{align}
while the Sudakov exponentials are given by 
\begin{align}
    U_{\mathcal{G}\, i}\left(\mu_i, \mu;\zeta\right) & = \exp\left[\int_{\mu_i}^\mu\frac{d\mu'}{\mu'} \gamma^{\mu\, {\rm axis}}_{J_i}\left(\mu, \zeta\right)\right]\,,
    \\
    Z_{\mathcal{G}\, i}\left(\zeta_i, \zeta;\mu_i\right) & = \exp\left[\int_{\zeta_i}^\zeta\frac{d\zeta'}{\zeta'} \frac{1}{2}\gamma^{\nu\, {\rm axis}}_{C_i}\left(\kappa, \mu_i\right)\right]\,.
\end{align}
This coupled evolution can be represented graphically in Fig.~\ref{fig:Jet-evo}. 

\subsection{Evolution of the semi-inclusive jet broadening function}
To compute the evolution equations of the semi-inclusive jet broadening functions, we need to consider the hard-collinear contributions. These contributions are precisely the same as the graphs (B) and (C) for the semi-inclusive jet function \cite{Kang:2017mda,Kang:2017glf}. Here we list the expressions
\begin{align}
    \hat{h}_{qq}\left(z, \omega_J, R, \mu\right) = & \delta(1-z)+\frac{\alpha_s C_F}{2\pi}\Bigg[\left(-\frac{1}{\epsilon^2}+\frac{1}{\epsilon}L_R -\frac{1}{2}L_R^2+\frac{\pi^2}{12}\right)\delta(1-z)-(1-z) \nn \\
    & +\left(\frac{1}{\epsilon}-L_R-2 \left(\ln(1-z)\right)_+\right) P_{qq}(z) \Bigg] +\mathcal{O}\left(\alpha_s^2\right),  \\
   \hat{h}_{gq}\left(z, \omega_J, R, \mu\right) = & \frac{\alpha_s C_F}{2\pi}\left[-z+\left(\frac{1}{\epsilon}-L_R-2\left(\ln{\left(1-z\right)}\right)_+\right)P_{gq}(z)\right] +\mathcal{O}\left(\alpha_s^2\right),  \\
   \hat{h}_{qg}\left(z, \omega_J, R, \mu\right) = & \frac{\alpha_s n_f}{2\pi}\left[-1+\left(\frac{1}{\epsilon}-L_R-2\left(\ln{\left(1-z\right)}\right)_++1 \right)P_{gq}(z)\right] +\mathcal{O}\left(\alpha_s^2\right), \\
   \hat{h}_{gg}\left(z, \omega_J, R, \mu\right) = & \delta(1-z)+\frac{\alpha_s C_A}{2\pi}\Bigg[\left(-\frac{1}{\epsilon^2}+\frac{1}{\epsilon}L_R -\frac{1}{2}L_R^2+\frac{\pi^2}{12}+\frac{\beta_0}{2}\left(L_R-\frac{1}{\epsilon}\right)\right)\delta(1-z) \nn \\
    & +\left(\frac{1}{\epsilon}-L_R-2 \left(\ln(1-z)\right)_+\right) P_{gg}(z) \Bigg] +\mathcal{O}\left(\alpha_s^2\right). 
\end{align}
We see from these expressions that the double poles of the hard-collinear modes and the soft-collinear modes precisely cancel one another. Thus, we find that the evolution equation for the semi-inclusive jet function is governed by the DGLAP type evolution equation
\begin{align}
\label{eq:DGLAP3}
\frac{d}{d \ln \mu} \mathcal{G}_{i\, {\rm alg}}^{\rm axis} (N,\tau, \omega_J,R, \mu) = \frac{\alpha_s}{\pi} c_{ij}\, \mathcal{P}_{ij}(N)\, \mathcal{G}_{j\, {\rm alg}}^{\rm axis} (N,\tau, \omega_J,R, \mu) \, , 
\end{align}
where we have simplified the evolution equation of the semi-inclusive jet broadening function by working in Mellin space.

In the case of semi-inclusive jet functions, the UV divergences at $q_\perp = 0$ cancel against those at $q_\perp \rightarrow \infty$ and $x = 1$. The remaining UV divergences are those that generate a DGLAP evolution. Thus, semi-inclusive jet functions follow the evolution equation
\begin{align}
    \frac{d}{d \ln \mu}G_{i\, {\rm alg}}^{\rm axis}\left(z, \varphi, \tau, \omega_J, R, \mu, \zeta\right) = \frac{\alpha_s}{\pi} c_{ij}\, P_{ij}(z)\otimes G_{j\, {\rm alg}}^{\rm axis}\left(z, \varphi, \tau, \omega_J, R, \mu, \zeta\right)\,.
\end{align}
To simplify the evolution equation, it is convenient to work in Mellin moment space 
\begin{align}
    \mathcal{G}_{i\, {\rm alg}}^{\rm axis}\left(N,\varphi, \tau, \omega_J, R, \mu, \zeta\right) & = \int_0^1 z^{N-1} G_{i\, {\rm alg}}^{\rm alg}\left(z, \varphi, \tau, \omega_J, R, \mu, \zeta\right)\,,
    \\
    \mathcal{P}_{ji}\left(N\right) & = \int_0^1 z^{N-1} P_{ji}\left(z\right)\,, 
\end{align}
where the calligraphic font is used to denote a transformed function. Under this transformation, the coupled evolution equation becomes
\begin{align}
    \frac{d}{d \ln \mu}\mathcal{G}_{i\, {\rm alg}}^{\rm axis}\left(N,\varphi, \tau, \omega_J, R, \mu, \zeta\right) = \frac{\alpha_s}{\pi} c_{ij}\, \mathcal{P}_{ij}(N)\, \mathcal{G}_{j\, {\rm alg}}^{\rm axis}\left(N,\varphi, \tau, \omega_J, R, \mu, \zeta \right)\,.
\end{align}
The evolution of the semi-inclusive jet angularity function involves careful consideration. Emissions near the scale $\omega_J \lambda$ are resummed by solving the coupled evolution equations of the collinear and soft-collinear modes. While emissions near the scale $\omega_J R$ are resummed through the DGLAP evolution outlined here.

\section{The $r$ and $\varphi$ dependent jet shapes}\label{sec:shape}
The traditional integrated and differential jet shapes can be defined by the expressions
\begin{align}
    \psi_{\rm alg}^{\rm axis}(r,R) = \sum_{i\in j} \bar{n}_J\cdot p_i / \sum_{i\in J} \bar{n}_J\cdot p_i\,,
    \qquad
    \rho_{\rm alg}^{\rm axis}(r,R) = \frac{\partial}{\partial r} \psi(r,R)\,.
\end{align}
In the numerator of the integrated jet shape, all particles in a cone of size $r$ are included, while in the denominator all particles of the jet of radius $R$ are summed over. From the definition of these observables, it is clear that they provide information for the distribution of energy as a function of the sub-jet radius $r$. We will refer to these functions as the $r$ integrated and $r$ differential jet shapes, and will refer to them collectively as the $r$ dependent jet shapes.

In this paper, we introduce several generalizations of the $r$ dependent jet shapes. First, we note that  rather than defining the jet shape to measure the energy as a function of the sub-jet radius $r$, one can in principle measure the energy as a function of the azimuthal angle of a wedge of the jet. We denote these the $\varphi$ dependent jet shapes and define them as
\begin{align}
    \psi_{\rm alg}^{\rm axis}(\varphi,R) = \sum_{i\in J_{\varphi}} \bar{n}_J\cdot p_i / \sum_{i\in J} \bar{n}_J\cdot p_i\,,
    \qquad
    \rho_{\rm alg}^{\rm axis}(\varphi,R) = \frac{\partial}{\partial \varphi} \psi(\varphi,R)\,.
\end{align}
Unlike the jet broadening, where the jet wedge is constructed using the slicing method, the azimuthal-dependent jet shape is constructed using the iterative method. Starting from the parent parton, at each point in the jet's history, the azimuthal angle of the emitted particles are considered. If the azimuthal angle of the emission is within the azimuthal range governed by $\varphi$, the particle is considered to be within the jet wedge. As we will later demonstrate, the definition of this observable requires careful consideration of non-perturbative effects. In this paper, we demonstrate that the energy weighting of the jet shape eliminates these non-perturbative effects.

\begin{figure}
    \centering
    \includegraphics[width = \textwidth]{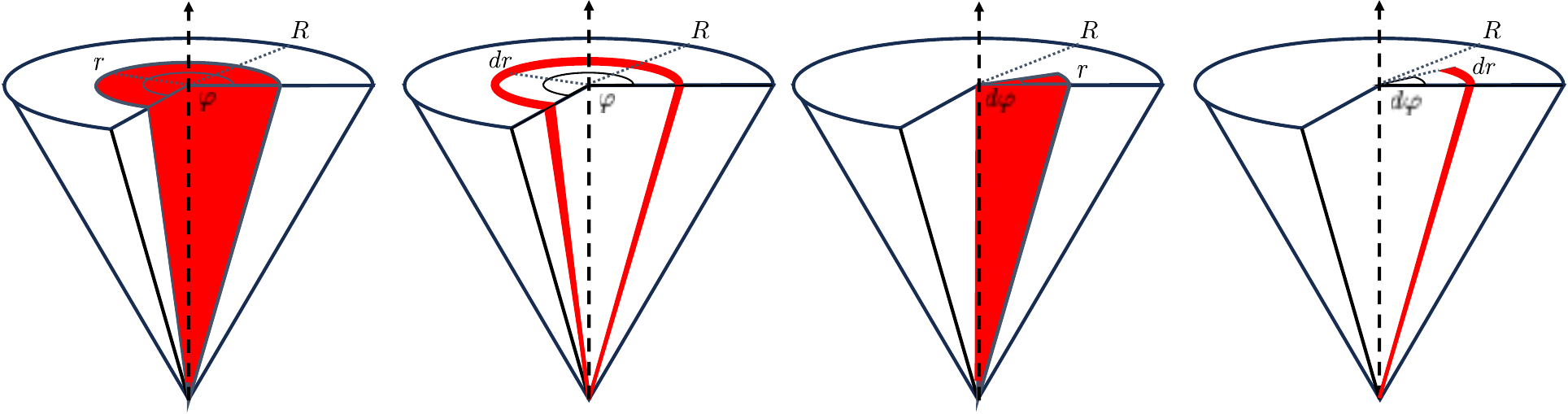}
    \caption{The four azimuthal-dependent jet shapes. From left to right, these are the double integrated jet shape, the $\varphi$ integrated $r$ differential jet shape, the $r$ integrated $\varphi$ differential jet shape, and the double differential jet shape.}
    \label{fig:Jet-shape}
\end{figure}

By introducing $\varphi$ dependence jet shapes, we can now also consider $r$ and $\varphi$ dependent jet shapes, which are defined as
\begin{align}
    \psi_{\rm alg}^{\rm axis}(\varphi,r,R) = \sum_{i\in j_\varphi} \bar{n}_J\cdot p_i / \sum_{i\in J} \bar{n}_J\cdot p_i\,,
\end{align}
where we use the notation that $j_\varphi$ represents a wedge of the jet with radius $r$ and azimuthal angle $\varphi$. By considering both $r$ and $\varphi$ dependence, this allows us to define the two single-differential jet shapes
\begin{align}
    \rho_{\rm alg}^{\rm axis}(\varphi,r,R) = \frac{\partial}{\partial r} \psi_{\rm alg}^{\rm axis}(\varphi,r, R)\,,
    \qquad
    \varPhi_{\rm alg}^{\rm axis}(\varphi,r,R) = \frac{\partial}{\partial \varphi}\psi_{\rm alg}^{\rm axis}(\varphi,r,R)\,,
\end{align}
while we can also define a double-differential jet shape
\begin{align}
    \Delta_{\rm alg}^{\rm axis}(\varphi,r,R) = \frac{\partial}{\partial r}\frac{\partial}{\partial \varphi} \psi_{\rm alg}^{\rm axis}(\varphi,r,R)\,.
\end{align}
These jet shapes are represented graphically in Fig.~\ref{fig:Jet-shape}.

The $r$ dependent jet shape and the $\varphi$ dependent jet shape can be obtained from the $r$ and $\varphi$ dependent jet shape by taking the limit that $\varphi = 2\pi$ or $r = R$. Thus, all information relating the jet shape to the cross section can be written succinctly as
\begin{align}
\label{shape0}
    \psi_{\rm alg}^{\rm axis}(\varphi,r,R) = & \int dz_w z_w \frac{d\sigma_c}{d\mathcal{PS}} \, \hat{j}_{c\, {\rm alg}}^{\rm axis}\left(\varphi, z_w, r, R, \omega_J,\mu, \zeta\right)/ \frac{d\sigma_c}{d\mathcal{PS}} \hat{J}_{c\, {\rm alg}}^{\rm axis}\left(R,\omega_J,\mu\right)\,, 
    \\
    \label{eq:shape}
    \psi_{\rm alg}^{\rm axis}(\varphi,r,R) = & \int dz_w z_w \frac{d\sigma_c}{dz\, d\mathcal{PS}} \otimes \hat{j}_{c\, {\rm alg}}^{\rm axis}\left(\varphi, z, z_w, r, R, \omega_J,\mu, \zeta\right)/ \nn \\
    & \frac{d\sigma_c}{dz\, d\mathcal{PS}} \otimes \hat{J}_{c\, {\rm alg}}^{\rm axis}\left(z, R,\omega_J,\mu\right) \,, 
\end{align}
where the first and second line represents the exclusive and semi-inclusive jet shapes. In this expression $\otimes$ stands for a collinear convolution in the variable $z$, while the momentum fraction of the wedge with respect to the jet is $z_w$, and in the limit that $\varphi \rightarrow 2\pi$ we have $z_w = z_r$ where $z_r$ is the energy fraction of the sub-jet. Additionally, in this expression $j$ is the sub-jet wedge function, while when $r = R$ we refer to this as the wedge function. In the following sections we will discuss the factorization and resummation of the $r$ and $\varphi$ dependent sub-jet functions.

\subsection{Power counting the jet shapes}
The wedge that defines the azimuthal-dependent sub-jet function extends from $R$ to $0$, this function will therefore contain contributions at the three critical angles $\theta_{\rm NP} \sim \Lambda_{\rm QCD}/\omega_J$, $\theta_{r}\sim r$, and $\theta_{R} \sim R$. The contributions of the order $\theta_{\rm NP}$ capture the non-perturbative contributions associated with energy flow. We will demonstrate that the azimuthal distribution of energy flow is governed by the `wedge function'.

The traditional jet shape is integrated over the azimuthal angle and thus does not depend on the wedge function. In the region where $\Lambda_{\rm QCD} \ll \omega_J r \ll \omega_J R$, the exclusive sub-jet functions take on the factorized forms
\begin{align}
    \hat{j}_{i\, {\rm alg}}^{\rm SJA}\left(z_r, r, R, \omega_J,\mu, \zeta\right) = & \hat{H}_{i\, {\rm alg}}^{\rm SJA}(\nu_s, \nu_c) \int d^{2-2\epsilon}k_\perp  \nn \\
    & \times  \hat{C}_i^{\rm SJA}\left(z_r, \bm{k}_\perp, r, \mu,\frac{\zeta}{\nu^2}\right)\, \hat{S}_i\left(-\bm{k}_\perp, R, \omega_J, \mu,\nu\right)\,,
    \\
    \hat{j}_{i\, {\rm alg}}^{\rm WTA}\left(z_r, r, R, \omega_J,\mu, \zeta\right) & = \hat{H}_{i\, {\rm alg}}^{\rm WTA}(\nu_s, \nu_c) \left[\hat{\mathcal{J}}_{ij} \otimes \hat{J}_j^{\rm WTA}\right]\left(z_r, r, R, \omega_J,\mu, \zeta\right)\,,
\end{align}
while the semi-inclusive sub-jet function are given by
\begin{align}
    \hat{j}_{i\, {\rm alg}}^{\rm axis}\left(z, z_w, r, R, \omega_J,\mu, \zeta\right) = \hat{h}_{ij}\left(z, \omega_J, R, \mu\right)\, \hat{j}_{j\, {\rm alg}}^{\rm axis}\left(z_w, r, R, \omega_J,\mu, \zeta\right)\,.
\end{align}
In the second expression, the capital $J$ is the semi-inclusive WTA jet function at a sub-jet radius of $r$ while the function $\mathcal{J}$ represents a matching coefficient function.

The $\varphi$ dependent sub-jet function probes the distribution of energy down to $\theta_{\rm NP}$ and thus must be perturbatively matched onto a non-perturbative function, namely the wedge function. This matching takes on the form
\begin{align}
    \hat{j}_{i\, {\rm alg}}^{\rm axis}\left(\varphi, z_w, R, \omega_J, \mu, \zeta\right) &= \left[\hat{\mathcal{K}}_{i\, {\rm alg}}^{{\rm axis}} \otimes \hat{w}_i^{\rm axis}\right]\left(\varphi, z_w, R, \omega_J, \mu, \zeta\right)\,,
\end{align}
and the semi-inclusive sub-jet function can be computed by taking into consideration the hard-collinear modes. In this expression, $\mathcal{K}$ represents the matching function that we will compute to one-loop in the following subsection.

Lastly, we examine the $r$ and $\varphi$ dependent jet shape. In the region where $\theta_r \gg \theta_{\rm NP}$, the azimuthal-dependent sub-jet function can be perturbatively matched onto the wedge function as
\begin{align}
    \hat{j}_{i\, {\rm alg}}^{\rm axis}\left(\varphi, z_w, r, R, \omega_J, \mu\right) &= \left[\hat{\mathcal{M}}_{i\, {\rm alg}}^{{\rm axis}} \otimes \hat{w}_i^{\rm axis}\right]\left(\varphi, z_w, r, R, \omega_J, \mu\right)\,,
\end{align}
where the matching function for the $r$ and $\varphi$ dependent sub-jet function is related to the matching function for the $\varphi$ dependent jet function. In the region where $r \ll R$, the jet functions contains logs of $r/R$ that require resummation. In this case, the QCD modes associated with emissions within the jet split into a hyper-collinear mode and a soft-collinear mode
\begin{align}
    p_{c}^\mu \sim \omega_J\left(r^2, 1, r\right)\,,
    \qquad
    p_{s}^\mu \sim \omega_J\frac{r}{R}\left(R^2, 1, R\right)\,.
\end{align}
We note that the soft-collinear mode is not energetic enough to contribute to the observable. However, using a SJA, the soft-collinear emissions will cause the hyper-collinear mode to recoil. In the case of the exclusive jet functions, the matching coefficient functions in this region take on the form
\begin{align}
    \label{eq:Mfact}
    \hat{\mathcal{M}}_{i\, {\rm alg}}^{{\rm SJA}} \left(\varphi, z_w,r, R,\omega_J, \mu, \zeta\right) & = \hat{H}_{i\, {\rm alg}}^{\rm SJA}(\nu_s, \nu_c) \int d^{2-2\epsilon}k_\perp  \nn \\
    & \hspace{-1cm} \times  \hat{C}_i^{\rm SJA}\left(\varphi, z_w, \bm{k}_\perp, r, \mu,\frac{\zeta}{\nu^2}\right)\, \hat{S}_i\left(\varphi, -\bm{k}_\perp, R, \omega_J, \mu,\nu\right)\,,
    \\
    \hat{\mathcal{M}}_{i\, {\rm alg}}^{{\rm SJA}} \left(z, \varphi, z_w,r, R,\omega_J, \mu, \zeta\right) &= \hat{h}_{ij}\left(z,\omega_J R, \mu\right) \, \hat{\mathcal{M}}_{j\, {\rm alg}}^{{\rm SJA}} \left(\varphi, z_w,r, R,\omega_J, \mu, \zeta\right) \,.
\end{align}
The functions $C$ and $S$ here denote the contributions of the hyper-collinear and soft-collinear modes with $\varphi$ dependence. The case of a WTA axis is drastically more simple. Since the direction of the jet axis is insensitive to soft emissions, no soft-collinear function enters into the factorization. In this case, the matching function factorizes as
\begin{align}
    \hat{\mathcal{M}}_{i}^{{\rm WTA}} \left(\varphi, z_w,r, R,\omega_J, \mu, \zeta\right) & = \hat{H}_{i}^{\rm WTA}(\nu_s, \nu_c)\, \left[\hat{\mathcal{J}}_{ij} \otimes \hat{j}_{j}^{\rm WTA} \right] \left(\varphi, z_w,r, R,\omega_J, \mu, \zeta\right) \,, \\
    \hat{\mathcal{M}}_{i}^{{\rm WTA}} \left(z, \varphi, z_w,r, R,\omega_J, \mu, \zeta\right) &= \hat{h}_{ij}\left(z,\omega_J R, \mu\right) \, \hat{\mathcal{M}}_{j}^{{\rm WTA}} \left(\varphi, z_w,r, R,\omega_J, \mu, \zeta\right) \,.
\end{align}
The matching coefficient $\mathcal{J}$ can be easily calculated but we leave this for a later study.

\subsection{The wedge function}
In this section, we calculate the expressions for the wedge function and the matching coefficients for the jet wedge functions at one-loop. Using these expressions, we obtain the resummation equations for these functions.

The non-perturbative contribution to the $\varphi$ dependent jet shapes are determined by the wedge function. To obtain the wedge function, an event must be reconstructed. Starting from the parent parton, an emission is within the wedge so long as the azimuthal angle of the emission is within a range determined by $\varphi$. Emissions outside of that this azimuthal range are removed from the wedge, while emissions from the remaining particles are considered. If the wedge extends over the full azimuthal angle, momentum conservation enforces that
\begin{align}
    \hat{w}_i^{\rm axis}\left(\varphi,z, \mu\right)|_{\varphi = 2\pi} = \delta\left(1-z\right)\,.
\end{align}
We now note that this expression holds to all orders in perturbation theory since all loop corrections to this will always yield a zero result. 

For values with $\varphi < 2\pi$, radiative emissions which are outside of the wedge will result in an energy loss of the wedge at each step in the reconstruction, leading to evolution in the scale $\mu$. As the function is non-perturbative, we will need to parameterize the function at the initial scale. We take the natural parameterization that
\begin{align}
    \hat{w}^{{\rm axis}\, (0)}_i\left(\varphi,z, \mu_0\right) = \frac{\varphi}{2\pi}\delta\left(1-z\right)\,,
\end{align}
where $\mu_0 \sim \Lambda_{\rm QCD}$ is the initial scale of the wedge function. The physical interpretation of this expression is that a parton that travels along the direction of the wedge will have a probability $\varphi/2\pi$ of being clustered into the wedge. This parameterization enforces that $\langle z \rangle = \varphi/2\pi$ at the initial scale, and that the initial parameterization is isotropic. The isotropic condition, however, breaks down  with the introduction of spin correlations, which can introduce non-trivial azimuthal dependence. A transversely polarized quark, for instance, will result in an anisotropic pattern of radiation in the final-state. By momentum conservation, the wedge function in the $\varphi \rightarrow 2\pi$ region for a transversely polarized quark will remain unchanged, regions in the azimuthal plane will contain more collinear momentum than others. In this paper, we study the observable and discuss complications associated with these functions while we leave spin dependence for another study. 

At one-loop, the expression for the SJA and WTA wedge function can be written down  as
\begin{align}
    \hat{w}^{{\rm SJA}\, (1)}_i\left(\varphi,z\right) & = \operatorname{min}\left(\frac{\varphi}{2\pi}, \frac{2\pi-\varphi}{2\pi}\right)\frac{\alpha_s}{2\pi} \sum_j c_{ji} P_{ji}(z)\left(\frac{1}{\epsilon_{\rm UV}}-\frac{1}{\epsilon_{\rm IR}}\right)\,,
    \\
    \hat{w}^{{\rm WTA}\, (1)}_i\left(\varphi,z\right) & = \frac{\varphi}{2\pi}\left(\frac{2\pi-\varphi}{2\pi}\right)\frac{\alpha_s}{2\pi} \sum_j c_{ji} P_{ji}(z)\left(\frac{1}{\epsilon_{\rm UV}}-\frac{1}{\epsilon_{\rm IR}}\right)\,,
\end{align}
where we exploited the fact that contributions with both partons entering the wedge vanish, as will contributions where no partons enter the jet wedge. The additional factor of $\varphi/2\pi$ on the expression with the WTA axis enters because the parton that remains within the wedge only contributes a portion $\varphi/2\pi$ to the energy of the wedge. From these expressions, one can easily see that the renormalized wedge function obeys a modification of the standard DGLAP evolution equation
\begin{align}
    \frac{d}{d\ln\mu}w^{\rm SJA}\left(\varphi, z, \mu\right) & = \operatorname{min}\left(1, \frac{2\pi-\varphi}{\varphi}\right) \frac{\alpha_s}{2\pi}\sum_j c_{ji} \left[P_{ji}\otimes w_i \right]\left(\varphi, z, \mu\right)\,,
    \\
    \frac{d}{d\ln\mu}w^{\rm WTA}\left(\varphi, z, \mu\right) & = \left(\frac{2\pi-\varphi}{2\pi}\right) \frac{\alpha_s}{2\pi}\sum_j c_{ji} \left[P_{ji}\otimes w_i \right]\left(\varphi, z, \mu\right)\,.
\end{align}
The evolution equation for the SJA satisfies the condition that it obeys the standard DGLAP below $\varphi < \pi$, while as $\varphi$ approaches $2\pi$, the anomalous dimension vanishes, which is required by momentum conservation. Additionally, the anomalous dimension of the WTA wedge function vanishes in this limit as well.

\subsection{The $\varphi$ dependent sub-jet function}
In the region where $\omega_J R \gg \Lambda_{\rm QCD}$, the jet wedge function can be written as
\begin{align}
    \hat{j}_{i\, {\rm alg}}^{\rm axis}\left(\varphi, z_w, R, \omega_J, \mu, \zeta\right) = \left[\hat{\mathcal{K}}_{i\, {\rm alg}}^{{\rm axis}} \otimes \hat{w}_i^{\rm axis}\right]\left(\varphi, z_w, R, \omega_J, \mu, \zeta\right)\,.
\end{align}
At tree level, we have the trivial matching coefficient 
\begin{align}
    \hat{\mathcal{K}}_{i\, {\rm alg}}^{{\rm axis} (0)}\left(z_w, R, \omega_J, \mu\right) = \delta\left(1-z_w\right)\,.
\end{align}
To obtain the matching function  at one-loop, we now note that the graphs for the jet wedge function contain IR divergences that are identical to those of the wedge function. These IR divergences occur due to IR emissions outside of the jet wedge but inside of the jet radius. At one loop, we must consider two configurations. The first configuration is associated with both partons being inside the jet but only one parton is within the wedge. The second configuration is associated with both partons being inside the jet and inside the wedge. The latter are identical to exclusive jet functions except for the angular integrations. Upon considering the azimuthal angles of the emissions, the matching functions take on the form
\begin{align}
    \hat{\mathcal{K}}_{i\, {\rm alg}}^{{\rm SJA}\, (1)}\left(\varphi, z_w, R, \omega_J, \mu\right) = & \theta\left(\pi-\varphi\right) \frac{\varphi}{2\pi} \hat{\mathcal{K}}_{i\, {\rm alg}\, A_{\rm II}}^{{\rm SJA}\, (1)}\left(z_w, R, \omega_J, \mu\right) + \theta\left(\varphi-\pi\right) \nn \\
    & \times \Bigg[\left(\frac{\varphi-\pi}{\pi}\right) \delta\left(1-z_w\right)\hat{J}_i^{\rm SJA}\left(R, \omega_J, \mu\right) \nn \\
    & + \left(\frac{2\pi-\varphi}{2\pi}\right) \hat{\mathcal{K}}_{i\, {\rm alg}\, A_{\rm II}}^{{\rm SJA}\, (1)}\left(z_w, R, \omega_J, \mu\right) \Bigg]\,,
    \\
    \hat{\mathcal{K}}_{i}^{{\rm WTA}\, (1)}\left(\varphi, z_w, R, \omega_J, \mu\right) & = \frac{\varphi}{2\pi}\delta\left(1-z_w\right)\hat{J}_i^{\rm WTA}\left(R, \omega_J, \mu\right) \nn \\
    & + \left(\frac{2\pi-\varphi}{2\pi}\right) \hat{\mathcal{K}}_{i\, A_{\rm II}}^{{\rm WTA}\, (1)}\left(z_w, R, \omega_J, \mu\right)\,,
\end{align}
where the terms that are associated with the $A_{\rm II}$ contain the IR divergences. This naming convention will becomes clear in Sec.~\ref{subsec:rphi}. We find that the expression for these contributions are given by
\begin{align}
    \hat{\mathcal{K}}_{q\, {\rm Cone}\, A_{\rm II}}^{{\rm SJA}\, (1)}\left(z_w, R, \omega_J, \mu\right) = & \frac{\alpha_s C_F}{2\pi}\Bigg[\delta\left(1-z_w\right)\left(\frac{1}{\epsilon^2}-\frac{L_R}{\epsilon}+\frac{3}{2\epsilon}+\frac{1}{2}L_R^2-\frac{3}{2}L_R-\frac{\pi^2}{12}+2\right) \nn \\
    & +2\left(P_{qq}(z_w)+P_{gq}(z_w)\right)\left(L_R+\ln\left(z_w\left(1-z_w\right)\right)\right)+2\Bigg]
    \\
    \hat{\mathcal{K}}_{g\, {\rm Cone}\, A_{\rm II}}^{{\rm SJA}\, (1)}\left(z_w, R, \omega_J, \mu\right) = &\frac{\alpha_s}{2\pi} \Bigg\{ C_A \Bigg[\delta\left(1-z_w\right)\left(\frac{1}{\epsilon^2}-\frac{L_R}{\epsilon}+\frac{11}{6\epsilon}+\frac{1}{2}L_R^2-\frac{11}{6}L_R-\frac{\pi^2}{12} \right) \nn \\
    & +P_{gg}(z_w)\left(L_R+2\ln(1-z_w)\right)\Bigg] 
    + n_f\left[\delta\left(1-z_w\right)\left(-\frac{1}{3\epsilon}+\frac{L_R}{3}\right)  \right. \nn \\  
    &   +P_{gq}(z_w)\left(L_R+2\ln(1-z_w)-1\right)+1\Bigg] \Bigg\}\, ,
\end{align}
\begin{align}
    \hat{\mathcal{K}}_{q\, {\rm k_T}\, A_{\rm II}}^{{\rm SJA}\, (1)}  \left(z_w, R, \omega_J, \mu\right) =  &\frac{\alpha_s C_F}{2\pi}\Bigg[\delta\left(1-z_w\right)\left(\frac{1}{\epsilon^2}-\frac{L_R}{\epsilon}+\frac{3}{2\epsilon}+\frac{1}{2}L_R^2-\frac{3}{2}L_R-\frac{\pi^2}{12}+2\right) \nn \\
    & +2\left(P_{qq}(z_w)+P_{gq}(z_w)\right)\left(L_R+\ln\left(z_w\left(1-z_w\right)\right)\right)+2\Bigg] \, ,
    \\
    \hat{\mathcal{K}}_{g\, {\rm k_T}\, A_{\rm II}}^{{\rm SJA}\, (1)} \left(z_w, R, \omega_J, \mu\right) = & \frac{\alpha_s}{2\pi} \Bigg\{ C_A \Bigg[\delta\left(1-z_w\right)\left(\frac{1}{\epsilon^2}-\frac{L_R}{\epsilon}+\frac{11}{6\epsilon}+\frac{1}{2}L_R^2-\frac{11}{6}L_R-\frac{\pi^2}{12} \right) \nn \\
    & +P_{gg}(z_w)\left(L_R+2\ln\left(z_w (1-z_w)\right)\right)\Bigg]  
    + n_f\left[\delta\left(1-z_w\right)\left(-\frac{1}{3\epsilon}+\frac{L_R}{3}\right)    \right. \nn \\  &    +P_{gq}(z_w)\left(L_R+2\ln\left(z_w(1-z_w)\right)+2z_w(1-z_w)\right)\Bigg] \Bigg\} \, ,
\end{align}
for the SJA. For the WTA 
\begin{align}
    \hat{\mathcal{K}}_{q\, A_{\rm II}}^{{\rm WTA}\, (1)}\left(z_w, R, \omega_J, \mu\right) &= \hat{\mathcal{K}}_{q\, {\rm Cone}\, A_{\rm II}}^{{\rm SJA}\, (1)}\left(z_w, R, \omega_J, \mu\right) \, , 
    \\
    \hat{\mathcal{K}}_{g\, A_{\rm II}}^{{\rm WTA}\, (1)}\left(z_w, R, \omega_J, \mu\right) &= \hat{\mathcal{K}}_{g\, {\rm Cone}\, A_{\rm II}}^{{\rm SJA}\, (1)}\left(z_w, R, \omega_J, \mu\right) \, .
\end{align}
The semi-inclusive matching functions can be obtained by considering the hard-collinear contributions. From this result, we obtain the one loop expressions for the total matching functions
\begin{align}
    \hat{\mathcal{K}}_{q\, {\rm Cone}}^{{\rm SJA}\, (1)} & \left(\varphi, z_w, R, \omega_J, \mu\right) = \frac{\alpha_s C_F}{2\pi}\Bigg\{\delta\left(1-z_w\right) \nn \\
    &\times \Bigg[\frac{\varphi}{2\pi} \left(\frac{1}{\epsilon^2}-\frac{L_r}{\epsilon}+\frac{3}{2\epsilon}+\frac{1}{2}L_r^2-\frac{3}{2}L_r-\frac{3\pi^2}{4}+7+6\ln(2)\right) +\frac{\pi^2}{3}-\frac{7}{2}-3\ln(2)\Bigg] \nn \\
    & +\left(\frac{2\pi-\varphi}{\pi}\right)\Big[ \left(P_{qq}(z_w)+P_{gq}(z_w)\right) \left(L_r+2 \ln\left(z_w(1-z_w)\right)_+\right)+2\Big]\Bigg\}\,,
    \\
    \hat{\mathcal{K}}_{g\, {\rm Cone}}^{{\rm SJA}\, (1)} & \left(\varphi, z_w, R, \omega_J, \mu\right) = \frac{\alpha_s}{2\pi}\Bigg\{\delta\left(1-z_w\right) \nn \\
    &\times \Bigg[\frac{\varphi}{2\pi} C_A \left(\frac{1}{\epsilon^2}-\frac{L_r}{\epsilon}+\frac{11}{6\epsilon}+\frac{1}{2}L_r^2-\frac{11}{6}L_r-\frac{3\pi^2}{4}+\frac{137}{18}+\frac{22}{3}\ln(2)\right) \nn \\
    & + \frac{\varphi}{2\pi} n_f \left(-\frac{1}{3\epsilon}+\frac{1}{3}L_r-\frac{23}{18}-\frac{4}{3}\ln(2)\right) \Bigg] \nn \\
    & +\left(\frac{2\pi-\varphi}{\pi}\right) \Big[\left(C_A P_{qq}(z_w)+n_f P_{qg}(z_w)\right)\left[L_r+2\ln(1-z_w)\right]+2n_f z_w(1-z_w) \Big]\nn \\
    & + \left[C_A\left(\frac{\pi^2}{3}-\frac{11}{3}\ln(2)-\frac{137}{36}\right)+n_f\left(\frac{23}{36}+\frac{2}{3}\ln(2)\right)\right] \Bigg\}\,, 
    \\
    \hat{\mathcal{K}}_{q\, {\rm k_T}}^{{\rm SJA}\, (1)} & \left(\varphi, z_w, R, \omega_J, \mu\right) = \frac{\alpha_s C_F}{2\pi}\Bigg\{\delta\left(1-z_w\right) \nn \\
    &\times \Bigg[\frac{\varphi}{2\pi} \left(\frac{1}{\epsilon^2}-\frac{L_r}{\epsilon}+\frac{3}{2\epsilon}+\frac{1}{2}L_r^2-\frac{3}{2}L_r-\frac{17}{12}\pi^2+13\right) +\frac{2}{3}\pi^2-\frac{13}{2}\Bigg] \nn \\
    & +\left(\frac{2\pi-\varphi}{\pi}\right)\Big[ \left(P_{qq}(z_w)+P_{gq}(z_w)\right) \left(L_r+2 \ln\left(z_w(1-z_w)\right)_+\right)+1\Big]\Bigg\}\,,
    \\
    \hat{\mathcal{K}}_{g\, {\rm k_T}}^{{\rm SJA}\, (1)} & \left(\varphi, z_w, R, \omega_J, \mu\right) = \frac{\alpha_s}{2\pi}\Bigg\{\delta\left(1-z_w\right) \nn \\
    &\times \Bigg[\frac{\varphi}{2\pi} C_A \left(\frac{1}{\epsilon^2}-\frac{L_r}{\epsilon}+\frac{11}{6\epsilon}+\frac{1}{2}L_r^2-\frac{11}{6}L_r-\frac{17}{12}\pi^2+\frac{134}{9}\right) \nn \\
    & + \frac{\varphi}{2\pi} n_f \left(-\frac{1}{3\epsilon}+\frac{1}{3}L_r-\frac{23}{9}\right) \Bigg] \nn \\
    & +\left(\frac{2\pi-\varphi}{\pi}\right) \Big[\left(C_A P_{qq}(z_w)+n_f P_{qg}(z_w)\right)\left[L_r+2\ln(z_w(1-z_w))_+\right]+2n_f z_w(1-z_w) \Big]\nn \\
    & + \left[C_A\left(\frac{2}{3}\pi^2-\frac{67}{9}\right)+n_f\left(\frac{23}{18}\right)\right] \Bigg\}\,, 
\end{align}
for the SJA, and for the WTA
\begin{align}
    \hat{\mathcal{K}}_{q}^{{\rm WTA}\, (1)} & \left(\varphi, z_w, R, \omega_J, \mu\right) = \hat{\mathcal{K}}_{q\, {\rm Cone}}^{{\rm SJA}\, (1)} \left(\varphi, z_w, R, \omega_J, \mu\right)\,,
    \\
    \hat{\mathcal{K}}_{g}^{{\rm WTA}\, (1)} & \left(\varphi, z_w, R, \omega_J, \mu\right) = \hat{\mathcal{K}}_{g\, {\rm Cone}}^{{\rm SJA}\, (1)} \left(\varphi, z_w, R, \omega_J, \mu\right)\,.
\end{align}

The $\varphi$ dependent jet shape can be determined by the azimuthal-dependent jet wedge from Eqs.~\eqref{shape0},~\eqref{eq:shape}. At one-loop, these expressions reduce to 
\begin{align}
    \psi_{i\, {\rm alg}}^{{\rm axis}}(\varphi,R) = &\frac{\varphi}{2\pi} +  \int dz_w z_w j_c^{(1)}\left(\varphi, z_w, R, \omega_J,\mu\right) - J_c^{(1)}\left(R,\omega_J,\mu\right) + \mathcal{O}\left(\alpha_s^2\right) \, .
\end{align}
The one-loop results for the azimuthal-dependent jet shapes are then simply given by the $z_w$ weighted expressions for the matching coefficients. We now, however, note an important point. In the expressions for the jet shape we saw the appearance of IR divergences that require the introduction of the wedge function. However, by introducing the energy weighing the IR divergence terms will go like 
\begin{align}
    \hat{j}_{i\, {\rm alg\, IR}}^{{\rm axis}} \sim \frac{1}{\epsilon_{\rm IR}}\int d z_w z_w \sum_j P_{ji}(z_w) = 0\, .
\end{align}
Consequently, the energy weighting results in the IR divergences vanishing. For every expression for the jet shape, we then obtain the result that
\begin{align}
    \psi_{i\, {\rm alg}}^{{\rm axis}}(\varphi,R) = \frac{\varphi}{2\pi} + \left(\frac{\varphi-2\pi}{2\pi}\right)J_{i\, {\rm alg}}^{{\rm axis}\, (1)}\left(R, \omega_J, \mu\right)+\mathcal{O}\left(\alpha_s^2\right)\, ,
\end{align}
exactly as expected. 

\subsection{The $r$ and $\varphi$ dependent sub-jet function}\label{subsec:rphi}
\begin{figure}
    \centering
    \includegraphics[width = 0.49\textwidth, valign = c]{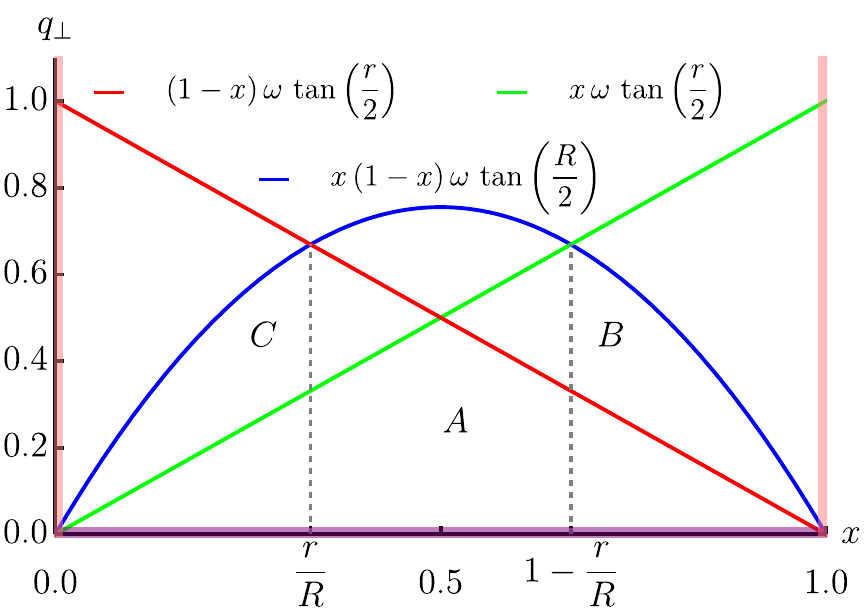}
    \includegraphics[width = 0.49\textwidth, valign = c]{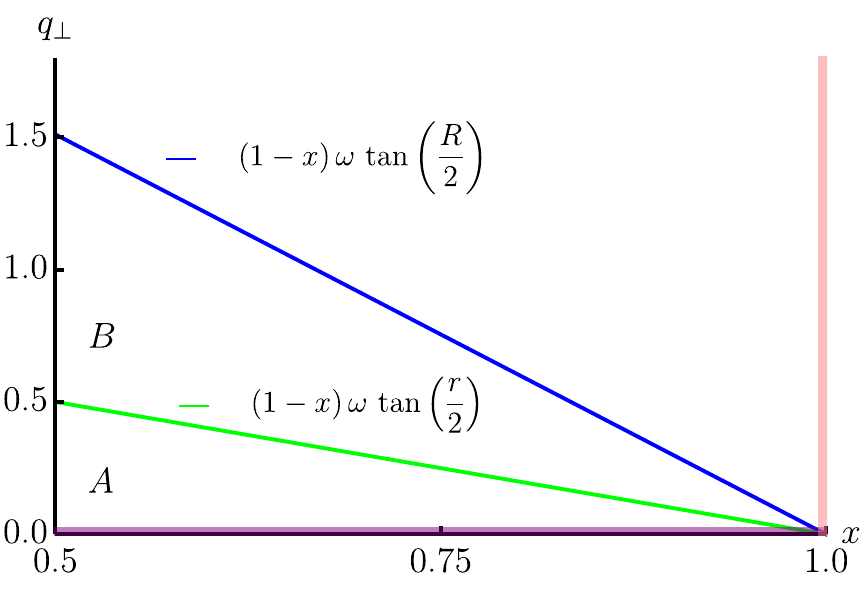}
    \caption{Left: Kinematic regions for the jet shape for the SJA. Right: Same but for the WTA axis. The blue curves represent the restriction that the emission is inside of the jet while the red and green curve represent that the restrictions associated with the subjet radius. Each region is labelled (A, B, C). The purple lines along $q_\perp = 0$ represents the collinear divergences while the pink lines at $x = 0$ and $1$ represent soft divergences.}
    \label{fig:Regions}
\end{figure}
To compute the $r$ and $\varphi$ dependent sub-jet function, we must first compute the central sub-jet wedge function. To achieve this, we need to consider emissions that are outside of the sub-jet wedge and jet, emissions that are inside the jet but outside of the sub-jet wedge, and emissions that are inside the jet and sub-jet wedge. The relevant configuration depends on the particular values of $x$, $q_\perp$, and $\varphi$. At one-loop, the semi-inclusive jet shape with a SJA contains five contributions, while the exclusive jet shape contains three contribution with a SJA. The case of the jet shape is simpler for the WTA jet axis, in which the semi-inclusive jet shape contains four contributions while the exclusive jet shape contains only two contributions. Each contribution is associated with integrating a partonic contribution over some region in $x$ and $q_\perp$ and is represented in Fig.~\ref{fig:Regions}. Thus, up to one-loop we will write
\begin{align}
    \hat{j}_{i\, {\rm alg}}^{{\rm axis}} & \left(\varphi,z_w,r, R,\mu, \zeta\right)  =  \hat{w}_i^{\rm axis}\left(\varphi, z_w, \mu\right) \nn \\
    & + \sum_G \left[\hat{\mathcal{M}}_{i\, {\rm alg}}^{{\rm axis}\, (G)\, (1)}\otimes \hat{w}_i^{\rm axis} \right] \left(\varphi, z_w, r, R, \mu, \zeta\right) + \mathcal{O}\left(\alpha_s^2\right) \, , 
\end{align}
where regions $A$ through $C$ are depicted in Fig.~\ref{fig:Regions}. The semi-inclusive sub-jet function contains contributions which cannot be depicted on the left side of Fig.~\ref{fig:Regions} due to the requirement of a different frame choice for emissions outside of the jet. To compute the sub-jet wedge function, we now note that for the SJA the contributions from $B$ and $C$ contain only a single parton within the sub-jet and are therefore related to the azimuthally integrated jet shape via the relations
\begin{align}
    \hat{\mathcal{M}}_{i\,  {\rm alg}}^{{\rm axis}\, (g)\, (1)} \left(\varphi, z_w,r, R,\mu, \zeta\right) & = \frac{\varphi}{2\pi} \hat{j}_{i\, {\rm alg}}^{{\rm axis} (g)\, (1)} \left(z_w,r, R,\mu, \zeta\right)\,.
\end{align}
The expressions for the azimuthally integrated jet shapes are given in the appendix, and $g$ is an element of $\left(B, C\right)$. However for the A contributions, we find that
\begin{align}
    \hat{\mathcal{M}}_{i\,  {\rm alg}}^{{\rm axis}\, ({\rm A})\, (1)} \left(\varphi, z_w,r, R,\mu, \zeta\right) = \hat{\mathcal{K}}_{i\,  {\rm cone}}^{{\rm axis}\, (1)} \left(\varphi, z_w, r,\mu, \zeta\right) \, .
\end{align}
The matching coefficients for graph $A$, however, are precisely given by the matching coefficient functions for the wedge jet function with a cone algorithm onto the wedge function. 
\begin{align}
    \hat{\mathcal{M}}_{q\, {\rm k_T}}^{{\rm SJA}\, (1)}  &\left(\varphi, z_w,r, R,\mu, \zeta\right) =  \frac{\alpha_s C_F}{2\pi} \Bigg\{ \delta\left(1-z_w\right) \Bigg[ \frac{\varphi}{2\pi} \Bigg(\frac{1}{\epsilon^2}+\frac{3}{2 \epsilon}-\frac{L_r}{\epsilon}-\frac{L_{r/R}}{\epsilon} \nn \\
    & + L_R L_{r/R}+\frac{L_{r}^2}{2}-\frac{3 L_{r}}{2}-\frac{L_{r/R}^2}{2}-\frac{3 L_{r/R}}{2}-\frac{3 \pi ^2}{4}\Bigg) \nn \\
    & + \left(\frac{\pi-\varphi}{2\pi}\right) \left(-7 - 6\ln(2)\right) + \frac{\pi^2}{3}\Bigg] + \frac{\varphi}{\pi} \nn \\
    & + \frac{\varphi}{2\pi} (P_{qq}(z_w)+P_{gq}(z_w)) \Bigg[(L_{r/R}+2 \ln (z_w)) \theta \left(\frac{r}{R}+z_w-1\right) \nn \\
    & +2 (L_r+\ln (1-z_w)+\ln
   (z_w))\Bigg]\Bigg\} \, , 
   \\
    \hat{\mathcal{M}}_{g\, {\rm k_T}}^{{\rm SJA}\, (1)} & \left(\varphi, z_w,r, R,\mu, \zeta\right) =  \frac{\alpha_s}{2\pi} \Bigg[ \delta\left(1-z_w\right) \Bigg\{ \frac{\varphi}{2\pi} C_A \Bigg(\frac{1}{\epsilon^2}+\frac{11}{6 \epsilon}-\frac{L_r}{\epsilon}-\frac{L_{r/R}}{\epsilon} \nn \\
    & + L_R L_{r/R}+\frac{L_{r}^2}{2}-\frac{11 L_{r}}{6}-\frac{L_{r/R}^2}{2}-\frac{11 L_{r/R}}{6}-\frac{3 \pi ^2}{4}\Bigg) \nn \\
    & + \frac{\varphi}{2\pi} n_f \Bigg(-\frac{1}{3\epsilon}+\frac{L_r}{3}+\frac{L_{r/R}}{2}\Bigg) \nn \\
    & + \left(\frac{\pi-\varphi}{2\pi}\right) \left[C_A \left(-\frac{137}{18}-\frac{22}{3}\ln(2)\right)+n_f\left(\frac{23}{18}+\frac{4}{3}\ln(2)\right)\right] + C_A \frac{\pi^2}{3}\Bigg\} \nn \\
    & + \frac{\varphi}{\pi} n_f z_w\left(1-z_w\right) \nn \\
    & + \frac{\varphi}{2\pi} (C_A P_{gg}(z_w)+n_f P_{qg}(z_w)) \Bigg[(L_{r/R}+2 \ln (z_w)) \theta \left(\frac{r}{R}+z_w-1\right) \nn \\
    & +2 (L_r+\ln (1-z_w)+\ln
   (z_w))\Bigg]\Bigg] \, , 
\end{align}
\begin{align}
    \hat{\mathcal{M}}_{q\, {\rm Cone}}^{{\rm SJA}\, (1)} & \left(\varphi, z_w,r, R,\mu, \zeta\right) =  \frac{\alpha_s C_F}{2\pi} \Bigg\{ \delta\left(1-z_w\right) \Bigg[ \frac{\varphi}{2\pi} \Bigg(\frac{1}{\epsilon^2}+\frac{3}{2 \epsilon}-\frac{L_r}{\epsilon}-\frac{L_{r/R}}{\epsilon} \nn \\
    & + L_R L_{r/R}+\frac{L_{r}^2}{2}-\frac{3 L_{r}}{2}-\frac{L_{r/R}^2}{2}-\frac{3 L_{r/R}}{2}-\frac{3 \pi ^2}{4}\Bigg) \nn \\
    & + \left(\frac{\pi-\varphi}{2\pi}\right) \left(-7 - 6\ln(2)\right) + \frac{\pi^2}{3}\Bigg] + \frac{\varphi}{\pi} \nn \\
    & + \frac{\varphi}{2\pi} (P_{qq}(z_w)+P_{gq}(z_w)) \Bigg[2\Bigg(\ln\left(\frac{z_w}{1-z_w}\right) \Theta\left(\frac{R}{r+R}-z_w, z_w-\frac{1}{2}\right) \nn \\
    & +L_r+\ln(1-z_w)+\ln(z_w)\Bigg) + L_{r/R} \Theta\left(z_w-\frac{R}{r+R}\right)\Bigg]\Bigg\} \, , 
   \\
    \hat{\mathcal{M}}_{g\, {\rm Cone}}^{{\rm SJA}\, (1)} &\left(\varphi, z_w,r, R,\mu, \zeta\right) =  \frac{\alpha_s}{2\pi} \Bigg[ \delta\left(1-z_w\right) \Bigg\{ \frac{\varphi}{2\pi} C_A \Bigg(\frac{1}{\epsilon^2}+\frac{11}{6 \epsilon}-\frac{L_r}{\epsilon}-\frac{L_{r/R}}{\epsilon} \nn \\
    & + L_R L_{r/R}+\frac{L_{r}^2}{2}-\frac{11 L_{r}}{6}-\frac{L_{r/R}^2}{2}-\frac{11 L_{r/R}}{6}-\frac{3 \pi ^2}{4}\Bigg) \nn \\
    & + \frac{\varphi}{2\pi} n_f \Bigg(-\frac{1}{3\epsilon}+\frac{L_r}{3}+\frac{L_{r/R}}{2}\Bigg) \nn \\
    & + \left(\frac{\pi-\varphi}{2\pi}\right) \left[C_A \left(-\frac{137}{18}-\frac{22}{3}\ln(2)\right)+n_f\left(\frac{23}{18}+\frac{4}{3}\ln(2)\right)\right] + C_A \frac{\pi^2}{3}\Bigg\} \nn \\
    & + \frac{\varphi}{\pi} n_f z_w\left(1-z_w\right) \nn \\
    & + \frac{\varphi}{2\pi} (C_A P_{gg}(z_w)+n_f P_{qg}(z_w)) \Bigg[ 2\ln\left(\frac{z_r}{1-z_r}\right) \Theta\left(\frac{R}{r+R}-z_r, z_r-\frac{1}{2}\right) \nn \\
    & + L_{r/R}\Theta\left(z_r-\frac{R}{r+R}\right)+ L_r+2 \ln\left(1-z_r\right)\Bigg]\Bigg] \, , 
\end{align}
\begin{align}
    \hat{\mathcal{M}}_{q}^{{\rm WTA}\, (1)}& \left(\varphi, z_w,r, R,\mu, \zeta\right) =  \frac{\alpha_s C_F}{2\pi} \Bigg\{ \delta\left(1-z_w\right) \Bigg[ \frac{\varphi}{2\pi} \Bigg(\frac{1}{\epsilon^2}+\frac{3}{2 \epsilon}-\frac{L_r}{\epsilon}-\frac{L_{r/R}}{\epsilon} \nn \\
    & + L_R L_{r/R}+\frac{L_{r}^2}{2}-\frac{3 L_{r}}{2}-\frac{L_{r/R}^2}{2}-\frac{3 L_{r/R}}{2}-\frac{3 \pi ^2}{4}\Bigg) \nn \\
    & + \left(\frac{\pi-\varphi}{2\pi}\right) \left(-7 - 6\ln(2)\right) + \frac{\pi^2}{3}\Bigg] + \frac{\varphi}{\pi} + \frac{\varphi}{2\pi} (P_{qq}(z_w)+P_{gq}(z_w)) \Bigg[\theta \left(z_w-\frac{1}{2}\right) L_{r/R}  \nn \\
    & +2 (L_r+\ln (1-z_w)+\ln
   (z_w))\Bigg]\Bigg\} \, , 
   \\
    \hat{\mathcal{M}}_{g}^{{\rm WTA}\, (1)} & \left(\varphi, z_w,r, R,\mu, \zeta\right) =  \frac{\alpha_s}{2\pi} \Bigg[ \delta\left(1-z_w\right) \Bigg\{ \frac{\varphi}{2\pi} C_A \Bigg(\frac{1}{\epsilon^2}+\frac{11}{6 \epsilon}-\frac{L_r}{\epsilon}-\frac{L_{r/R}}{\epsilon} \nn \\
    & + L_R L_{r/R}+\frac{L_{r}^2}{2}-\frac{11 L_{r}}{6}-\frac{L_{r/R}^2}{2}-\frac{11 L_{r/R}}{6}-\frac{3 \pi ^2}{4}\Bigg) \nn \\
    & + \frac{\varphi}{2\pi} n_f \Bigg(-\frac{1}{3\epsilon}+\frac{L_r}{3}+\frac{L_{r/R}}{2}\Bigg) \nn \\
    & + \left(\frac{\pi-\varphi}{2\pi}\right) \left[C_A \left(-\frac{137}{18}-\frac{22}{3}\ln(2)\right)+n_f\left(\frac{23}{18}+\frac{4}{3}\ln(2)\right)\right] + C_A \frac{\pi^2}{3}\Bigg\} \nn \\
    & + \frac{\varphi}{\pi} n_f z_w\left(1-z_w\right) + \frac{\varphi}{2\pi} (C_A P_{gg}(z_w)+n_f P_{qg}(z_w)) \Bigg[\theta \left(z_w-\frac{1}{2}\right) L_{r/R} \nn \\
    & +2 (L_r+\ln (1-z_w)+\ln
   (z_w))\Bigg]\Bigg] \, .
\end{align}
In the previous section, we saw that the energy weighting removed the IR divergences. After performing the $z_w$ weighting, we obtain the expression for the $r$ and $\varphi$ dependent jet shape
\begin{align}
    \psi_{q\, {\rm k_T}}^{\rm SJA}\left(\varphi, r, R\right)  = &\frac{\varphi}{2\pi}+\frac{\alpha_s C_F}{2\pi} \frac{\varphi}{2\pi}\left[-\frac{1}{2}L_{r/R}^2+\frac{3}{2}L_{r/R}-\frac{9}{2}+6\frac{r}{R}-\frac{3}{2}\frac{r^2}{R^2}\right] \nn \\
    & + \left(\frac{\varphi-2\pi}{2\pi}\right)J_{q\, {\rm k_T}}^{{\rm SJA}\, (1)}\left(z, R, \omega\right) + \mathcal{O}\left(\alpha_s^2\right)\,,
    \\
    \psi_{g\, {\rm k_T}}^{\rm SJA}\left(\varphi, r, R\right)  = &\frac{\varphi}{2\pi}+ \frac{\alpha_s}{2\pi} \frac{\varphi}{2\pi} \Bigg[ C_A\left(- \frac{1}{2}L_{r/R}^2  -\frac{203}{36} +8\frac{r}{R} - 3\frac{r^2}{R^2} + \frac{8}{9}\frac{r^3}{R^3} - \frac{1}{4}\frac{r^4}{R^4}\right) 
    \nn \\ &
    + \frac{\beta_0}{2} L_{r/R} + n_f\left(\frac{41}{36} - 2\frac{r}{R} + \frac{3}{2}\frac{r^2}{R^2} - \frac{8r^3}{9R^3} + \frac{r^4}{4R^4}\right)\Bigg] \nn 
    \\
    & + 
    \left(\frac{\varphi-2\pi}{2\pi}\right)J^{{\rm SJA}\, (1)}_{g\, {\rm k_T}}(z, \omega_R, \mu) + \mathcal{O}\left(\alpha_s^2\right)
    \,, \\
    \psi_{q\, {\rm cone}}^{\rm SJA}\left(\varphi, r, R\right)  = &\frac{\varphi}{2\pi}+\frac{\alpha_s C_F}{2\pi}\frac{\varphi}{2\pi} \Bigg[-\frac{1}{2}L_{r/R}^2+\frac{3}{2}L_{r/R}+\frac{3}{2} -3\ln(2)-\frac{\pi^2}{3} \nn\\ & 
    +4{\rm Li}_2\left(\frac{r}{r+R}\right)+2\ln^2\left(1+\frac{r}{R}\right) +3\ln\left(1+\frac{r}{R}\right)+\frac{3r}{r+R}\Bigg] \nn \\
    & + \left(\frac{\varphi-2\pi}{2\pi}\right)J_{q\, {\rm cone}}^{{\rm SJA}\, (1)}\left(z, R, \omega\right) + \mathcal{O}\left(\alpha_s^2\right)\,, \\
    \psi_{g\, {\rm cone}}^{\rm SJA}\left(\varphi, r, R\right) = &\frac{\varphi}{2\pi}  + \frac{\alpha_s}{2\pi} \frac{\varphi}{2\pi} \Bigg[C_A\Bigg( -\frac{L_{r/R}^2}{2} +2\ln^2\left(1+\frac{r}{R}\right) \nn \\
    &  +4 \,{\rm Li}_2\left(\frac{r}{r+R}\right) -\frac{\pi^2}{3}+ \frac{R-r}{6(r+R)^3}\left(11r^2+22 rR+12R^2\right) \Bigg)
     \nn\\ &
    +\beta_0\left(\frac{L_{r/R}}{2}+\ln\left(1+\frac{r}{R}\right)-\ln(2) \right) + n_f \frac{R-r}{3(r+R)^3} \left(r^2+2rR+\frac{3}{2}R^2\right) \Bigg] \nn 
    \\
    & + 
    \left(\frac{\varphi-2\pi}{2\pi}\right)J^{{\rm SJA}\, (1)}_{g\, {\rm cone}}(z, \omega_R, \mu) + \mathcal{O}\left(\alpha_s^2\right)
    \,, 
    \\
    \psi_{q}^{\rm WTA}\left(\varphi, r, R\right)  = &\frac{\varphi}{2\pi}+\frac{\alpha_s C_F}{2\pi} \frac{\varphi}{2\pi} L_{r/R}\left(\frac{3}{8}-2\ln(2)\right) \nn \\
    & + \left(\frac{\varphi-2\pi}{2\pi}\right)J_{q}^{{\rm WTA}\, (1)}\left(z, R, \omega\right) + \mathcal{O}\left(\alpha_s^2\right)\,,
    \\
    \psi_{g}^{\rm WTA}\left(\varphi, r, R\right)  =  &\frac{\varphi}{2\pi}+ \frac{\alpha_s}{2\pi} \frac{\varphi}{2\pi} L_{r/R}\bigg[C_A\Big(\frac{43}{96}-2\ln(2)\Big)-n_f \frac{7}{96}\bigg] \nn \\
    & + 
    \left(\frac{\varphi-2\pi}{2\pi}\right)J^{{\rm WTA}\, (1)}_{g\, {\rm k_T}}(z, \omega_R, \mu) + \mathcal{O}\left(\alpha_s^2\right)
    \,.
\end{align}
The results recover exactly the $r$ dependent jet shape in the limit that $\varphi = 2\pi$, while they recover the $\varphi$ dependent jet shape in the limit that $r$ approaches $R$. In these expressions, we have used the short-hand that $L_{r/R} = -2\ln{r/R}$.
\subsection{The soft-collinear function}
We now derive the expression for the soft-collinear function that enters into the factorization of the matching function in the resummed region in Eq.~\eqref{eq:Mfact}. The soft-collinear function for the sub-jet function with a SJA is given by the expression
\begin{align}
    \hat{S}_i^{(1)} \left(\bm{k}_\perp, R, \omega_J, \mu, \nu\right)  =  &2 g^2 C_i \left(\frac{\mu^2 e^{\gamma_E}}{4\pi}\right)^\epsilon \int \frac{d^dl}{\left(2\pi\right)^d}\frac{n_J \cdot \bar{n}_J}{n_J\cdot l\, \bar{n}_J\cdot l}2\pi \delta\left(l^2\right) \left(\frac{2 l_0}{\nu}\right)^{\eta} \nn \\
    & \times \delta^{2-2\epsilon}\left(\bm{l}_\perp-\bm{k}_\perp\right) \Theta \left(\tan^2{\frac{R}{2}}-\frac{l^+}{l^-}\right)\,.
\end{align}
The one-loop integration yields
\begin{align}
    \hat{S}_i^{(1)} \left(\bm{k}_\perp, R, \omega_J, \mu, \nu\right) = \frac{\alpha_s C_i}{\pi^2}\frac{1}{\eta}\left(\mu^2 \pi e^{\gamma_E}\right)^\epsilon \frac{\left(\nu \tan{\frac{R}{2}}\right)^\eta}{k_\perp^{2+\eta}}\,,
\end{align}
which gives the momentum space expression for the soft-collinear contribution
\begin{align}
    \hat{S}_i^{(1)} \left(\bm{k}_\perp, R, \omega_J, \mu, \nu\right) = \frac{\alpha_s C_i}{2\pi}\Bigg[ & \delta^{2-2\epsilon}\left(\bm{k}_\perp\right)\left(\frac{1}{\epsilon^2}-\frac{2}{\eta \epsilon} - \frac{1}{\epsilon} L_{\nu s}-\frac{\pi^2}{12}\right) \nn \\
    & + \left(\frac{1}{k_\perp}\right)_+ \left(\frac{4}{\eta}+2L_{\nu s} + \ln\left(\frac{\mu}{k_\perp}\right)_+\right)\Bigg]\,.
\end{align}
To obtain the soft contribution to the jet shape, we now note that if the soft region has a transverse momentum $\bm{k}_\perp$, that the collinear mode has a momentum of
\begin{align}
    p_c^\mu = \left\{ \frac{k_\perp^2}{\omega_J}, \omega_J, \bm{k}_\perp\right\}\,.
\end{align}
The one-loop expression for the soft-collinear function is then given by
\begin{align}
    \hat{\mathscr{C}}^{\rm SJA}   \left[C_i^{{\rm SJA}\, (0)}\, S_i^{(1)}\right] &\left(z_w, R, \omega_J, \mu, \nu\right) \nn \\
    & = \frac{\alpha_s C_i}{2\pi}\left(\frac{1}{\epsilon^2}-\frac{2}{\eta \epsilon}+\frac{L_{\nu s}}{\epsilon}+\frac{2 L_{\mu s}^{\rm SJA}}{\eta}-\frac{1}{2}{L_{\mu s}^{\rm SJA}}^2-L_{\mu s}^{\rm SJA}L_{\nu s}-\frac{\pi^2}{12}\right)\,,
\end{align}
where we have introduced the logs
\begin{align}
    L_{\mu s}^{\rm SJA} = 2\left(\frac{\omega_J \tan{\frac{R}{2}}}{\mu}\right)\,,
    \qquad
    L_{\nu s} = 2\left(\frac{\mu}{\nu \tan{\frac{R}{2}}}\right)\,.
\end{align}
The expressions for the evolution equation of the soft contribution are then given by
\begin{align}
    \frac{d}{d \ln \mu} \ln \mathscr{C}^{\rm SJA}  \left[C_i^{{\rm SJA}\, (0)}\, S_i^{(1)}\right] \left(z_w, R, \omega_J, \mu, \nu\right) & = \Gamma_i^{\rm cusp} L_{\mu s}\,,
    \\
    \frac{d}{d \ln \nu} \ln \mathscr{C}^{\rm SJA}  \left[C_i^{{\rm SJA}\, (0)}\, S_i^{(1)}\right] \left(z_w, R, \omega_J, \mu, \nu\right) & = \Gamma_i^{\rm cusp} L_{\nu s}\,.
\end{align}
\subsection{The hyper-collinear function}
We now derive the expression for the hyper-collinear function that enters into the factorization of the matching function in the resummed region in Eq.~\eqref{eq:Mfact}. We decompose the collinear function in terms of the contributions of the different graphs
\begin{align}
    \hat{C}_i^{\rm SJA}\left(\varphi, z_w, \bm{k}_\perp, r, \omega_J, \mu,\frac{\zeta}{\nu^2}\right) & =  
    \frac{\varphi}{2\pi}\delta\left(1-z_w\right) \nn \\
    & + \sum_G \hat{C}_i^{\rm SJA\, (G)}\left(\varphi, z_w, \bm{k}_\perp, r, \omega_J, \mu,\frac{\zeta}{\nu^2}\right) + \mathcal{O}\left(\alpha_s^2\right) \,,
\end{align}
where here $G \in \left(A, B, C\right)$ and the regions are the same as the fixed order region except that one takes $R \rightarrow \infty$. 

Without the effects of soft-recoil, the expressions for these graphs are
\begin{align}
    \hat{C}_i^{\rm SJA\ (A)}\left(\varphi, z_w, \bm{0}_\perp, r, \omega_J, \mu,\frac{\zeta}{\nu^2}\right)  = &\hat{\mathcal{K}}_{i\, {\rm Cone}}^{{\rm SJA}} \left(\varphi, z_w, R, \omega_J, \mu, \zeta\right)\,, \\
    \hat{C}_i^{\rm SJA\ (B)}\left(\varphi, z_w, \bm{0}_\perp, r, \omega_J, \mu,\frac{\zeta}{\nu^2}\right)  =  &\sum_j \int d\Phi\, \delta(x-z_w) \hat{P}_{ji}\left(x,q_\perp, \epsilon\right) \Theta_{((k)q)}^{\rm SJA} \nn \\
    &\times \left(\frac{\nu}{(1-x)\sqrt{\zeta}}\right)^{\eta}  \, , \\ 
    \hat{C}_i^{\rm SJA\ (C)}\left(\varphi, z_w, \bm{0}_\perp, r, \omega_J, \mu,\frac{\zeta}{\nu^2}\right)  = & \sum_j \int d\Phi\, \delta\left(x-(1-z_w)\right)\hat{P}_{ji}\left(x,q_\perp, \epsilon\right)  \Theta_{((q)k)}^{\rm SJA} \,, 
\end{align}
where $d\Phi = dx\, d^{2-2\epsilon}q_\perp$. Here, $\bm{0}_\perp$ denotes that the soft-recoil transverse momenta are taken to be zero. In these expressions, the jet algorithm constraints are given by
\begin{align}
    \Theta_{((k)q)}^{\rm SJA} & = \Theta\left[\omega_J x \tan{\frac{r}{2}}-q_\perp\right]\, \Theta\left[q_\perp-\omega_J (1-x) \tan{\frac{r}{2}}\right]\,, \\
    \Theta_{((q)k)}^{\rm SJA} & = \Theta\left[q_\perp-\omega_J x \tan{\frac{r}{2}}\right]\, \Theta\left[\omega_J (1-x) \tan{\frac{r}{2}}-q_\perp\right]\,.
\end{align}
The contributions from graphs (B) and (C) are the same as those for the azimuthally integrated case except for the factor of $\varphi/2\pi$. Their exact expressions were derived in Ref.~\cite{Kang:2017mda}. The one-loop expressions for the hyper-collinear functions are given by
\begin{align}
    \hat{C}_q^{\rm SJA\ (1)} & \left(\varphi, z_w, \bm{0}_\perp, r, \omega_J, \mu,\frac{\zeta}{\nu^2}\right) = \frac{\alpha_s C_F}{2\pi} \Bigg\{ \delta\left(1-z_w\right) \nn \\
    & \times \Bigg[ \frac{\varphi}{2\pi} \Bigg(\frac{2}{\eta \epsilon}-\frac{2}{\eta}L_{\mu c}+\frac{3}{2 \epsilon}-\frac{L_{\nu c}}{\epsilon} +L_{\nu c} L_{\mu c} -\frac{3}{2}L_{\mu c} -\frac{2\pi^2}{3}\Bigg) \nn \\
    & + \left(\frac{\pi-\varphi}{2\pi}\right) \left(-7 - 6\ln(2)\right) + \frac{\pi^2}{3}\Bigg] + \frac{\varphi}{\pi} + \frac{\varphi}{\pi} (P_{qq}(z_w)+P_{gq}(z_w)) \nn \\
    & \times \Bigg(\theta \left(z_w-\frac{1}{2}\right) \ln
   \left(\frac{z_w}{1-z_w}\right) +L_{\mu c}+2 \ln (1-z_w)+\ln
   \left(\frac{z_w}{1-z_w}\right)\Bigg)\Bigg\} \, ,
   \\
    \hat{C}_g^{\rm SJA\ (1)} & \left(\varphi, z_w, \bm{0}_\perp, r, \omega_J, \mu,\frac{\zeta}{\nu^2}\right) = \frac{\alpha_s}{2\pi} \Bigg[ \delta\left(1-z_w\right)  \nn \\
    & \times \Bigg\{ \frac{\varphi}{2\pi} C_A \Bigg( \frac{2}{\eta \epsilon}-\frac{2}{\eta}L_{\mu c}+\frac{11}{6 \epsilon}-\frac{L_{\nu c}}{\epsilon} +L_{\nu c} L_{\mu c}-\frac{11}{6}L_{\mu c}-\frac{2\pi^2}{3} \Bigg) \nn \\
    & + \frac{\varphi}{2\pi} n_f \Bigg(-\frac{1}{3\epsilon}+\frac{L_{\mu c}}{3}\Bigg) \nn \\
    & + \left(\frac{\pi-\varphi}{2\pi}\right) \left[C_A \left(-\frac{137}{18}-\frac{22}{3}\ln(2)\right)+n_f\left(\frac{23}{18}+\frac{4}{3}\ln(2)\right)\right] + C_A \frac{\pi^2}{3}\Bigg\} \nn \\
    & + \frac{\varphi}{\pi} n_f z_w\left(1-z_w\right) \nn \\
    & + \frac{\varphi}{2\pi} (C_A P_{gg}(z_w)+n_f P_{qg}(z_w)) \left(L_{\mu c}+2 \ln (1-z_w)+2 \ln \left(\frac{z_w}{1-z_w}\right)\right)\Bigg] \, ,
\end{align}
where we have introduced the logs
\begin{align}
    L_{\mu c} = 2\left(\frac{\omega \tan{\frac{r}{2}}}{\mu}\right)\,,
    \qquad
    L_{\nu c} = \ln\left(\frac{\zeta}{\nu^2}\right)\,.
\end{align}
We find that the evolution equation for the collinear functions are given by
\begin{align}
    \frac{d}{d \ln \mu} \ln C_i^{\rm SJA} \left(z_w, \bm{0}_\perp, \omega_J, r, \mu, \frac{\zeta}{\nu^2}\right) & = C_i \gamma_{C i}^{\mu\, {\rm SJA}}\left(\mu, \frac{\mu}{\nu}\right)\,,
    \\
    \frac{d}{d \ln \nu} \ln C_i^{\rm SJA} \left(z_w, \bm{0}_\perp, \omega_J, r, \mu, \frac{\zeta}{\nu^2}\right) & = C_i \gamma_{C i}^{\nu\, {\rm SJA}}\left(\mu_r, \mu\right)\,,
\end{align}
where the anomalous dimensions that enter into these expressions are given by
\begin{align}
    \gamma_{C i}^{\mu\, {\rm SJA}}\left(\mu, \frac{\mu}{\nu}\right) & = C_i \gamma^{\rm cusp} \mathcal{L}_{\nu c}+\gamma^i \,, 
    \qquad
    \gamma_{C i}^{\nu\, {\rm SJA}}\left(\kappa, \mu\right) = C_i \gamma^{\rm cusp} \mathcal{L}_{\mu c} \,.
\end{align}

\section{Example application in DIS}\label{sec:Model}
Lepton-jet correlations in the back-to-back limit have been used to probe the non-perturbative structure of TMD PDFs using a SJA in \cite{Liu:2018trl,Liu:2020dct}. In \cite{Fang:2023thw,Fang:2024auf}, it was demonstrated that employing a WTA axis can massively improve the perturbative framework, particularly the precision or range at which these structures can be studied using lepton-jet correlations. Notably, this improvement was achieved at N$^3$LL accuracy. However, this observable faced a significant limitation: the integration over the azimuthal angle in the jet function restricted the number of spin structures that could be probed.

This limitation was addressed in \cite{Kang:2021ffh}, where it was shown that considering hadron-in-jet formation substantially increases the number of accessible spin-dependent non-perturbative structures. However, the use of hadron-in-jet observables comes with a trade-off. While the number of probed spin structures increases, these observables rely on spin-dependent fragmentation functions, thereby introducing additional non-perturbative functions required for their description.

The collinear transversity PDF is particularly challenging to probe due to its chiral-odd nature. For instance, in traditional DIS, the transversity PDF would naturally couple to a chiral-even jet function, causing the observable to vanish. Recently, it was shown in \cite{Liu:2021ewb,Lai:2022aly} that the WTA jet function contains a chiral-odd component. Specifically, hadronization effects correlate the direction of the jet with the spin of the struck quark. These studies demonstrated that this observable can be used to probe the transversity TMD PDF. Since the transversity TMD PDF can be collinearly matched onto the collinear transversity PDF, this observable serves as an indirect probe of the latter. Nevertheless, directly accessing the collinear transversity PDF remains a significant challenge.

In this section, we demonstrate that azimuthal-dependent jet broadening in lepton-jet correlations in the collinear limit can serve as a direct probe of the collinear transversity PDF. Moreover, we show that the number of non-perturbative structures required to probe this distribution is very small, enabling a clean and precise determination of the collinear transversity PDF.

\subsection{Kinematics}
In this section, we will study lepton-jet production in DIS
\begin{align}
    e\left(\ell\right)+p\left(P\right) \rightarrow e\left(\ell'\right) + j\left(P_J\right)+X\,,
\end{align}
where $e$, $p$, $j$, and $X$ denote the electron, proton, jet, and unobserved states, and the momenta are labeled in the parenthesis of this expression. 

In the collider CM frame, the incoming beams have momenta
\begin{align}
    P^\mu = \sqrt{S}\,\frac{n^\mu}{2} + \mathcal{O}\left(\frac{M^2}{S}\right)\,,
    \qquad
    \ell^\mu = \sqrt{S}\,\frac{\bar{n}^\mu}{2}+ \mathcal{O}\left(\frac{m_\ell^2}{S}\right)\,,
\end{align}
where $M$ and $m_\ell$ denote the masses of the proton and the electron, respectively. The momentum of the final-state lepton can be written as
\begin{align}
    {\ell'}^\mu = \sqrt{S}\left\{1-y, \frac{{\ell_T'}^2}{S\left( 1-y\right)}, \frac{\ell'_x}{\sqrt{S}}, \frac{\ell'_y}{\sqrt{S}} \right\}\,,
\end{align}
where the event inelasticity is defined by $y = 1-P\cdot \ell'/P\cdot \ell$ and we chose the frame such that the final-state jet moves in the $x-z$ plane. The photon's momentum is given by $q = \ell-\ell'$. From this parameterization, we can define the kinematic variables
\begin{align}
    Q^2 = \frac{{\ell_T'}^2}{1-y}\,,
    \qquad
    x_B = \frac{Q^2}{2 P\cdot q}\,,
    \\
    \hat{t} = -Q^2\,,
    \qquad
    \hat{s} = x_B S\,,
    \qquad
    \hat{u} = -\hat{s}-\hat{t}\,.
\end{align}
Finally, the transverse spin vector of the incoming proton can be written as
\begin{align}
    S_T^\mu = S_T \cos{\phi_S}\,\hat{x}^\mu+S_T \sin{\phi_S}\,\hat{y}^\mu\,,
\end{align}
where $S_T$ and $\phi_S$ denote the magnitude of the transverse spin of the proton and the azimuthal angle of the polarized proton.

\subsection{Power counting and factorization}
The spin-dependent differential cross section for this process can be inferred from \cite{Kang:2021ffh}, where a TMD formalism was used. We find
\begin{align}
    \frac{d\sigma}{d^2 \ell_T'\, dy\, dz\,  d\varphi\, d\tau} =  F_{UU} + S_T \sin\left(\phi_S-\varphi\right) F_{TU}^{\sin\left(\phi_S-\varphi\right)}+...\,,
\end{align}
where $z$, $\varphi$, and $\tau$ denote the energy fraction, azimuthal angle, and broadening of the observed jet. The azimuthal angle is defined such that $\varphi = 0$ is oriented in the $x$ direction. We note, however, that for $\theta_J = \pi/2$, $\varphi$ is not defined. Additionally, the ellipsis corresponds to additional spin structures that we do not consider in this paper. Lastly, we note that the labeling conventions for the structure functions correspond to the polarization of the incoming proton and electron. Namely, the second term is associated with an unpolarized electron and a transversely polarized proton.

To obtain this expression, we have written the matrix elements for the initial state as
\begin{align}
    & \frac{1}{2 N_C} \left \langle p, S\left | \bar{\chi}_{n j}(0) \delta\left(Q- n\cdot \mathcal{P}\right) \chi_{n i}(0) \right | p, S \right \rangle \nn \\
    & \hspace{0.5in} = f_1\left(x\right)\frac{\slashed{\bar{n}}_{ij}}{4}+ g_1\left(x\right)\frac{\slashed{\bar{n}}_{ij}}{4}\gamma^5+ h_1\left(x\right)\frac{1}{4}\left(\sigma^{\mu\nu} \gamma^5 \right)_{ij} S_{T \mu} \bar{n}_\nu + \mathcal{O}\left(\frac{M}{Q}\right)\,,
\end{align}
where $f_1$ and $h_1$ are the unpolarized and transversity PDFs, while $g_1$ is the helicity PDF that we will not discuss in this paper.

\begin{figure}
    \centering
    \includegraphics[width = 0.42\textwidth,valign = c]{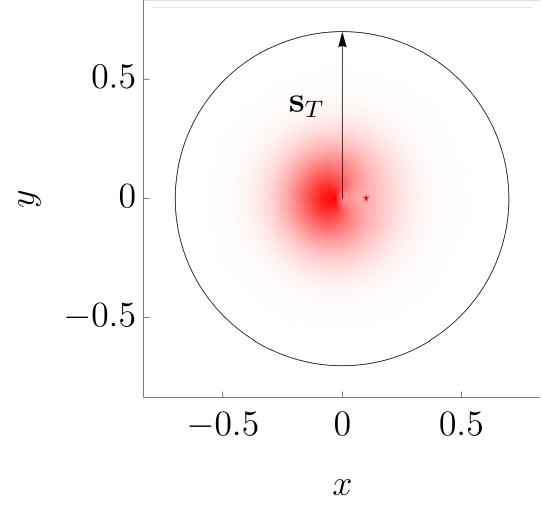}
    \hfill
    \includegraphics[width = 0.49\textwidth,valign = c]{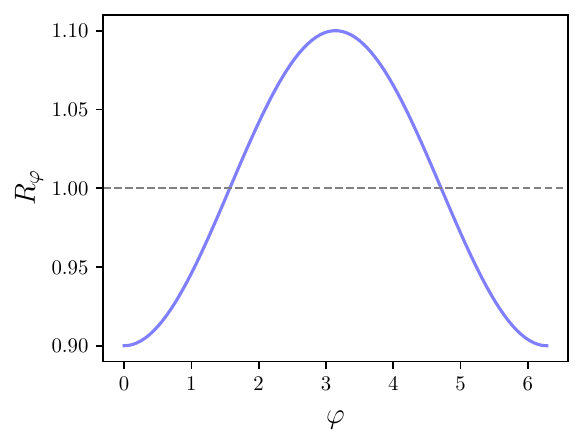}
    \caption{Left: Example of the density of hadrons in a jet with a SJA that was initiated from a transversely polarized quark. The saturation of the red indicated the density of hadrons in a particular region. The black circle represents the boundary of the jet. While transverse momentum is conserved in the jet, the Collins effect results in a higher multiplicity of hadrons on one side of the jet. The red star on the right side of the plot indicates the hardest hadron that determines the direction of the jet axis for a WTA axis. Right: Example of the broadening of the jet as a function of the azimuthal angle $\varphi$ of the jet.}
    \label{fig:Sub}
\end{figure}

The matrix elements for the final-state can be written as
\begin{align}
    \label{eq:jet-decomp}
    & \frac{1}{2 N_C} \left \langle 0 \left | \delta\left(Q - n\cdot \mathcal{P}\right)\, \delta\left(\tau- \hat{\tau}_j\right) \delta\left(z-\frac{\omega_J}{\omega}\right) \chi_{\bar{n}\, i}(0)  \right | j_\varphi \right \rangle\left \langle j_\varphi \left | \bar{\chi}_{\bar{n}\, j}(0)  \right | 0 \right \rangle \nn \\
    & = G_q\left(z,\varphi, \tau, R, \omega_J, \mu, \zeta\right) \left(\frac{\slashed{\bar{n}}_J}{4}\right)_{ij} + G_{T q}\left(z,\varphi, \tau, R, \omega_J, \mu, \zeta\right) \frac{1}{4} \left(\sigma^{\mu \nu}\right)_{ij} \bar{n}_{J \mu} \hat{\rho}_\nu + ... \,,
\end{align}
where $\hat{\rho}$ represents a vector in the azimuthal plane which can be thought of as an infinitely small wedge. A transversely polarized quark moving in the $-z$ direction will generate a pattern of hadronic radiation that resembles the left panel of Fig.~\ref{fig:Sub}. In this figure, the polarized quark is taken to be in the $y$ direction. This leads to a higher multiplicity of soft particles being generated in the $-x$ direction, while less hadron that are harder move in the positive $x$ direction. For a SJA the transverse momenta will be balanced and, thus, the anisotropic broadening will not be affected by the spin effects. For a WTA axis, however, the jet axis will be determined by the  hardest particle in the jet.  Consequently, the transverse momentum is not balanced within the jet in this case. This allows us to correlate the direction of the broadening with the spin of the proton. The second term in Eq.~(\ref{eq:jet-decomp}) does not mean that transverse momentum flows preferentially in one direction, that would violate transverse momentum conservation. Rather, it means that while transverse momentum is conserved, one side of the jet will have a higher number of hadrons, each containing a smaller transverse momentum on average, while the other side of the jet will contain a smaller number of hadrons, each containing a higher transverse momentum on average. The dots denote terms that are associated with polarized hadrons in the final-state or power suppressed terms. Lastly, we note an important point. Unlike the Collins effect, this jet function measures sums over the hadron multiplicity and is insensitive to information associated with non-perturbative collinear splitting and is, thus, less sensitive to non-perturbative effects than the Collins function.

From this expression, we can now define an asymmetry. We first want to take $\varphi = \pi$ and center the wedge in the $x$ direction. The expression for the asymmetry can be written in terms of the cross section
\begin{align}
    A_{TU}^{\sin\left(\phi_S-\varphi\right)} = \frac{F_{TU}^{\sin\left(\phi_S-\varphi\right)}}{F_{UU}}\,.
\end{align}
The unpolarized and transversely polarized structure functions take on the factorized form
\begin{align}
    F_{UU} & = \left[\hat{\sigma}_U\otimes f_1 \otimes G\right]\left(x,z,\varphi, \tau, R\right) \,,
    \\
    F_{TU}^{\sin\left(\phi_S-\varphi\right)} & = \left[\hat{\sigma}_T \otimes h_1 \otimes G_T\right] \left(x,z,\varphi, \tau, R\right) \,.
\end{align}
In these expressions the $\sigma$ terms are the partonic cross sections, their tree-level expressions are given in terms of the partonic Mandelstam variables as
\begin{align}
    \hat{\sigma}_U = \frac{\alpha_{\rm em}^2}{s {\ell_T'}^2} \frac{2\left(\hat{u}^2+\hat{s}^2\right)}{\hat{t}^2}\,,
    \qquad
    \hat{\sigma}_T = \frac{\alpha_{\rm em}^2}{s {\ell_T'}^2} \left(\frac{-4 \hat{u} \hat{s}}{\hat{t}^2} \right)\,,
\end{align}
while beyond tree-level they obey evolution equations which are complementary to those of the PDF and jet functions. Note the traditional expressions for the partonic cross sections are typically inversely proportional to $Q^2$ while the expressions listed here are inversely proportional to ${\ell_T'}^2$. This difference enters from the  $Q^2$ to ${\ell_T'}^2$ Jacobian.

In the region where $\tau \ll R$, the jet contains both hyper-collinear and soft-collinear contributions. In the region where $\tau \sim \Lambda_{\rm QCD}/Q$, there are non-perturbative contributions that we will need to parameterize. For these reasons, the jet function will take on the factorized form
\begin{align}
    G_q\left(z,\varphi, \tau, R, \mu, \zeta\right)  = & h_{jq}\left(z, \omega_J, R, \mu\right) \int d\tau_s\, d\tau_c\, d\tau_{\rm NP}\, \delta\left(\tau-\tau_c-\tau_s-\tau_{\rm NP}\right) \nn \\
    &\times C_j\left(\varphi, \tau_c, \mu, \frac{\zeta}{\nu^2}\right)\, S_j\left(\varphi, \tau_s, R, \mu, \nu\right)\, F_j\left(\varphi, \tau_{\rm NP}\right)\,.
\end{align}
The hyper-collinear and soft-collinear functions in this expression are the same as those for that were calculated previously, while the non-perturbative function $F$ will be parameterized in the next section.

For the case of a transversely polarized quark, the jet function factorizes as
\begin{align}
    G_{T q}\left(z,\varphi, \tau, R, \mu, \zeta\right)  = & h_{qq}^T\left(z, \omega_J, R, \mu\right) \int d\tau_s\, d\tau_c\, d\tau_{\rm NP}\, \delta\left(\tau-\tau_c-\tau_s-\tau_{\rm NP}\right) \nn \\
    &\times C_{T q}\left(\varphi, \tau_c, \mu, \frac{\zeta}{\nu^2}\right)\, S_q\left(\varphi, \tau_s, R, \mu, \nu\right)\, F_{T q}\left(\varphi, \tau_{\rm NP}\right)\,,
\end{align}
where we note that only $q \rightarrow q$ splitting can contribute to the observable since $q \rightarrow g$ splitting would lead to an unpolarized gluon in the jet and the asymmetry would vanish. As we will discuss in the next section, this jet function will follow a different evolution than the standard DGLAP of the unpolarized jet.

\subsection{Numerical parameterization}\label{subsec:numerics}
We follow the parameterization in Ref.~\cite{Kang:2015msa} for the transversity PDF
\begin{align}\label{eq:trans-param}
    h_{q/p}(x, Q_0) & = N_q^h x^{\alpha_q^h}(1-x)^{\beta_q^h} \frac{(\alpha_q^h+\beta_q^h)^{\alpha_q^h+\beta_q^h}}{{\alpha_q^h}^{\alpha_q^h}{\beta_q^h}^{\beta_q^h}} \frac{1}{2}\left[ f_{q/p}\left(x,Q_0 \right)+g_{q/p}\left(x,Q_0 \right)  \right] \,,
\end{align}
where $N_q^h$, $\alpha_q^h$, $\beta_q^h$ are fit parameters which were obtained in this reference for the $u$ and $d$ quarks, while the contributions of the sea quarks were set to zero. Here $Q_0 = 1$ GeV is the initial scale of the parameterization. This function evolves from the initial scale to the hard scale $Q$ by solving the evolution equation  Ref.~\cite{Stratmann:2001pt}
\begin{align}\label{eq:evo-trans}
    \frac{d }{d \ln \mu^2} h_{q/p}(x, \mu) = \frac{\alpha_s}{2\pi} \int_x^1 \frac{d\hat{x}}{\hat{x}} P_{q q}^{h}(\hat{x}) \, h_{q/p}\left(\frac{x}{\hat{x}}, \mu\right)\,,
\end{align}
where the splitting kernel for the transversity PDF is given by
\begin{align}
    P_{q q}^{h}(x) = C_F \left[\frac{2 \hat{x}}{(1-\hat{x})_+}+\frac{3}{2}\delta(1-\hat{x})\right]\,.
\end{align}
The evolution equation in Eq.~\eqref{eq:evo-trans} can be significantly simplified by taking the Mellin transform of this expression. Because this parameterization for the transversity PDF resulted in polynomial dependence on $x$ at $Q_0 = 1$ GeV, the Mellin transform for the transversity PDF initial condition can be performed analytically. As a result, evolving our parameterization for the transversity PDF from $Q_0$ can be accomplished by performing a single numerical integral which is associated with an inverse Mellin transformation.

To parameterize the non-perturbative contribution to the jet broadening, we use the simple form 
\begin{align}
    F_i\left(\varphi, u\right) = N\left(a, b, \Lambda\right) \frac{1}{\Lambda}\left(\frac{u}{\Lambda}\right)^{(a-1)}\, \exp\left[-\frac{(u-b)^2}{\Lambda^2}\right]\,,
\end{align}
where the parameter values are taken such that the non-perturbative shape function peaks in the region where $\tau \sim \Lambda_{\rm QCD}/Q$, such that in the perturbative region this contribution is small. Here, $N$ represents a normalization parameter that conserves unitarity. In this paper, we will use the parameter values from Ref.~\cite{Cao:2024ota} that $a = 1$, $b = 0.45$, and $\Lambda = 0.5$.

For the anisotropic contribution to the jet broadening, we will simply take the parameterization
\begin{align}
    F_{T q}\left(\varphi, u\right) = c_T F_i\left(\varphi, u\right)\, ,
\end{align}
where $c_T$ is some non-perturbative constant that must satisfy $c_T < 1$ by positivity. Additionally, we find that the perturbative contribution to the hyper-collinear function is identical to the one-loop matching for the Collins function.

Lastly, we note now that the jet function for the polarized case follows the same evolution as the transversity PDF. Namely, we have
\begin{align}\label{eq:evo-trans-G}
    \frac{d}{d \ln \mu^2} G_{T q}\left(z,\varphi, \tau, R, \mu, \zeta\right) = \frac{\alpha_s}{2\pi} \int_z^1 \frac{d\hat{z}}{\hat{z}} P_{qq}^{h}(\hat{z}) \, G_{T q}\left(\frac{z}{\hat{z}},\varphi, \tau, R, \mu, \zeta\right)\,.
\end{align}

\subsection{Predictions}\label{subsec:Predictions}

In the right panel of Fig.~\ref{fig:Sub}, we plot the ratio
\begin{align}
    R_\varphi = \frac{G_q\left(z,\varphi, \tau, R, \omega_J, \mu, \zeta\right) - \cos{\varphi}\, G_{T q}\left(z,\varphi, \tau, R, \omega_J, \mu, \zeta\right)}{G_q\left(z,\varphi, \tau, R, \omega_J, \mu, \zeta\right)},
\end{align}
for a quark polarized in the $y$-direction. This plot is presented as a function of the jet's azimuthal angle for an energy scale of $Q = 10$ GeV and jet radius $R = 0.8$, computed at next-to-leading logarithmic (NLL) accuracy. To illustrate the modulation effect, we use a non-perturbative constant $c_T = 0.1$, corresponding to a 10\% modulation of jet broadening in this region. However, the exact value of $c_T$ will need to be determined through a comprehensive analysis of spin-correlation experimental data, including deep inelastic scattering (DIS) and $e^+e^-$ collision data. While the Collins fragmentation function is a complex, multi-dimensional non-perturbative function that depends on both transverse and collinear dynamics, the anisotropic broadening we observe here primarily reflects transverse momentum effects. Therefore, the modulation in this ratio is driven primarily by the constant $c_T$, rather than requiring the full collinear fragmentation function.

\begin{figure}
    \centering
    \includegraphics[width = \textwidth]{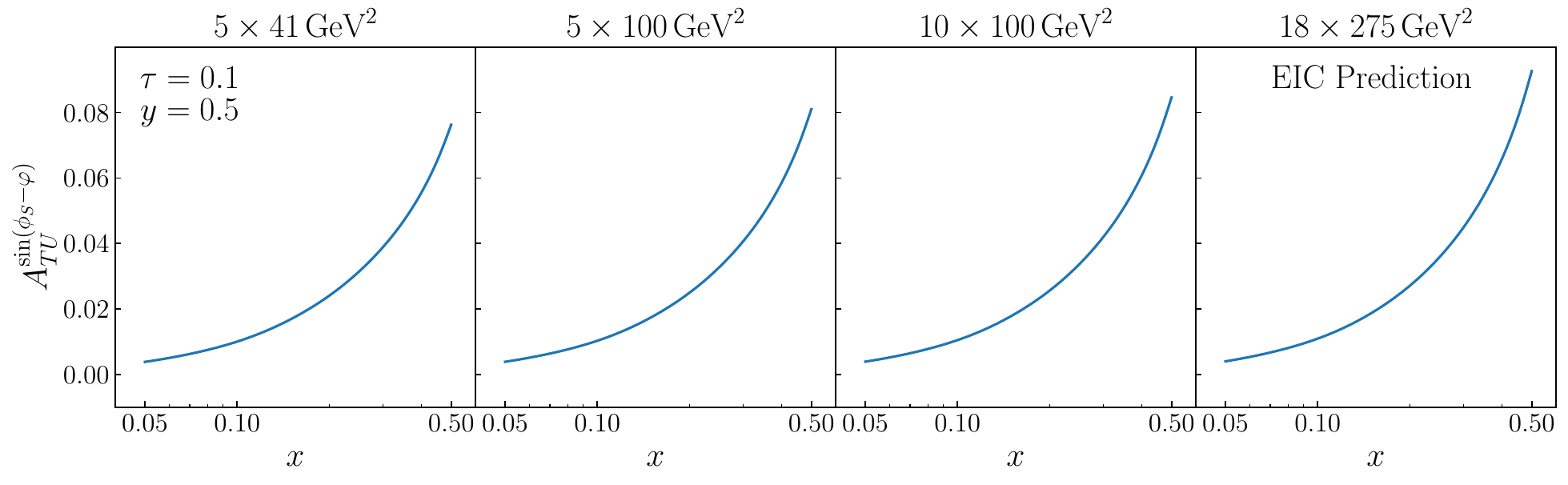}
    \caption{The transverse spin asymmetry as a function of $x$ at different lepton and proton energy configurations at the EIC.}
    \label{fig:ATU}
\end{figure}

In Fig.~\ref{fig:ATU}, we present the transverse spin asymmetry  expected to be observed at the Electron-Ion Collider (EIC) for various lepton and proton beam energy configurations. To highlight the utility of this observable in probing the transversity parton distribution function (PDF), we plot the asymmetry as a function of the Bjorken variable $x$. The asymmetries are shown for four collider setups: $5 \times 41~\mathrm{GeV}^2$, $5 \times 100~\mathrm{GeV}^2$, $10 \times 100~\mathrm{GeV}^2$, and $18 \times 275~\mathrm{GeV}^2$. For simplicity, we assume the event inelasticity to be $y = 0.5$ and the unpolarized contribution to the jet broadening to be $\tau = 0.1$. The ratios, shown from left to right, correspond to different hard scales. The behavior of the curves remains consistent across the panels, as the primary difference arises from changes in $Q$, which are driven by evolution effects. However, the differences in the DGLAP evolution between the transversity PDF and the unpolarized PDF are small, resulting in minor variations. This consistency underscores the value of this observable as a clean probe of the non-perturbative structure of the transversity PDF.

\section{Conclusions}\label{sec:Conclusions}
The primary motivation for this study was to develop a formalism for azimuthal-dependent jet substructure observables to better understand anisotropic effects in jet dynamics. Traditional analyses of jet substructure focus on isotropic observables, which average over azimuthal information and thus miss potentially critical directional dependencies. By extending jet broadening and jet shape observables into the azimuthal plane, we aim to probe phenomena such as spin correlations and medium-induced anisotropies, offering a more complete picture of jet properties and QCD dynamics.

In the course of this work, several by-products emerged for the azimuthally integrated case. We derived the semi-inclusive jet function for the Winner-Take-All axis, providing a new contribution that adds to the cases documented in the literature. Additionally, we calculated jet broadening in the fixed-order region, offering insights into the behavior of jets across different energy scales. We also developed the exclusive jet shape with the WTA axis, enhancing our understanding of isotropic angular energy distributions. These results address long-standing gaps in the theoretical description of jet substructure and establish a baseline for future studies. 

For the azimuthal-dependent case, we performed extensive calculations for jet broadening and jet shape observables. For both, we derived the jet functions in the resummed and fixed-order regions for both the Standard Jet Axis  and the WTA axis. Furthermore, we demonstrated that the jet function for the azimuthal-dependent jet shape contains non-perturbative contributions due to infrared divergences that arise in the perturbative expression. However, we showed that energy weighting removes these divergences, allowing for a well-defined theoretical treatment. Together, these calculations provide a comprehensive framework for studying azimuthal dependence in jet substructure.

As an application, we demonstrated that the azimuthal-dependent jet broadening can serve as a direct probe of the transversity parton distribution function in deep inelastic scattering. This showcases the potential of azimuthal-dependent observables to provide insight into fundamental QCD dynamics, particularly in relation to spin correlations and non-perturbative fragmentation functions, such as the Collins Fragmentation Function. This example underscores the utility of the formalism developed in this work for both theoretical exploration and experimental applications.

Looking forward, there are numerous avenues to extend and refine this work. Future studies could incorporate higher-order corrections to improve the precision and accuracy of predictions. Non-perturbative effects could be explored further, particularly in the context of medium-induced modifications in heavy-ion collisions. Additionally, extending the framework to multi-jet observables and heavy-flavor jets could broaden its applicability. Investigating the interplay between jet substructure and multi-scale phenomena, such as jet-medium interactions, represents another promising direction. These advancements will enhance the utility of azimuthal-dependent jet substructure observables, contributing to a deeper understanding of QCD and the properties of strongly interacting matter.

\section*{Acknowledgements}
We thank Zhong-bo Kang, Christopher Lee, Kyle Lee, and Duff Neill for useful discussions. We also thank the Institute for Nuclear Theory at the University of Washington for their hospitality. The work of J.T. and I.V. is supported by the US Department of Energy through the Los Alamos National Laboratory. Los Alamos National Laboratory is operated by Triad National Security, LLC, for the National Nuclear Security Administration of the U.S. Department of Energy (Contract No. 89233218CNA000001). This research is funded by LANL’s Laboratory Directed Research and Development (LDRD) program under project numbers 20220715PRD1, 20230072ER and 20240131ER. The work of W.K. is supported by the LDRD program at LANL.

\appendix
\section{Review of the exclusive and semi-inclusive jet functions with a SJA}
\begin{figure}
    \centering
    \includegraphics[width = 0.7\textwidth]{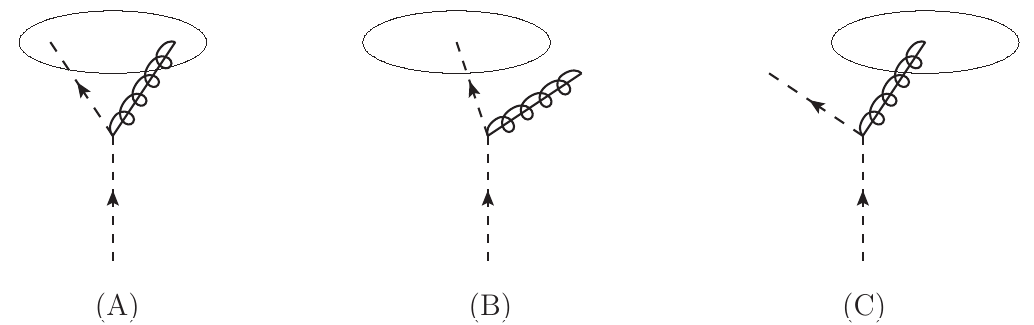}
    \caption{The graphs for the semi-inclusive quark jet function.}
    \label{fig:SiJF}
\end{figure}
In this section, we review the computation of the exclusive and semi-inclusive jet functions, providing pedagogical details and the discussing the renormalization group consistency for the jet substructure observables.

The semi-inclusive jet function can be expressed in a perturbative series as
\begin{align}
    J_{i\, {\rm alg}}^{\rm SJA}\left(z, \omega_J, R, \mu\right) = \sum_l \left(\frac{\alpha_s}{2\pi}\right)^l J_{i\, {\rm alg}}^{{\rm SJA}\, (l)}\left(z, \omega_J, R, \mu\right)\,.
\end{align}
In this paper, we are interested in performing the computation at one-loop. At one-loop, there are three combination associated with the permutations of the final-state partons enter or exiting the jet. The one-loop Feynman diagrams associated with the semi-inclusive jet function are given in Fig.~\ref{fig:SiJF}. Thus, the one-loop expression for the jet function read
\begin{align}
    J_{i\,\rm{alg}}^{{\rm SJA}(1)}\left(z, \omega_J, R, \mu\right) = \sum_{G} J_{i\,\rm{alg}}^{{\rm SJA}(G)}\left(z, \omega_J, R, \mu\right)\,,
\end{align}
where $\rm{G}$ is the contribution associated with the graphs in Fig.~\ref{fig:SiJF}. Following the notation of Sec.~\ref{subsec:kinematics}, the exact expressions for these graphs are given in this frame by 
\begin{align}
    J_{i\,\rm{alg}}^{{\rm SJA}(A)}\left(z, \omega_J, R, \mu\right) & = \delta\left(1-z\right)\int \Phi_{l q}\, \delta\left(\omega_J-\bar{n}_J\cdot l\right)\,\delta^{d-2}\left(\bm{l}_\perp\right)\, \left | \overline{\mathcal{M}}\right|_i^2\,  \Theta_{{\rm alg}\, (jk)}^{\rm SJA}\,, \\
    J_{i\,\rm{alg}}^{{\rm SJA}(B)}\left(z, \omega_J, R, \mu\right) & = \int \Phi_{l q}\, \delta\left(x-z\right)\, \delta\left(\omega_J-\bar{n}_J\cdot l\right)\,\delta^{d-2}\left(\bm{l}_\perp\right)\, \left | \overline{\mathcal{M}}\right|_i^2\,  \Theta_{{\rm alg}\, (j)}^{\rm SJA}\,, \\
    J_{i\,\rm{alg}}^{{\rm SJA}(C)}\left(z, \omega_J, R, \mu\right) & = \int \Phi_{l q}\, \delta\left(1-x-z\right)\, \delta\left(\omega_J-\bar{n}_J\cdot l\right)\,\delta^{d-2}\left(\bm{l}_\perp\right)\, \left | \overline{\mathcal{M}}\right|_i^2\,  \Theta_{{\rm alg}\, (k)}^{\rm SJA}\,,
\end{align}
where the amplitudes are given in Eqs.~\eqref{eq:Mq} and \eqref{eq:Mg} and $i$ denotes the flavor of the initial parton, while $j$ and $k$ denote the flavors of the final-state partons. Here, $x$ denotes the momentum fraction of parton $j$, while the jet constraints are given in the Heaviside theta functions. In the case of a jet cone algorithm, the angle of each final-state parton is compared against the jet radius. Their exact expressions are given by
\begin{align}
    \Theta_{{\rm cone} (jk)}^{\rm SJA} &=  \Theta\left(\omega_J\, x\, \tan{\frac{R}{2}}-q_\perp\right)\,\Theta\left(\omega_J\, (1-x)\tan{\frac{R}{2}} - q_\perp\right)\,,\\
    \Theta_{{\rm cone} (j)}^{\rm SJA} &=  \Theta\left(\omega_J\, x\, \tan{\frac{R}{2}} - q_\perp\right)\,\Theta\left(q_\perp-\omega_J\, (1-x)\tan{\frac{R}{2}}\right)\,,\\
    \Theta_{{\rm cone} (k)}^{\rm SJA} &=  \Theta\left(q_\perp - \omega_J\, x\, \tan{\frac{R}{2}}\right)\,\Theta\left( \omega_J\, (1-x)\tan{\frac{R}{2}} - q_\perp\right)\,,
\end{align}
where $q_\perp$ denotes the transverse momentum of each parton with respect to the jet axis. For an anti-$\rm{k_T}$ algorithm, the angle of the final-state partons must be smaller than the jet radius. This amounts to the jet constraints
\begin{align}
    \Theta_{{\rm{k_T}} (jk)}^{\rm SJA} &= \Theta\left(\omega_J\, x\, (1-x)\tan{\frac{R}{2}} - q_\perp\right)\,,\\
    \Theta_{{\rm{k_T}} (j)}^{\rm SJA} &= \Theta_{{\rm{k_T}} (k)}^{\rm SJA} = 1-\Theta_{{\rm{k_T}} (jk)}^{\rm SJA}\, .
\end{align}
The expressions for the graphs can then be shown to take on the form
\begin{align}
J_{i\,\rm{alg}}^{{\rm SJA}(A)}(z, \omega_J, R, \mu) & = \delta(1-z) \sum_{j,k} \int d\Phi \hat P_{(jk)i}(x, q_\perp, \epsilon) \Theta_{{\rm alg}\, (jk)}^{\rm SJA}\,,
\\
J_{i\,\rm{alg}}^{{\rm SJA}(B)}(z, \omega_J, R, \mu) & = \sum_{j,k} \int d\Phi \hat P_{(jk)i}(x, q_\perp, \epsilon) \delta\left(z-x\right)\Theta_{{\rm alg}\, (j)}^{\rm SJA}\,,
\\
J_{i\,\rm{alg}}^{{\rm SJA}(C)}(z, \omega_J, R, \mu) & = \sum_{j,k} \int d\Phi \hat P_{(jk)i}(x, q_\perp, \epsilon) \delta\left(1-z-x\right) \Theta_{{\rm alg}\, (k)}^{\rm SJA}\, . 
\end{align}
The sum in these expressions is used to account for the fact that the observable is insensitive to the type of radiation into the final-state. We have also used the short hand notation for the two dimensional integration
\begin{align}
    \int d\Phi = \int d^{d-2}q_\perp dx\,,
\end{align}
and we define the splitting kernels to be
\begin{align}
\hat P_{(q'g)q}(x, q_\perp, \epsilon) =& \delta_{q' q}\frac{C_F}{\pi} \left(\pi \mu^2 e^{\gamma_E}\right)^\epsilon \frac{1}{q_\perp^2} \left[\frac{1+x^2}{1-x} - \epsilon (1-x)\right]\,  \, ,\\
\hat P_{(gg)q}(x, q_\perp, \epsilon) =& 0 \, , \\
\hat P_{(q\bar{q})g}(x, q_\perp, \epsilon) =& \frac{n_f}{\pi} \left(\pi \mu^2 e^{\gamma_E}\right)^\epsilon \frac{1}{q_\perp^2} \left[\frac{1}{2}-\frac{x(1-x)}{1-\epsilon}\right] \, ,  
\\
\hat P_{(gg)g}(x, q_\perp, \epsilon) =& \frac{C_A}{\pi} \left(\pi \mu^2 e^{\gamma_E}\right)^\epsilon \frac{1}{q_\perp^2} \left[\frac{x}{1-x} + \frac{1-x}{x} + x(1-x)\right]\,,
\end{align}
where the splitting function for $q \rightarrow gg$ does not enter until two loops. The exclusive jet function can be computed from graph (A) via the relation
\begin{align}
    J_{i\,{\rm alg}}^{\rm SJA}(\omega_J, R, \mu) = \int_0^1 dz J_{i\,{\rm alg}}^{{\rm SJA} (A)}(z, \omega_J, R, \mu)\,.
\end{align}
The simplified expressions for the quark initiated jet function using an anti-$k_T$ algorithm are given by
\begin{align}
    J_{q\,{\rm alg}}^{{\rm SJA} (A)}\left(z,\omega_J, R, \mu\right) = & C_F\delta\left(1-z\right)\left[\frac{1}{\epsilon ^2}-\frac{L_R}{\epsilon}+\frac{3}{2
   \epsilon}+\frac{1}{2} L_R^2 -\frac{3}{2}L_R+c_{q\, {\rm alg}}^{\rm SJA}\right]\,, \\
    J_{q\,{\rm alg}}^{{\rm SJA} (B)}\left(z, \omega_J, R, \mu\right)  = & C_F\Bigg[-\delta(1-z)\left(\frac{1}{\epsilon^2}-\frac{L_R}{\epsilon}+\frac{3}{2 \epsilon}+\frac{1}{2}L_R^2-\frac{3}{2}L_R-\frac{\pi^2}{12}\right) \nn \\
    & +P_{qq}(z)\left(\frac{1}{\epsilon}-L_R-2\left(\ln(1-z)\right)_+\right) +z-1\Bigg] \,,  \\
   J_{q\,{\rm alg}}^{{\rm SJA} (C)}\left(z, \omega_J, R, \mu\right)  = & C_F\left[P_{gq}(z) \left(\frac{1}{\epsilon}-L_R-2\left(\ln{\left(1-z\right)}\right)_+\right)-z\right]  \,,  
\end{align}
where the algorithmic dependence of the jet function is encoded into the coefficients
\begin{align}
    c_{q\, {\rm k_T}}^{\rm SJA} = \frac{13}{2}-\frac{3 \pi^2}{4}\,,
    \qquad
    c_{q\, {\rm Cone}}^{\rm SJA} = \frac{7}{2}-\frac{5\pi^2}{12}+3\ln(2)\, .
\end{align}
We  have  introduced for notational convenience the logarithm
\begin{align}
    L_R = \ln\left(\frac{\omega_J^2 \tan^2{\frac{R}{2}}}{\mu^2}\right)\,,
\end{align}
and we define the splitting functions as
\begin{align}
    P_{qq}(x) & = \left[\frac{1+x^2}{(1-x)_+}+\frac{3}{2}\delta(1-z) \right]\,,
    \\
    P_{gq}(x) & = \left[\frac{1+(1-z)^2}{z}\right]\,,
    \\
    P_{qg}(x) & = \left[1-2x(1-x) \right]\,, 
    \\
    P_{gg}(x) & = 2\left[x(1-x)+\frac{1-x}{x}+\frac{x}{1-x}\right]+\frac{\beta_0}{2 C_A}\delta(1-x)\,. 
\end{align}
For the gluon initiated jets, we have
\begin{align}
    J_{g\,{\rm alg}}^{{\rm SJA} (A)}\left(z, \omega_J, R, \mu\right) 
     =  &C_A\delta\left(1-z\right)\left[\frac{1}{\epsilon^2}-\frac{L_R}{\epsilon}+\frac{\beta_0}{2 C_A \epsilon}+\frac{L_R^2}{2}-\frac{\beta_0}{2 C_A}L_R\right] \nn \\
    &  +\frac{\alpha_s}{2\pi}\delta(1-z) c_{g\, {\rm alg}}^{\rm SJA} \,, \\
    J_{g\,{\rm alg}}^{{\rm SJA} (B)}\left(z, \omega_J, R, \mu\right)  = & \frac{C_A}{2}\Bigg[-\delta\left(1-z\right)\left(\frac{1}{\epsilon ^2}-\frac{L_R}{\epsilon}+\frac{\beta_0}{2 C_A \epsilon}+\frac{L_R^2}{2}-\frac{\beta_0}{2 C_A}L_R-\frac{\pi^2}{12}\right) \nn \\ 
    & + P_{gg}(z)\left(\frac{1}{\epsilon}-L_R-2 \ln(1-z)_+\right)\Bigg] \nn \\
    & + \frac{n_f}{2}\Bigg[P_{qg}(z)\left(\frac{1}{\epsilon }-L_R-2\ln (1-z)\right)-z(1-z)\Bigg] \, ,  \\
    J_{g\,{\rm alg}}^{{\rm SJA} (C)}\left(z, \omega_J, R, \mu\right) = & J_{g\,{\rm alg}}^{(B)}\left(z, \omega_J,R, \mu\right) \, . 
\end{align}
The algorithm-dependent coefficients are given by
\begin{align}
    c_{g\, {\rm k_T}}^{\rm SJA} = & C_A\left(-\frac{3\pi^2}{4}+\frac{67}{9}\right)-n_f \left(\frac{23}{18}\right)\,,
    \\
    c_{g\, {\rm Cone}}^{\rm SJA} = & C_A\left(-\frac{5\pi^2}{12}+\frac{137}{36}+\frac{11}{3}\ln(2)\right)-n_f \left(\frac{23}{36}+\frac{2}{3}\ln(2)\right)\,.
\end{align}
For the WTA jet, we find
\begin{align}
    J_i^{{\rm WTA}\, (G)}\left(z, \omega_J, R, \mu\right) = J_{i\, {\rm Cone}}^{{\rm SJA}\, (G)}\left(z, \omega_J, R, \mu\right)\,,
\end{align}
where $G \in (A, B, C)$. 
The renormalized exclusive jet functions contain double poles that lead to a Sudakov resummation. Thus, the renormalized exclusive jet function obeys the evolution equation
\begin{align}
    \frac{d}{d\ln\mu}\ln J_{i\, {\rm alg}}^{{\rm axis}}\left(\omega_J, R, \mu\right) = \gamma^J_i\left(\mu\right)\,, \nn
\end{align}
where the anomalous dimensions are given by
\begin{align}
    \label{eq:anom-jet}
    \gamma^J_i\left(\mu\right) = - C_i \gamma^{\rm cusp}_i \left[\alpha_s(\mu)\right]L_R -\gamma^i\left[\alpha_s(\mu)\right]\,.
\end{align}
For the semi-inclusive jet function, the double poles of graph (A) are cancelled against those of graphs (B) and (C). In this case, the divergences which generate the evolution of the jet function are multiplied by the splitting functions, resulting in a DGLAP evolution equation. In Mellin space, this function obeys the equation
\begin{align}
    \frac{d}{d \ln \mu} \mathcal{J}_{i\, {\rm alg}}^{{\rm axis}}\left(N,\omega_J, \mu\right) = \frac{\alpha_s}{\pi} c_{ij} \mathcal{P}_{ij}(N) \mathcal{J}_{j\, {\rm alg}}^{\rm axis}\left(N,\omega_J, \mu\right)\,,
\end{align}
where the color factors are given by
\begin{align}
    c_{qq} = C_F\,,
    \qquad
    c_{gq} = C_F\,,
    \qquad
    c_{qg} = n_f\,,
    \qquad
    c_{gg} = C_A\,.
\end{align}
\section{Central sub-jet functions}
In Ref.~\cite{Kang:2017mda} the authors derived the expressions for the semi-inclusive central sub-jet function. However we found that the exclusive central sub-jet functions were not present in the literature. We include them here
\begin{align}
    j_{q\, {\rm k_T}}^{{\rm SJA}} \left(z_r,r, R,\mu, \zeta\right) = & \frac{\alpha_s C_F}{2\pi} \Bigg[ \delta\left(1-z_r\right) \Bigg( \frac{1}{\epsilon ^2} +\frac{3}{2 \epsilon } -\frac{L_r}{\epsilon } -\frac{L_{r/R}}{\epsilon }\nn \\
    & +L_R L_{r/R}+\frac{L_r^2}{2}-\frac{3 L_r}{2}-\frac{L_{r/R}^2}{2}-\frac{3
   L_{r/R}}{2}-\frac{5 \pi ^2}{12}+\frac{7}{2}+3 \ln (2)\Bigg) \nn \\
   & + (P_{gq}(z_r)+P_{qq}(z_r)) \Bigg(2 \ln \left(\frac{z_r}{1-z_r}\right) \theta \left(-\frac{r}{R}-z_r+1,z_r-\frac{1}{2}\right) \nn \\
   & +(L_{r/R}+2 \ln (z_r)) \theta \left(\frac{r}{R}+z_r-1\right)\Bigg) \Bigg]  \,,
   \\
    j_{g\, {\rm k_T}}^{{\rm SJA}} \left(z_r,r, R,\mu, \zeta\right) = & \frac{\alpha_s}{2\pi} \Bigg[ C_A \delta\left(1-z_r\right) \Bigg( \frac{1}{\epsilon ^2} +\frac{11}{6 \epsilon }-\frac{L_{r/R}}{\epsilon } -\frac{L_r}{\epsilon } \nn \\
    & +L_R L_{r/R}+\frac{L_r^2}{2}-\frac{11 L_r}{6}-\frac{L_{r/R}^2}{2}-\frac{11 L_{r/R}}{6}-\frac{5 \pi ^2}{12}+\frac{137}{36}+\frac{11 \ln (2)}{3}\Bigg) \nn \\
    & + n_f \delta\left(z_r\right) \Bigg( -\frac{1}{3\epsilon} +\frac{L_r}{3}+\frac{L_{r/R}}{3}-\frac{22}{36}-\frac{2}{3}\ln(2)\Bigg) \nn \\
   & + (C_A P_{gg}(z_r)+n_f P_{qg}(z_r)) \Bigg( L_{r/R}+2\ln{z_r}\Bigg)\Theta\left(\frac{r}{R}+z_r-1\right) \Bigg] \,,
\end{align}
\begin{align}
    j_{q\, {\rm Cone}}^{{\rm SJA}} \left(z_r,r, R,\mu, \zeta\right) = & \frac{\alpha_s C_F}{2\pi} \Bigg[ \delta\left(1-z_r\right) \Bigg( \frac{1}{\epsilon ^2} +\frac{3}{2 \epsilon } -\frac{L_r}{\epsilon } -\frac{L_{r/R}}{\epsilon }\nn \\
    & +L_R L_{r/R}+\frac{L_r^2}{2}-\frac{3 L_r}{2}-\frac{L_{r/R}^2}{2}-\frac{3
   L_{r/R}}{2}-\frac{5 \pi ^2}{12}+\frac{7}{2}+3 \ln (2)\Bigg) \nn \\
   & + (P_{qq}(z_r)+P_{gq}(z_r)) \Bigg( L_{r/R} \left(z_r-\frac{R}{r+R}\right) \nn \\
   & + 2\ln\left(\frac{z_r}{1-z_r}\right) \Theta\left(\frac{R}{r+R}-z_r, z_r-\frac{1}{2}\right)\Bigg) \Bigg]  \,,
   \\
    j_{g\, {\rm Cone}}^{{\rm SJA}} \left(z_r,r, R,\mu, \zeta\right) = & \frac{\alpha_s}{2\pi} \Bigg[ C_A \delta\left(1-z_r\right) \Bigg( \frac{1}{\epsilon ^2} +\frac{11}{6 \epsilon }-\frac{L_{r/R}}{\epsilon } -\frac{L_r}{\epsilon } \nn \\
    & +L_R L_{r/R}+\frac{L_r^2}{2}-\frac{11 L_r}{6}-\frac{L_{r/R}^2}{2}-\frac{11 L_{r/R}}{6}-\frac{5 \pi ^2}{12}+\frac{137}{36}+\frac{11 \ln (2)}{3}\Bigg) \nn \\
    & + n_f \delta\left(z_r\right) \Bigg( -\frac{1}{3\epsilon} +\frac{L_r}{3}+\frac{L_{r/R}}{3}-\frac{23}{36}-\frac{2}{3}\ln(2)\Bigg) \nn \\
   & + (C_A P_{gg}(z_r)+n_f P_{qg}(z_r)) \Bigg( L_{r/R} \Theta\left(z_r-\frac{R}{r+R}\right) \nn \\
   & +2\ln\left(\frac{z_r}{1-z_r}\right)\Theta\left(\frac{R}{r+R}-z_r, z_r-\frac{1}{2}\right)\Bigg) \Bigg]  \,,
\end{align}
\begin{align}
    j_{q}^{{\rm WTA}} \left(z_r,r, R,\mu, \zeta\right) = & \frac{\alpha_s C_F}{2\pi} \Bigg[ \delta\left(1-z_r\right) \Bigg( \frac{1}{\epsilon ^2} +\frac{3}{2 \epsilon } -\frac{L_r}{\epsilon } -\frac{L_{r/R}}{\epsilon }\nn \\
    & +L_R L_{r/R}+\frac{L_r^2}{2}-\frac{3 L_r}{2}-\frac{L_{r/R}^2}{2}-\frac{3
   L_{r/R}}{2}-\frac{5 \pi ^2}{12}+\frac{7}{2}+3 \ln (2)\Bigg) \nn \\
   & + (P_{gq}(z_r)+P_{qq}(z_r)) \Theta\left(z_r-\frac{1}{2}\right)L_{r/R} \Bigg]  \,,
   \\
    j_{g}^{{\rm WTA}} \left(z_r,r, R,\mu, \zeta\right) = & \frac{\alpha_s}{2\pi} \Bigg[ C_A \delta\left(1-z_r\right) \Bigg( \frac{1}{\epsilon ^2} +\frac{11}{6 \epsilon }-\frac{L_{r/R}}{\epsilon } -\frac{L_r}{\epsilon } \nn \\
    & +L_R L_{r/R}+\frac{L_r^2}{2}-\frac{11 L_r}{6}-\frac{L_{r/R}^2}{2}-\frac{11 L_{r/R}}{6}-\frac{5 \pi ^2}{12}+\frac{137}{36}+\frac{11 \ln (2)}{3}\Bigg) \nn \\
    & + n_f \delta\left(z_r\right) \Bigg( -\frac{1}{3\epsilon} +\frac{L_r}{3}+\frac{L_{r/R}}{3}-\frac{22}{36}-\frac{2}{3}\ln(2)\Bigg) \nn \\
   & + (C_A P_{gg}(z_r)+n_f P_{qg}(z_r)) \Theta\left(z_r-\frac{1}{2}\right)L_{r/R} \Bigg]  \,.
\end{align}
\section{Anomalous dimensions up to NNLL}
In this section, we provide the expressions for the cusp, non-cusp, and rapidity anomalous dimensions up to NNLL. These can be expressed as a perturbative series in the strong coupling as
\begin{align}
    \gamma^{\rm cusp}\left[\alpha_s(\mu)\right] = \sum_{n = 0}^\infty \left(\frac{\alpha_s}{4\pi}\right)^{n+1}\gamma^{\rm cusp}_n \, ,  
    \qquad
    \gamma^i\left[\alpha_s(\mu)\right] = \sum_{n = 0}^\infty \left(\frac{\alpha_s}{4\pi}\right)^{n+1}\gamma_n^i\,.
\end{align}
At NNLL, the cusp and non-cusp terms are given by
\begin{align}
\label{eq:c1}
  \gamma_{0}^{\rm cusp} = & \, 4 \,,
  \\
  \gamma_{1}^{\rm cusp} = & \, C_A  \left(\frac{268}{9}-8 \zeta_2\right)-\frac{40  n_f}{9} \,,
  \\
  \gamma_{2}^{\rm cusp} = & \, C_A^2  \left(-\frac{1072 \zeta_2}{9}+\frac{88
      \zeta_3}{3}+88 \zeta_4+\frac{490}{3}\right) 
    +C_A  n_f \left(\frac{160 \zeta_2}{9}-\frac{112
    \zeta_3}{3}-\frac{836}{27}\right) \nn \\
    & + C_F n_f \left(32 \zeta_3-\frac{110}{3}\right)-\frac{16
                 n_f^2}{27} \,.
\end{align}
The quark and gluon anomalous dimensions up to two loops are
\begin{align}
  \gamma^q_0 = & \, -3 C_F \,,
  \\
  \gamma^q_q = & \, C_A C_F \left(-11 \zeta_2+26
    \zeta_3-\frac{961}{54}\right)
+C_F^2 \left(12 \zeta_2-24 \zeta_3-\frac{3}{2}\right)+C_F n_f \left(2 \zeta_2+\frac{65}{27}\right) \,,
\\
  \gamma^g_0 = &\, - \beta_0 \,,
\\
  \gamma^g_1 = &\, \,C_A^2 \left(\frac{11 \zeta_2}{3}+2 \zeta_3-\frac{692}{27}\right)+C_A n_f \left(\frac{128}{27}-\frac{2 \zeta_2}{3}\right)+2 C_F n_f \,.
\end{align}
Similarly, the Collins-Soper anomalous dimension of the TMDs can be written as
\begin{align}
    \gamma_\zeta(\mu,b) = -2 C_F \int_{\mu_b}^\mu \frac{d\mu'}{\mu'}\gamma_{\rm cusp}\left[\alpha_s\left(\mu'\right)\right]-C_F \gamma^r\left[\alpha_s(\mu_b)\right]\,,
\end{align}
where $\gamma^r$ is the rapidity anomalous dimension. This anomalous dimension can be expressed at a perturbative series as
\begin{align}
    \gamma^r\left[\alpha_s(\mu)\right] = \sum_{i = 0}^\infty \left(\frac{\alpha_s}{4\pi}\right)^{i+1}\gamma^r_i\,,
\end{align}
where at NNLL, the anomalous dimension are while the soft anomalous dimensions are
\begin{align}
    \gamma_{0}^r = & \, 0 \,, \\
    \gamma_{1}^r = & \, C_A \left(\frac{22 \zeta_2}{3}+28 \zeta_3-\frac{808}{27}\right)+ n_f \left(\frac{112}{27}-\frac{4 \zeta_2}{3}\right)- 2 \zeta_2 \beta_0 \,. 
\end{align}

\bibliography{main}

\bibliographystyle{JHEP}
\end{document}